\documentclass[a4paper,12pt]{article}

\usepackage[colorlinks=true,allcolors=black]{hyperref}

\makeatletter
	\@addtoreset{equation}{section}
	\renewcommand{\thefigure}{\thesection.\arabic{figure}}\@addtoreset{figure}{section}
	\renewcommand{\thetable}{\thesection.\arabic{table}}\@addtoreset{table}{section}
\makeatother

\makeatletter
\let\MYcaption\@makecaption
\makeatother

\usepackage{subcaption}

\makeatletter
\let\@makecaption\MYcaption
\makeatother

\usepackage{amsmath,amsthm,amssymb,mathrsfs,amsfonts,mathtools,color,physics,url,listings,graphicx,tikz,here,multirow,array,ulem,bm,cite,hyperref}

\usetikzlibrary{calc,decorations.markings}
\tikzset{->-/.style={decoration={markings,mark=at position .3 with {\arrow[scale=1.35]{latex}}},postaction={decorate}}}
\tikzset{-<-/.style={decoration={markings,mark=at position .5 with {\arrow{latex reversed}}},postaction={decorate}}}

\allowdisplaybreaks[4]

\begin{document}

\begin{titlepage}
\begin{flushright}
TIT/HEP-687\\
November,  2021
\end{flushright}
\vspace{0.5cm}
\begin{center}
{\Large \bf 
    Wall-crossing of TBA equations and WKB periods for the third order ODE
}

\lineskip .75em
\vskip 2.5cm

{\large  Katsushi Ito$^{a,}$\footnote{ito@th.phys.titech.ac.jp}, Takayasu Kondo$^{a,}$\footnote{t.kondo@th.phys.titech.ac.jp} 
and  Hongfei Shu$^{b,c,}$\footnote{shuphy124@gmail.com } 
}
\vskip 2.5em
 {\normalsize\it 
 $^{a}$Department of Physics, Tokyo Institute of Technology,
Tokyo, 152-8551, Japan\\
$^{b}$Beijing Institute of Mathematical Sciences and Applications (BIMSA), Beijing, 101408, China\\
$^{c}$Yau Mathematical Sciences Center (YMSC), Tsinghua University, Beijing, 100084, China
}

\vskip 3.0em
\end{center}
\begin{abstract}
We study the WKB periods for the third order ordinary differential equation (ODE) with polynomial potential, which is obtained by the Nekrasov-Shatashvili limit of ($A_2,A_N$) Argyres-Douglas theory in the Omega background. In the minimal chamber of the moduli space, we derive the Y-system and the  thermodynamic Bethe ansatz (TBA) equations by using the ODE/IM correspondence. The exact WKB periods are identified with the Y-functions. Varying the moduli  parameters of the potential, the wall-crossing of the TBA equations occurs. 
We study the process of the wall-crossing from the minimal chamber to the maximal chamber for $(A_2,A_2)$ and $(A_2,A_3)$. When the potential is a monomial type, we show the TBA equations obtained from the ($A_2, A_2$) and ($A_2, A_3$)-type ODE lead to the $D_4$ and $E_6$-type TBA equations respectively.

\end{abstract}

\end{titlepage}

\newpage
\baselineskip=0.6cm

\section{Introduction}
The ODE/IM correspondence \cite{Dorey:1998pt,Bazhanov:1998wj} shows a non-trivial relation between the spectral problem of the ordinary differential equation (ODE) and the functional relations in the quantum integrable model (IM).
This has been first studied for the Schr\"odinger equation with monomial potential.
A generalization to the higher order ODE with monomial potential has been studied in\cite{Dorey:1999pv,Dorey:2000ma,Suzuki:1999hu,Bazhanov:2001xm,Dorey:2006an}.
The Wronskians of the basis of the solutions at different singularities of the ODE provide the Baxter's Q-functions, while the solutions at infinity define the T-functions as well as the Y-functions as the cross ratios of the T-functions.

The higher order ODE studied in this correspondence is also obtained from the conformal limit of the linear problem associated with the modified affine Toda field equation\cite{Lukyanov:2010rn,Dorey:2012bx,Ito:2013aea,Adamopoulou:2014fca}. 
In particular, for a simply-laced Lie algebra, the Bethe ansatz equations obtained from the ODE can be transformed into the Non-linear integral equation (NLIE), from which we can identify the effective central charge of the corresponding CFT\cite{Dorey:2006an,Ito:2018wgj,Ito:2020htm}.

The ODE/IM correspondence has been recently generalized to the second order ODE of the Schr\"odinger type with generic polynomial potential \cite{Ito:2018eon}.
In particular, the Y-functions of the integrable model can be identified with the exponential of the exact WKB periods (the Voros symbols), which share the same asymptotics and the discontinuity in the complexified Plank constant plane. One can determine the quantum periods directly by 
the Thermodynamic Bethe ansatz (TBA) equations \cite{zamo-ADE} satisfied by the Y-functions. See also \cite{Gaiotto:2014bza,Grassi:2019coc,Fioravanti:2019vxi,Ito:2020ueb,Imaizumi:2020fxf,Emery:2020qqu,Imaizumi:2021cxf,Grassi:2021wpw,gabai2021exact}, for more general potential.

The WKB periods of the second order ODE with a polynomial potential also appear as the quantum Seiberg-Witten (SW) periods of the Argyres-Douglas (AD) theories defined in the Nekrasov-Shatashvili limit of the 
Omega background\cite{Nekrasov:2009rc}, where the SW curve is quantized with the Omega background parameter as the Planck constant.
The TBA equation is a very useful tool \cite{Ito:2017ypt} to study the non-perturbative structure of the strong-coupled field theories and their relation to 2d theories \cite{Beem:2013sza,Cordova:2015nma,Xie:2016evu}.
The wall-crossing of the BPS spectrum is quite important for studying strong coupling physics.
Interestingly, the TBA equations also change according to the wall-crossing and  describe different integrable models in each chamber of the moduli space. Then we need to explore the whole structure of the TBA systems in the modulis space. In particular, the superconformal point of the AD theory is in the maximal chamber, where the corresponding integrable model is that obtained from  the 4d/2d correspondence \cite{Ito:2017ypt}.
Based on the singularity structure at the superconformal point in the moduli space, the equivalences between $(A_1, A_r)\sim (A_r, A_1)$, $(A_2, A_2)\sim D_4$ and $(A_2, A_3)\sim E_6$ AD theories have been found in \cite{Cecotti:2010fi}. However, the quantum SW curves of these equivalent AD theories at Omega background appear in a very different form.  It is not obvious to see these dualities in the Omega background. In this work, we will see this equivalence by investigating the TBA equations which describe the quantum periods. 

The wall-crossing of the TBA equations were also studied in the context of four-dimensional gauge theories \cite{Gaiotto:2009hg}, the cluster algebra \cite{Cecotti:2010fi} and the gluon scattering amplitudes \cite{Alday:2010vh,Hatsuda:2010cc}.
This phenomenon is also understood as the cluster mutation of the Voros symbols
for the Schr\"odinger type ODE \cite{iwaki2014exact}. 

So far, the wall-crossing in the context of gauge theory is most well studied in the rank one case \cite{Ferrari:1996sv,Gaiotto:2009hg,Maruyoshi:2013fwa,Longhi:2016rjt}\footnote{See also \cite{Chen:2011gk,Gaiotto:2012db} for the higher rank case.}, while the TBA equations and their wall-crossing have been studied for the second order ODE.  It is interesting to explore the similar relation for the higher order ODE in order to study more general AD theories and the integrable models. For recent developments, see \cite{Neitzke:2017yos,Hollands:2019wbr,Hollands:2021itj,Fioravanti:2019awr, Yan:2020kkb, Dumas:2020zoz}\footnote{In \cite{Dumas:2020zoz}, the authors have studied the $(A_2, A_2)$ case numerically using the approach of \cite{Gaiotto:2009hg}.}. In a previous paper \cite{Ito:2021boh}, we studied the WKB expansion of the higher order ODE and found the relation between the WKB periods  and the Y-functions associated with the TBA equations.
In particular, we considered the case of the quadratic potential, which is found to correspond to the $(A_r,A_1)$-type TBA system. We have also shown that the wall-crossing phenomena occur for the third order ODE with cubic potential. 

In the case of the Schr\"odinger type ODE, the correspondence has been studied in \cite{Ito:2018eon}, where the TBA equations change when the parameters of the potential cross the wall of marginal stability. 
The approach of the ODE/IM correspondence provides a concrete method to study the 
wall-crossing phenomena and its relation to integrable models characterized by the TBA equations.
In this paper, we will work out the third order ODE in detail by taking the examples of cubic and quartic polynomial potential, which show an essential feature of the wall-crossing of general polynomial potential. When the potential is a monomial type, we show the TBA equations obtained from the ($A_2, A_2$) and ($A_2, A_3$)-type ODE lead to the $D_4$ and $E_6$-type TBA equations, respectively, by tracking the wall-crossing. In the context ${\cal N}=2$ theories in four dimensions, this relation is understood as the duality of $(A_2, A_2)$ and $(D_4,A_1)$ or $(A_2,A_3)$ and $(E_6,A_1)$ theories which share the same singularity point  \cite{Cecotti:2010fi,Xie:2012hs}.

This paper is organized as follows:
In section \ref{sec:Third_order_ODE}, we study the WKB solution of the third order ODE and the WKB periods based on the differential operators. We also comment on the Borel summability and the discontinuity of the WKB period.
In section \ref{sec:Yfunc_and_period}, we introduce the Y-functions from the Wronskians of the subdominant solutions of the ODE and compute their classical limit from the Stokes graph.
We then propose a relation between the Y-functions and the WKB periods.
In section \ref{sec:TBA_in_minimal}, we construct the TBA equations satisfied by the Y-functions, where 
the associated WKB periods are in the minimal chamber. We will check the relation numerically for cubic and quartic potentials.
In section \ref{sec:wc}, we will consider the wall-crossing of the TBA equations.
We will work out numerically for the cubic and quartic potentials to test the identification between WKB periods and Y-functions. In particular, we construct the TBA equations in the maximal chamber, which includes the monomial potential.
We find that these TBA equations at monomial potential are equivalent to the $D_4$ and $E_6$-type TBA equations, respectively. In section \ref{sec:con-dis}, we present the conclusions and discussion. In Appendix \ref{sec:D4_E6_TBA}, we present the $D_4$ and $E_6$ type TBA equations based on the scattering theories. In Appendix \ref{sec:newY-A2A3}, we show the definitions of the new Y-functions of $(A_2, A_3)$ case for completeness.
In Appendix \ref{sec:TBA-max}, we show the twelve TBA equations in the maximal chamber for $(A_2, A_3)$ case.

\section{Third order ODE}\label{sec:Third_order_ODE}
\subsection{WKB analysis}\label{sec:WKB_ana}
In this section, we  study the WKB analysis of the third order ODE defined in the complex plane:
\begin{align}
\left(\epsilon^3\frac{d^3}{dx^3}+p(x)\right)\psi(x)=0,
\label{eq:3rdode}
\end{align}
where $\epsilon$ is a complex parameter. $p(x)$ is a polynomial in $x$ of order $N+1$:
\begin{align}
    p(x)=u_0 x^{N+1}+u_1 x^{N}+\dots +u_{N+1},
    \label{eq:polyn}
\end{align}
where $u_i$ $(i=0,\dots,N+1)$ are complex parameters.
The ODE (\ref{eq:3rdode}) is regarded as the quantum SW curve of the AD theory of $(A_2,A_{N})$-type defined in the Nekrasov-Shatashvili limit of the Omega-background. 
We also refer it as the $(A_2,A_N)$-type ODE. 
In a previous paper \cite{Ito:2021boh}, we have studied the $N=1$ case.
We consider the WKB solution of the (\ref{eq:3rdode}) of the form
\begin{align}
    \psi(x)&=\exp\left(\frac{1}{\epsilon} \int^x P(x') dx'\right),\quad
    P(x)=\sum_{n=0}^{\infty}\epsilon^n p_n(x).
\end{align}
$P(x)$ satisfies the Riccatti equation
\begin{align}
    p(x)+P^3+3\epsilon P P'+\epsilon^2 P^{(2)}=0.
    \label{eq:ode3ricatti}
\end{align}
Substituting the $\epsilon$-series of $P(x)$, 
we can determine $p_n$ $(n\geq 1)$ recursively by
\begin{equation}
    \begin{aligned}
        p_0&=(-p)^{\frac{1}{3}},\\
        p_n&=-{\frac{1}{3p_0^2}}\qty[ p_0 \sum_{i=1}^{n-1} p_{n-i}p_i +\sum_{i=1}^{n-1}\sum_{j=0}^{n-i} p_{n-i-j}p_i p_j+3 \sum_{i=0}^{n-1} p_{n-1-i}p'_i+p''_{n-2} ], \quad n\geq 1. 
    \end{aligned}
\end{equation}
The first few examples of $p_n$ are found to be
\begin{align}
    p_1&=-\frac{p_0'}{p_0},\\
    p_2&=-\frac{(p_0')^2}{p_0^3}+\frac{2}{3}\frac{p_0''}{p_0^2},\\
    p_3&=-\frac{2(p_0')^3}{p_0^5}+\frac{2 p'_0 p''_0}{p_0^4}-\frac{1}{3}\frac{p^{(3)}_0}{p_0^3}, \\
    p_4&=-\frac{4 (p'_0)^4}{p_0^7}+\frac{16}{3}\frac{(p'_0)^2p''_0}{p_0^6}
    -\frac{2}{3}\frac{(p''_0)^2}{p_0^5}-\frac{10}{9}\frac{p'_0p^{(3)}_0}{p_0^5}
    +\frac{1}{9}\frac{p^{(4)}_0}{p_0^4}.
\end{align}
Explicit calculation of $p_n$, we observe that $p_n$ for odd $n$ takes the form of total derivatives, which is the same as the Schr\"odinger equation. 
Moreover $p_{6i+4}$ ($i=0,1,2,\dots$) also become total derivatives.
We introduce the WKB curve $\Sigma$:
\begin{align}
    y^3&=-p(x).
    \label{eq:wkbcurve1}
\end{align}
This is the SW curve of $(A_2,A_N)$-type AD theory. On the curve \eqref{eq:wkbcurve1}, there is a basis of meromorphic differentials \cite{Farkas_1992}
\begin{align}
    \frac{x^{i-1} dx}{y} , \quad \frac{x^{i-1}dx}{y^2}, \quad i=1,\dots, N-1,
    \label{eq:basis1}
\end{align}
in which $\frac{x^{i-1}}{y}dx$ ($\frac{x^{i-1}}{y^2}dx$) for $i<\frac{N+1}{3}$ ($i<\frac{2N+2}{3}$) are holomorphic differentials. The number of the holomorphic differentials determines the genus of the WKB curve, which is equal to $N$ for $N\equiv 0,1\ ({\rm mod}\;3)$ or $N-1$ for $N\equiv 2\ ({\rm mod} \;3)$.
For an one-cycle $\gamma$ on the curve $\Sigma$, one defines the WKB period
\begin{align}
    \Pi_\gamma(\epsilon)&=\int_\gamma P(x)dx.
    \label{eq:wkbperiod1}
\end{align}
The WKB period is expanded in $\epsilon$:
\begin{align}
     \Pi_\gamma(\epsilon)&=\sum_{n=0}^{\infty}\epsilon^n \Pi_\gamma^{(n)},
     \label{eq:epseries1}
\end{align}
where
\begin{align}
    \Pi_\gamma^{(n)}=\int_\gamma p_n(x) dx.
\end{align}
The corrections $\Pi^{(n)}$ vanish for odd $n$ and $n=6i+4$ ($i=0,1,\dots$) since the corresponding $p_n$ is the total derivative.

Now we compute the quantum corrections to the periods by using the differential operators.
Since the differential $p_n dx$ for $n=6i$ or $n=6i+2$ defines a meromorphic differential on the WKB curve, one can express it as a linear combination of the basis \eqref{eq:basis1}:
\begin{align}
    p_n dx&=\sum_{a=1}^{2}\sum_{i=1}^{N-1} B^{(n)}_{ai}\frac{x^{i-1}}{y^a}dx+d(*),
    \label{eq:pf1}
\end{align}
where $B^{(n)}_{ai}$ is a function of $u_i$'s. $d(*)$ denotes a total derivative term\footnote{See \cite{Ito:2018eon} for the similar expression in the case of the second order ODE.}. In (\ref{eq:pf1}), $n$ dependence appears only in the coefficients, from which one can see the asymptotic series structure of the WKB periods. We can evaluate the period integrals of $p_n$ in terms of those of the basis (\ref{eq:basis1}). It is useful since in the direct integration of $p_n$ it is necessary to regularize divergence of $p_n$ near the branch points. 

For example, for the curve with $p(x)$ given by (\ref{eq:polyn}), it is found that
\begin{align}
    p_0 dx&=y dx=-\frac{1}{3+N+1}\sum_{i=2}^{N+1}
    \qty(  i u_i -\frac{(N+2-i) u_{i-1} u_1}{(N+1) u_0}) \frac{x^{N+1-i}}{y^{2}}dx+d(*).
    \label{eq:scaling1}
\end{align}
Here the last term is the total derivatives.
From the expansion \eqref{eq:pf1}, the quantum periods are expressed as
\begin{align}
    \Pi_\gamma(\epsilon)&=\sum_{a=1}^{2}\sum_{i=1}^{N-1} B_{ai}(\epsilon)
    (\Pi_{ai})_{\gamma},
\end{align}
where
\begin{align}
    B_{ai}(\epsilon)=\sum_{n=0}^{\infty} B^{(n)}_{ai}\epsilon^n
    \label{eq:epseries2}
\end{align}
and $(\Pi_{ai})_{\gamma}$ are the period integrals of the basis (\ref{eq:basis1}): 
\begin{align}
    (\Pi_{ai})_{\gamma}&=\int_\gamma \frac{x^{i-1}dx}{y^a}.
    \label{eq:pint_mdiff}
\end{align}
For example, from (\ref{eq:scaling1}), we find that $B_{1i}^{(0)}=0$ and
\begin{align}
B_{2 i}^{(0)}&=-\frac{1}{N+4} \qty( (N+2-i) u_{N+2-i} -\frac{i u_{N+1-i} u_1}{(N+1) u_0}).
\end{align}
We then introduce the Seiberg-Witten differentials $y^a dx$ ($a=1,2$), which generate the basis of meromorphic differentials:
\begin{align}
    \partial_{u_i}y^a dx&=-\frac{a}{3}\frac{x^{N+1-i}}{y^{3-a}}dx.
    \label{eq:swd2}
\end{align}
We also define the SW periods
\begin{align}
    (\hat{\Pi}_{a})_{\gamma}=\int_\gamma y^a dx.
    \label{eq:SW_period}
\end{align}
Here the classical WKB period corresponds to the SW period with $a=1$: $\Pi^{(0)}_\gamma=(\hat{\Pi}_1)_{\gamma}$.
From (\ref{eq:swd2}), we obtain
\begin{align}
    (\Pi_{ai})_\gamma&=-\frac{a}{3} \partial_{u_{N+2-i}} \hat{\Pi}_{3-a\gamma}.
\end{align}
Finally, the quantum correction is expressed by the classical SW periods by acting the differential operator with respect to the moduli $u_i$:
\begin{align}
    \Pi_\gamma^{(n)}&=\sum_{a=1}^{2} \mathcal{O}_{a}^{(n)} \hat{\Pi}_{a\gamma},
\end{align}
where
\begin{align}
    \mathcal{O}_{a}^{(n)} &=-\sum_{i=1}^{N-1}B^{(n)}_{3-a N+2-i}\frac{3}{a} \partial_{u_{N+2-i}}.
\end{align}
We refer $\mathcal{O}_{a}^{(n)}$'s as the Picard-Fuchs operators.
Then one can compute the higher order corrections in the WKB expansion from the classical SW periods.

Since the r.h.s. of \eqref{eq:epseries2} is an asymptotic series in $\epsilon$, the Borel resummation is necessary. The Borel resummed period defines an analytic function on the complex $\epsilon$-plane, which has singularities and discontinuities. 
The aim of the present work is to explore its analytic structure using the TBA equations, which we will study in the following sections.

We will explain quantum corrections by two examples: $(A_2,A_2)$ and $(A_2, A_3)$-type ODEs.
We first consider the $(A_2,A_2)$-type ODE \eqref{eq:3rdode} with the potential
\begin{align}
    p(x)=u_0 x^3+u_1 x^2+u_2 x+u_3.
\end{align}
The WKB curve has the genus $g=1$, where only $\frac{dx}{y^2}$ is a holomorphic differential.
One can compute the corrections as follows:
\begin{equation}
    \begin{aligned}
        \Pi_{\gamma}^{(2)}&=\mathcal{O}_{2}^{(2)}\hat{\Pi}_{2\gamma}=-\sum_{i=1}^{2}B^{(2)}_{2i}3\partial_{u_{4-i}} \hat{\Pi}_{2\gamma},\\ 
        \Pi_{\gamma}^{(6)}&=\mathcal{O}^{(6)}_{1}\hat{\Pi}_{1\gamma}=-B^{(6)}_{11}\frac{3}{2}\partial_{u_{3}} \hat{\Pi}_{1\gamma},
    \end{aligned}
\end{equation}
where the coefficients in the Picard-Fuchs operators can be expressed in $u_i$. They are simplified as follows:
\begin{align}
    B^{(2)}_{21}&=\frac{1}{18}\frac{D_0}{\Delta}\frac{u_1}{3},\qquad\qquad B^{(2)}_{22}=\frac{1}{18}\frac{D_0}{\Delta}u_0,\\
    B^{(6)}_{11}&=-\frac{1}{174960}\frac{D_0}{\Delta^4}\qty(21983 D_0^4 - 823446 u_0^2 D_0^2 \Delta + 6633171 u_0^4\Delta^2).
\end{align}
Here $\Delta$ is the discriminant of the WKB curve
\begin{equation}
    \Delta=- u_1^2u_2^2+4 u_3 u_1^3+4 u_0 u_2^3-18 u_0 u_2 u_3 u_1+27 u_0^2 u_3^2,
\end{equation}
$D_0$ is defined by
\begin{equation}
    D_0=2 u_1^3-9 u_0 u_2 u_1+27 u_0^2 u_3. 
    \label{eq:PF_D0}
\end{equation}
The PF operators contain a common factor $D_0$.
It is interesting to note that $D_0=0$ for $u_1=u_3=0$, i.e. $p(x)=u_0 x^3+u_2 x$ and hence the quantum corrections become zero. We have also confirmed that some higher order terms vanish.  This implies that the classical periods give the exact result, where a similar phenomenon happens in the harmonic potential for the Schr\"odinger equation. This result is shown to be consistent with  the TBA equations as we will see in sect. 4.2.1. 

Next we consider the $(A_2,A_3)$-type ODE with
\begin{equation}
   p(x)=u_0 x^4+u_1 x^3+u_2 x^2+u_3 x+u_4. 
\end{equation}
The second order corrections are given by
\begin{align}
    \Pi_{\gamma}^{(2)}&=-\sum_{i=1}^{3}B^{(2)}_{2i}3\partial_{u_{5-i}} \hat{\Pi}_{2\gamma},
\end{align}
where
\begin{align}
    B^{(2)}_{21}=&\frac{1}{18\Delta} \Bigl(2 u_1^2 u_2^4 - 8 u_0 u_2^5 - 9 u_1^3 u_2^2 u_3 + 40 u_0 u_1 u_2^3 u_3 + 4 u_0 u_1^2 u_2 u_3^2 - 78 u_0^2 u_2^2 u_3^2 \nonumber\\
    &+ 9 u_0^2 u_1 u_3^3 + 27 u_1^4 u_2 u_4 - 150 u_0 u_1^2 u_2^2 u_4 + 160 u_0^2 u_2^3 u_4 + 9 u_0 u_1^3 u_3 u_4 \nonumber\\  
    &+112 u_0^2 u_1 u_2 u_3 u_4+ 72 u_0^3 u_3^2 u_4 + 72 u_0^2 u_1^2 u_4^2 - 512 u_0^3 u_2 u_4^2\Bigr),\\
    B^{(2)}_{22}=&\frac{1}{18\Delta}\Bigl( 6 u_1^3 u_2^3 - 24 u_0 u_1 u_2^4 - 27 u_1^4 u_2 u_3 + 118 u_0 u_1^2 u_2^2 u_3 + 8 u_0^2 u_2^3 u_3 + 21 u_0 u_1^3 u_3^2 \nonumber\\ 
    &- 270 u_0^2 u_1 u_2 u_3^2 + 108 u_0^3 u_3^3 + 81 u_1^5 u_4 - 450 u_0 u_1^3 u_2 u_4 + 464 u_0^2 u_1 u_2^2 u_4 \nonumber\\ 
    & + 588 u_0^2 u_1^2 u_3 u_4 - 288 u_0^3 u_2 u_3 u_4 - 960 u_0^3 u_1 u_4^2\Bigr), \\
    B^{(2)}_{23}=&-\frac{5u_0}{18\Delta}\Bigl(  -2 u_1^2 u_2^3 + 8 u_0 u_2^4 + 9 u_1^3 u_2 u_3 - 40 u_0 u_1 u_2^2 u_3 - 3 u_0 u_1^2 u_3^2 + 72 u_0^2 u_2 u_3^2 \nonumber\\ 
    &- 27 u_1^4 u_4 + 144 u_0 u_1^2 u_2 u_4 - 128 u_0^2 u_2^2 u_4 - 192 u_0^2 u_1 u_3 u_4 + 384 u_0^3 u_4^2\Bigr).
\end{align}
Here $\Delta$ is the discriminant of the curve
\begin{equation}
    \begin{aligned}
        \Delta&=u_1^2 u_2^2 u_3^2-4 u_0 u_2^3 u_3^2-4 u_1^3 u_3^3+18 u_0 u_1 u_2 u_3^3-27 u_0^2 u_3^4 - 4 u_1^2 u_2^3 u_4 \\
        &+ 16 u_0 u_2^4 u_4+ 18 u_1^3 u_2 u_3 u_4 - 80 u_0 u1 u_2^2 u_3 u_4 - 6 u_0 u_1^2 u_3^2 u_4 + 144 u_0^2 u_2 u_3^2 u_4\\  
        &-  27 u_1^4 u_4^2 + 144 u_0 u_1^2 u_2 u_4^2 - 128 u_0^2 u_2^2 u_4^2 - 192 u_0^2 u_1 u_3 u_4^2 + 256 u_0^3 u_4^3.
    \end{aligned}
\end{equation}
Higher order corrections can be calculated in a similar way.

\subsection{Classical period}
\begin{figure}[htbp]
    \centering
        \begin{tikzpicture}
        \def\ra{0.6};
        \def\d{0.6};
        \def\dd{0.07};
        \def\lgth{-2.0};
        \def\dr{0.3};
        
        \coordinate[] (x7) at (-8*\d,0);
        \coordinate[] (x6) at ($({-6*\d},0)+({\dr*cos(-150)},{\dr*sin(-150)})$);
        \coordinate[] (x5) at ($({-3*\d},0)+({\dr*cos(-135)},{\dr*sin(-135)})$);
        \coordinate[] (x4) at ($({-1.5*\d},0)+({\dr*cos(135)},{\dr*sin(135)})$);
        \coordinate[] (x3) at ($({1.5*\d},0)+({\dr*cos(-30)},{\dr*sin(-30)})$);
        \coordinate[] (x2) at ($({3*\d},0)+({\dr*cos(0)},{\dr*sin(0)})$);
        \coordinate[] (x1) at ($({6*\d},0)+({\dr*cos(-45)},{\dr*sin(-45)})$);
        \coordinate[] (x0) at (8*\d,0);
        
        \coordinate[] (l7) at (-8*\d+\lgth/2,0);
        \coordinate[] (l6) at ({-6*\d+\dr*cos(-150)-1},\lgth);
        \coordinate[] (l5) at ({-3*\d+\dr*cos(-135)-0.5},\lgth);
        \coordinate[] (l4) at ({-1.5*\d+\dr*cos(135)-0.2},\lgth);
        \coordinate[] (l3) at ({1.5*\d+\dr*cos(-30)+0.2},\lgth);
        \coordinate[] (l2) at ({3*\d+\dr+0.5},\lgth);
        \coordinate[] (l1) at ({6*\d+\dr*cos(-45)+1},\lgth);
        \coordinate[] (l0) at (8*\d-\lgth/2,0);
        
        \coordinate[] (x3a) at ($(x3)+(0,\ra)$);
        \coordinate[] (x3r) at ($(x3)+(\ra,0)$);
        \coordinate[] (x4a) at ($(x4)+(0,\ra)$);
        \coordinate[] (x4l) at ($(x4)+(-\ra,0)$);
        
        \coordinate[] (x3p) at ($(x3)+({\ra*cos(-83.8)},{\ra*sin(-83.8)})+(\dd,0)$);
        \coordinate[] (x3n) at ($(x3)+({\ra*cos(-83.8)},{\ra*sin(-83.8)})+(-\dd,0)$);
        \coordinate[] (x4p) at ($(x4)+({\ra*cos(-96.4)},{\ra*sin(-96.4)})+(\dd,0)$);
        \coordinate[] (x4n) at ($(x4)+({\ra*cos(-96.4)},{\ra*sin(-96.4)})+(-\dd,0)$);

        \fill (x0) circle [radius=2pt];\draw (x0) node[above]{$x_0$};
        \fill (x1) circle [radius=2pt];\draw (x1) node[above]{$x_1$};
        \draw ({5*\d},-0.3) node[above]{$\cdots$};
        \fill (x2) circle [radius=2pt];\draw ($(x2)+(0.2,0)$) node[above]{$x_{k-2}$};
        \fill (x3) circle [radius=2pt];\draw (x3) node[above]{$x_{k-1}$};
        \fill (x4) circle [radius=2pt];\draw (x4) node[above]{$x_k$};
        \fill (x5) circle [radius=2pt];\draw ($(x5)+(-0.2,0)$) node[above]{$x_{k+1}$};
        \draw ({-4.8*\d},-0.43) node[above]{$\cdots$};
        \fill (x6) circle [radius=2pt];\draw (x6) node[above]{$x_{N-1}$};
        \fill (x7) circle [radius=2pt];\draw (x7) node[above]{$x_N$};
        
        \draw[line width=0.4mm, gray] (l7) -- (x7);
        \draw[line width=0.4mm, gray] (l6) -- (x6);
        \draw[line width=0.4mm, gray] (l5) -- (x5);
        \draw[line width=0.4mm, gray] (l4) -- (x4);
        \draw[line width=0.4mm, gray] (l3) -- (x3);
        \draw[line width=0.4mm, gray] (l2) -- (x2);
        \draw[line width=0.4mm, gray] (l1) -- (x1);
        \draw[line width=0.4mm, gray] (l0) -- (x0);
        
        \draw[] (0,0.5) node[above]{$\gamma_{a,k}$};
        
        \draw[line width=0.3mm, red] (x4n) to [out=172.6, in=270] (x4l);
        \draw[line width=0.3mm, red] (x4l) to [out=90, in=180] (x4a);
        \draw[line width=0.3mm, red,->-] (x4a) to [out=0, in=185.2] (x3n);
        
        \draw[line width=0.3mm, dashed, red] (x3p) to [out=7.2, in=270] (x3r);
        \draw[line width=0.3mm, dashed, red] (x3r) to [out=90, in=0] (x3a);
        \draw[line width=0.3mm, dashed, red,->-] (x3a) to [out=180, in=-5.6] (x4p);
        
    \end{tikzpicture}
    \caption{The choice of the branch cuts and the definition of the cycle $\gamma_{a,k}$, where the solid and dotted lines are on the $a$-th and $a+1$-th sheet respectively.}
    \label{fig:branchcut}
\end{figure}
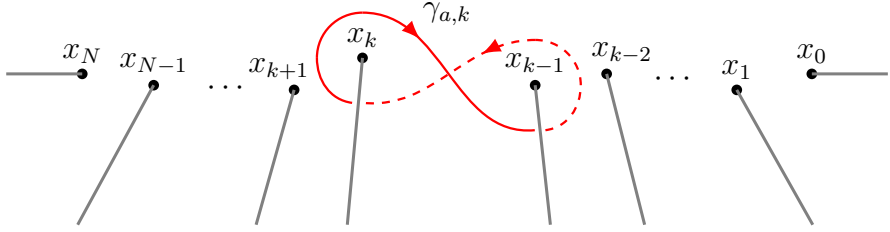

We have seen that the WKB periods are expressed as the linear combinations of the period integrals of the meromorphic differentials defined in \eqref{eq:pint_mdiff}, that are expressed by the first-order derivatives of the classical SW periods \eqref{eq:SW_period}.
The periods are also specified by one-cycles on the WKB curve, which are determined by the branch points $x_k$ ($k=0,1,\dots,N$), i.e. $p(x_k)=0$. 
In this paper, we consider the WKB curve whose branch points are distinct.
The classical SW periods \eqref{eq:SW_period} are then given by
\begin{equation}
    (\hat{\Pi}_{a})_{\gamma}=\int_\gamma u_0^{\frac{a}{3}}\prod_{k=0}^N(x-x_k)^{\frac{a}{3}} \dd x,\qquad a=1,2.
\end{equation}
To specify the cycle $\gamma$, we have to choose the branch cut.
We label the branch points such that $\Re(x_0)\geq\Re(x_1)\geq\cdots\geq\Re(x_N)$.
Then we choose the branch cut to be as in figure \ref{fig:branchcut}.
We introduce the one-cycle $\gamma_{l,k}$ ($l=1,2,3,$ $k=1,\dots,N$) on $\Sigma$, where $\gamma_{l,k}$ encircles the branch points $x_{k-1}$ anticlockwise and $x_k$ clockwise, respectively, on $l$-th and $(l+1)$-th sheets.
Here we defined the sheet $l$ on which $y=y_l$:
\begin{equation}
    y_l\coloneqq e^{\frac{2\pi i}{3}l}p(x)^{\frac{1}{3}},\qquad l=1,2,3.
\end{equation}
With this choice of the branch cuts and the other notation,  the classical SW period $(\hat{\Pi}_a)_{\gamma_{l,k}}$ $(a=1,2,\;l=1,2,3,\;k=1,2,\dots,N)$ is 
\begin{equation}
    \begin{aligned}
        (\hat{\Pi}_a)_{\gamma_{l,k}}&= \qty(e^{\frac{2\pi i}{3}l}-e^{\frac{2\pi i}{3}(l+1)})\oint_{\gamma_{l,k}} p(x)^{\frac{a}{3}}\dd x\\
        &= -2ie^{\frac{\pi i}{3}(2l+1)a}u_0^{\frac{a}{3}}\sin\frac{\pi a}{3} \int_{x_{k}}^{x_{k-1}} \prod_{i=0}^N(x-x_i)^{\frac{a}{3}}\dd x,
    \end{aligned}
    \label{eq:hypergeometric_int}
\end{equation}
which is the generalized hypergeometric integral.
Let us evaluate the integral of $(\hat{\Pi}_a)_{\gamma_{l,k}}$ for the $N=2$ and $N=3$ cases.

We first consider the $N=2$ case, where branch points are $x_0,x_1,x_2$ and
the cycles are $\gamma_{l,1}$ and $\gamma_{l,2}$.
By using the fractional linear transformation, the SW classical periods \eqref{eq:hypergeometric_int} for $N=2$ are evaluated in terms of hypergeometric function as
\begin{equation}
    \begin{aligned}
        (\hat{\Pi}_a)_{\gamma_{l,k}}=&-2ie^{\frac{\pi i}{3}(2l+2k)a+\delta(k,a)}d_k^{-2-\frac{2}{3}a}\qty(x_{k+1}-x_{k-1})^{\frac{a}{3}}u_0^{\frac{a}{3}}\sin\frac{\pi a}{3}\\
        &\qquad \cdot B\qty(1+\frac{a}{3},1+\frac{a}{3}){}_2F_1\qty(2+a,1+\frac{a}{3};2+\frac{2a}{3};z_k^{(1)}),\qquad a=1,2,
    \end{aligned}
    \label{eq:classical_period_A2A2}
\end{equation}
where 
\begin{equation}
    d_k=\sqrt{\frac{x_{k+1}-x_{k-1}}{(x_{k}-x_{k-1})(x_{k}-x_{k+1})}}.
\end{equation}
$z_k^{(1)}$ ($k=1,2$) are defined by
\begin{equation}
    z_1^{(1)}\coloneqq\frac{x_{1}-x_{0}}{x_{2}-x_{0}},\qquad
    z_2^{(1)}\coloneqq\frac{x_{2}-x_{1}}{x_{0}-x_{1}}.
    \label{eq:def_z_k^1}
\end{equation}
$\delta(k,a)$ is a phase factor that ensures the integrand of \eqref{eq:hypergeometric_int} to be the principal value. 
$B(a,b)$ and ${}_2F_1(a,b;c;z)$ represent the Beta function and the hypergeometric function, respectively, that are defined by
\begin{equation}
    \begin{aligned}
        &B(a,b)=\int_0^1 t^{a-1}(1-t)^{b-1}\dd t,\qquad \Re(a),\;\Re(b)>0,\\
        &{}_2F_1(a,b;c;z)=\frac{1}{B(b,c-b)}\int_0^1\frac{t^{b-1}(1-t)^{c-b-1}}{(1-zt)^a}\dd t,\qquad \abs{z}\leq1. 
    \end{aligned}
\end{equation}
Next, we consider the $N=3$ case where the branch points are $x_0,x_1,x_2,x_3$ and the cycles are $\gamma_{l,1},\gamma_{l,2}$ and $\gamma_{l,3}$.
In the same way, one can evaluate the SW classical periods:
\begin{equation}
    \begin{aligned}
        \hat{\Pi}_{\gamma_{l,k}}^{a(0)}=&-2ie^{\frac{\pi i}{3}(2l+2k)a+\delta(k,a)}d^{-2-\frac{4}{3}a}(x_{k+1}-x_{k-1})^{\frac{a}{3}}(x_{k}-x_{k+2})^{\frac{a}{3}} \\
        &\quad\cdot B\qty(1+\frac{a}{3},1+\frac{a}{3}) F_1\qty(1+\frac{a}{3},2+\frac{4a}{3},-\frac{a}{3},2+\frac{2a}{3};z_k^{(1)},z_k^{(2)}),\quad a=1,2,
    \end{aligned}
\end{equation}
where $z_k^{(1)}$ $(k=1,2,3)$ are defined by
\begin{equation}
    z_1^{(1)}\coloneqq\frac{x_{1}-x_{0}}{x_{2}-x_{0}},\qquad
    z_2^{(1)}\coloneqq\frac{x_{2}-x_{1}}{x_{3}-x_{1}},\qquad
    z_3^{(1)}\coloneqq\frac{x_{3}-x_{2}}{x_{0}-x_{2}}.
\end{equation}
$z_k^{(2)}$ is defined by
\begin{equation}
    z_k^{(2)}\coloneqq\frac{(x_{k}-x_{k-1})(x_{k+1}-x_{k+2})}{(x_{k+1}-x_{k-1})(x_k-x_{k+2})},\qquad k=1,2,3,
\end{equation}
and $F_1(a,b,b^\prime,x,z^{(1)},z^{(2)})$ is the Appell hypergeometric function which is given by
\begin{equation}
    F_1(a,b,b^{\prime},c;z^{(1)},z^{(2)}) = \frac{1}{B(a,c-a)}\int_0^1 \frac{t^{a-1}(1-t)^{c-a-1}}{(1-z^{(1)} t)^{b}(1-z^{(2)} t)^{b^{\prime}}}\dd t.
\end{equation}

\subsubsection{Classical periods for the monomial potential}\label{sec:period-mono}
In this subsection, we summarize the classical periods for the monomial potential of degree $N+1$, which is the most symmetric and suitable potential to see the dualities of ODE.
Since the potential is now expressed as
\begin{equation}
    p(x)=u_0x^{N+1}+u_{N+1}=u_0(x^{N+1}-u),\qquad u=-\frac{u_{N+1}}{u_0},
\end{equation}
we can label the zero points as
\begin{equation}
    \begin{cases}
        x_{2k-1}=u^{\frac{1}{N+1}}e^{\frac{2\pi ik}{N+1}},& k=1,2,\dots,\lfloor \frac{N+1}{2}\rfloor,\\
        x_{2k}=u^{\frac{1}{N+1}}e^{-\frac{2\pi ik}{N+1}}, & k=0,1,2,\dots,\lfloor \frac{N}{2}\rfloor.
    \end{cases}
\end{equation}
The period integrals along the cycles $\gamma_{a,k}$ take simpler form 
for these branch points, where
one can compute the period integral \eqref{eq:pint_mdiff} directly:
\begin{equation}
    \begin{cases}
        (\Pi_{ai})_{\gamma_{l,2k-1}}=\sum_{j=1}^{k-1}\qty(I_{l,j}^{a,i}+I_{l,N+2-j}^{a,i})+I_{l,k}^{a,i}, & k=1,2,\dots,\lfloor \frac{N+1}{2}\rfloor,\\
        (\Pi_{ai})_{\gamma_{l,2k}}=-\sum_{j=1}^{k}\qty(I_{l,j}^{a,i}+I_{l,N+2-j}^{a,i}), & k=1,2,\dots,\lfloor \frac{N}{2} \rfloor.
    \end{cases}
    \label{eq:classical_period_mono} 
\end{equation}
Here $I_{l,j}^{a,i}$ ($j=1,2,\dots,N$) is defined by the integral along the cycle which encircles the zero points at $u^{\frac{1}{N+1}}e^{\frac{2\pi i(j-1)}{N+1}}$ anti-clockwise and $u^{\frac{1}{N+1}}e^{\frac{2\pi ij}{N+1}}$ clockwise on the $a$-th and $a+1$-th sheets:
\begin{equation}
    \begin{aligned}
        I_{l,j}^{a,i}&=\qty( e^{-\frac{2\pi i}{3}al}-e^{-\frac{2\pi i}{3}(l+1)a} )
        \int_{u^{\frac{1}{N+1}}e^{\frac{2\pi i j}{N+1}}}^{u^{\frac{1}{N+1}}e^{\frac{2\pi i (j-1)}{N+1}}}
        \frac{x^{i-1}}{(u_0x^{N+1}-u)^\frac{a}{3}}dx\\
        &=\frac{4}{N+1}e^{2\pi i\qty(-\frac{1}{3}a(l+\frac{1}{2})+\frac{j-1}{N+1}i+\delta(l,j))}\sin\frac{\pi a}{3}\sin\frac{\pi i}{N+1}u_0^{-\frac{i}{N+1}}u_{N+1}^{\frac{i}{N+1}-\frac{a}{3}}B\qty(\frac{i}{N+1},1-\frac{a}{3}),
    \end{aligned}
\end{equation}
where the phase $\delta(l,j)=(\delta_{2,j}+\delta_{3,j})\frac{l}{3}$ which ensures the integrand to be principal value.

\subsection{Borel resummation and discontinuity}\label{sec:Borel_resum}
The WKB period \eqref{eq:epseries1} is an asymptotic formal series in $\epsilon$, which has a factorial growth. To promote the series, we consider the Borel transform of the WKB period
\begin{equation}
    {\cal B}[{\Pi}_{\gamma}](\xi)=\sum_{n\geq0}\frac{1}{n!}\Pi_{\gamma}^{(n)}\xi^{n}
\end{equation}
and the Borel resummation along the direction $\varphi$
\begin{equation}
   s_{\varphi}(\Pi_{\gamma})(\epsilon)=\frac{1}{\epsilon}\int_{0}^{\infty e^{i\varphi}}e^{-\xi/\epsilon} {\cal B}[{\Pi}_{\gamma}](\xi)d\xi.
\end{equation}
We also denote by $s(\Pi_\gamma)(\epsilon)$ the Borel resumed WKB period at real and positive $\epsilon$. The WKB period $\Pi_\gamma$ is said to be Borel summable when the resumed period $ s_{\varphi}(\Pi_{\gamma})(\epsilon)$ converges for small $\epsilon$. If ${\cal B}[{\Pi}_\gamma]$ has singularity along the direction $\varphi$, there arises a discontinuity for the Borel resummed WKB period
\begin{equation}
    \begin{aligned}
       {\rm disc}_{\varphi}\Pi_{\gamma}(\epsilon)&=s_{\varphi^{+}}(\Pi_{\gamma})(\epsilon)-s_{\varphi^{-}}(\Pi_{\gamma})(\epsilon)\big)\\
       &=\lim_{\delta\to0_{+}}\big(s(\Pi_{\gamma})(e^{i\varphi+i\delta}\epsilon)-s(\Pi_{\gamma})(e^{i\varphi-i\delta}\epsilon)\big).
    \end{aligned}
\end{equation}
\begin{figure}[htbp]
    \centering
    \includegraphics[width=60mm]{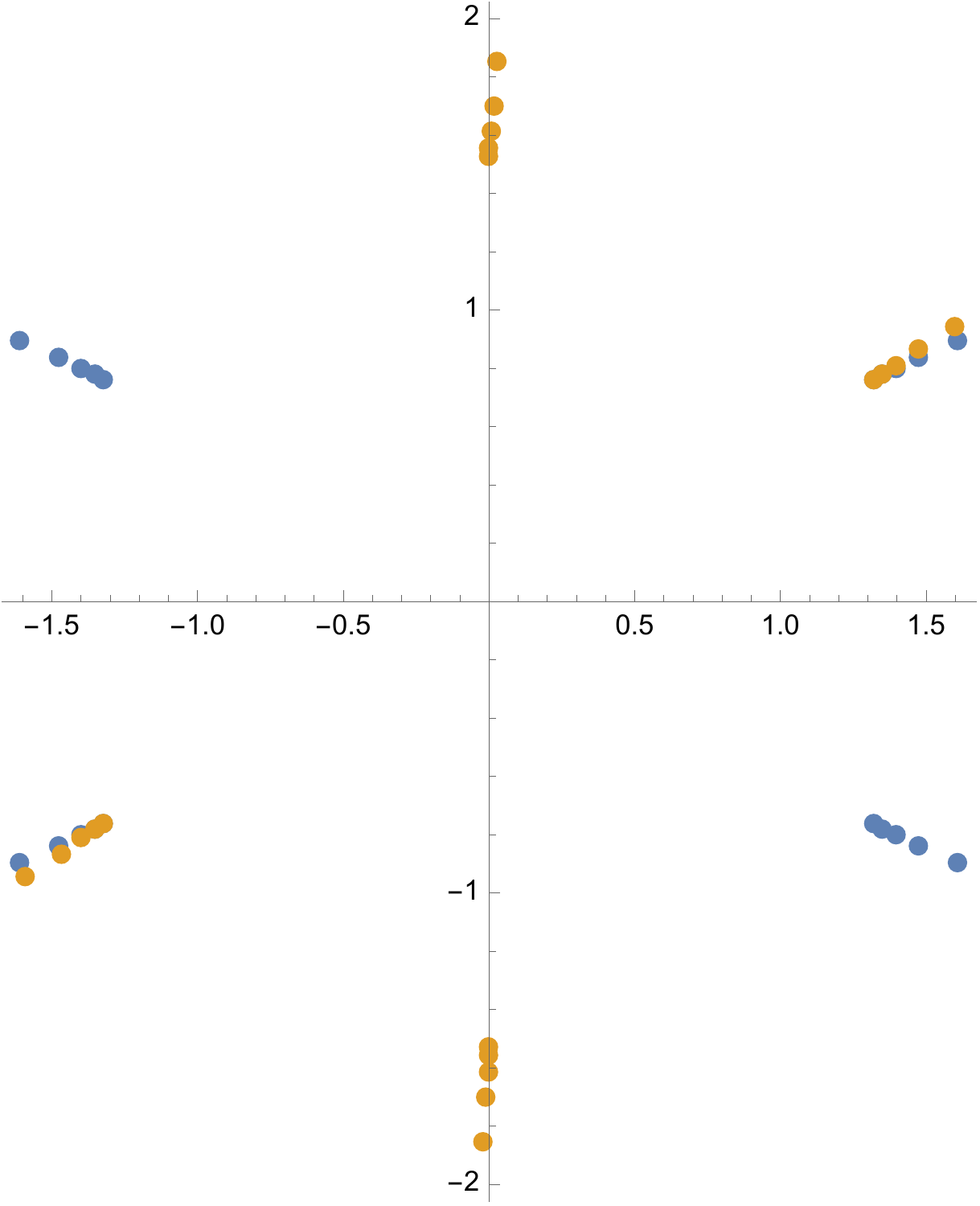}
    \caption{The singularity structure of the Borel transform of $\Pi_{\gamma_{1,1}}$ (blue) and $\Pi_{\gamma_{3,2}}$ (yellow) obtained by using the Borel-Pad\'e technique applied  to order $\epsilon^{160}$ terms of the formal power series. Here the potential is chosen to be $p(x)=-x^3+7x+6$.}
    \label{fig:discon}
\end{figure}
Moreover, the singularity of Borel transform can be approximated numerically by using the standard Borel-Pad\'e technique.
In Fig. \ref{fig:discon}, we plot the singularity structure of the Borel transform of $\Pi_{{\gamma}_{1,1}}$ (blue) and $\Pi_{\gamma_{3,2}}$ (yellow) obtained by using the Borel-Pad\'e technique (order $\epsilon^{160}$).

\section{Y-function and WKB period}\label{sec:Yfunc_and_period}
In this section, we will study the ODE/IM correspondence of the ODE \eqref{eq:3rdode}. 
From the solutions of the ODE, one can introduce the Y-functions, which satisfy Y-system and TBA equations in the integrable model.
In the case of Schr\"odinger equations, the Y-functions are identified with the WKB periods in the ODE.
A similar identification is also expected for the third order ODE \cite{Ito:2021boh}.
In this section, we first introduce the Y-functions for the $(A_2,A_N)$-type ODE.
Then in the sec. \ref{sec:Stokes}, we identify the Y-functions with WKB periods by the leading order of the WKB approximation.

\subsection{\texorpdfstring{Y-function for $(A_2,A_N)$-type ODE}{A2AN}}
We introduce the Y-function by using the ODE/IM correspondence\footnote{See \cite{Dorey:1999pv,Bazhanov:2001xm} for the case of the third order ODE with a monomial  potential}.
The $(A_2, A_N)$-type ODE \eqref{eq:3rdode} is invariant under the Symanzik-Sibuya rotation:
\begin{equation}
    x \to \omega^{-1}x, \quad u_i \to \omega^{-i}u_i, \qquad i = 0, 1, \dots, N + 1,
\end{equation}
where $\omega \coloneqq e^{2\pi i/(N + 4)}$.
This rotation is equal to the transformation $\epsilon \to e^{2\pi i/3}\epsilon$ since the rotated function
\begin{equation}
    \psi(\omega^{-1}x, \qty{\omega^{-i}u_i}; \epsilon) = \psi(x, \qty{u_i}; e^{2\pi i/3}\epsilon)
\end{equation}
satisfies the same ODE \eqref{eq:3rdode}.
Let us consider a solution $\phi_0$ whose asymptotic property is
\begin{equation}
    \phi_0(x, \qty{u_i}; \epsilon) \sim \frac{\epsilon}{i\sqrt{3}}x^{-\frac{N + 1}{3}}\exp(-\frac{1}{\epsilon}\frac{3}{N + 4}x^{\frac{N + 4}{3}}),\quad \abs{x} \to \infty,\quad  \abs{\arg x}<\frac{\pi}{N+4}
\end{equation}
and its rotated solutions $\phi_k$:
\begin{equation}
    \phi_k(x, \qty{u_i}; \epsilon) \coloneqq \phi_0(x, \qty{u_i}; e^{\frac{2\pi i}{3}k}\epsilon).
\end{equation}
Here we have set $u_0=1$ without loss of generality. The solution $\phi_k$ is called the subdominant solution since it decays fastest in the Stokes sector $\mathcal{S}_k$ $(k \in \mathbb{Z})$ defined by
\begin{equation}
    \mathcal{S}_k \coloneqq \qty{x \in \mathbb{C}; \abs{\arg{x} - \frac{2\pi k}{N + 4}} < \frac{\pi}{N + 4}}.
\end{equation}
Note that the solution has property:
\begin{equation}
    \phi_{k + N + 4}(x, \qty{u_i}; \epsilon) \propto \phi_k(x, \qty{u_i}; \epsilon).
\end{equation}
The set of the subdominant solutions $\qty{\phi_k, \phi_{k + 1}, \phi_{k + 2}}$ are the basis of the solutions of the ODE \eqref{eq:3rdode}, since the Wronskian of these solutions is equal to one:
\begin{equation}
    W[\phi_k, \phi_{k + 1}, \phi_{k + 2}] = 1,
\end{equation}
where the Wronskian of the functions $f$ is given by
\begin{equation}
    W[f_{k_1}, f_{k_2}, f_{k_3}] = \det\mqty[f_{k_1} & f_{k_2} & f_{k_3} \\ \partial_x f_{k_1} & \partial_x f_{k_2} & \partial_x f_{k_3} \\ \partial_x^2 f_{k_1} & \partial_x^2 f_{k_2} & \partial_x^2 f_{k_3}].
\end{equation}
We introduce the T-functions $T_{a, k}$ $(0 \leq a \leq 3, k \in \mathbb{Z})$ by
\begin{equation}
    \begin{aligned}
        &T_{0, k} = W[\phi_{-1}, \phi_0, \phi_1]^{[- k - 1]} = 1,&
        &T_{1, k} = W[\phi_{-1}, \phi_0, \phi_{k + 1}]^{[-k]},\\
        &T_{2, k} = W[\phi_{0}, \phi_{k + 1}, \phi_{k + 2}]^{[- k - 1]},&
        &T_{3, k} = W[\phi_k, \phi_{k + 1}, \phi_{k + 2}]^{[-k]} = 1,
    \end{aligned}
\end{equation}
where we used the notation
\begin{equation}
    f^{[k]}(\qty{u_i}; \epsilon) \coloneqq f(\qty{u_i}; e^{\frac{\pi i}{3}k}\epsilon).
\end{equation}
Using the Pl\"ucker relation
\begin{equation}
    W[f_0,f_1,f_2]W[f_0,f_3,f_4]=W[f_0,f_1,f_3]W[f_0,f_2,f_4]+W[f_0,f_1,f_4]W[f_0,f_3,f_2]
\end{equation}
and the property of the Wronskian
\begin{equation}
    W[\phi_{k_0}, \phi_{k_1}, \phi_{k_2}]^{[2k]} = W[\phi_{k_0 + k}, \phi_{k_1 + k}, \phi_{k_2 + k}],\qquad k \in \mathbb{Z},
\end{equation}
one finds that the T-functions satisfy the functional relations called T-system:
\begin{equation}
    T_{a, k}^{[+1]}T_{a, k}^{[-1]} = T_{a - 1, k}T_{a + 1, k} + T_{a, k - 1}T_{a, k + 1}
    \label{eq:T_system}
\end{equation}
with the boundary conditions:
\begin{equation}
    T_{a, 0} = T_{a, N + 1} = 1,\qquad a = 1, 2.
\end{equation}
The cross-ratios of the T-functions define the Y-functions:
\begin{equation}
    Y_{a, k} = \frac{T_{a - 1, k}T_{a + 1, k}}{T_{a, k - 1}T_{a, k + 1}}, \qquad a = 1, 2,\quad k = 1, \dots, N.
    \label{eq:Y-function_A2AN}
\end{equation}
From the T-system \eqref{eq:T_system}, one can derive the functional relations of the Y-functions:
\begin{equation}
    Y_{a, k}^{[+1]}Y_{a, k}^{[-1]} = \frac{(1 + Y_{a - 1, k})(1 + Y_{a + 1, k})}{(1 + Y_{a, k - 1}^{-1})(1 + Y_{a, k + 1}^{-1})}
    \label{eq:Y-system_A2AN}
\end{equation}
with the boundary conditions:
\begin{equation}
    Y_{0, k} = Y_{3, k} = 0, \quad Y_{a, 0} = Y_{a, N + 1} = \infty,\qquad a=1,2.
\end{equation}
The system \eqref{eq:Y-system_A2AN} is $(A_2,A_N)$-type Y-system \cite{zamo-ADE}.

\subsection{Stokes graph and WKB approximation of the Y-function }\label{sec:Stokes}
In this subsection, we discuss the asymptotic behaviors of the Y-functions.
Since the Y-functions are cross-ratios of T-functions that are the Wronskians of the solutions to the ODE, we first evaluate the Wronskians by the WKB approximation. 
We should choose the region where the WKB approximation is valid. 
These regions can be found by considering the imaginary part of the solutions.
The Stokes curve \cite{Honda_2015} starting from the turning points on the complex plane is defined by
\begin{equation}
    \Im e^{-i\vartheta}\int_{x_{\ast}}^{x}\qty(y_a(x) - y_b(x))\dd{x} = 0, \qquad a, b = 1, 2, 3, \quad  a \neq b,
    \label{eq:def_Stokes_curve}
\end{equation}
where $\vartheta$ is the phase of $\epsilon$, $a$ and $b$ are the labels of ordered pair of distinct three sheets of $\Sigma$, and $x_\ast$ is a turning point at which $y_a(x_\ast) = y_b(x_\ast)$.
We label the Stokes curve $(a,b)$ on which 
\begin{equation}
    \Re e^{-i\vartheta}\int_{x_{\ast}}^{x}\qty(y_a(x) - y_b(x))\dd{x} < 0
    \label{eq:curve_label}
\end{equation}
holds.
The Stokes curves will end on the other turning points or extend to infinity. Furthermore, if the curves contained in the Stokes graph intersect, we need to add a new curve depending on their intersection angles \cite{doi:10.1063/1.525467,Honda_2015}.

For general third order ODE with simple turning points, where the characteristic equation has simple zeros, three Stokes curves emerge from the point \cite{aoki1994,aoki1998exact}.
For the case of the ODE \eqref{eq:3rdode}, however,
all the zeros of $p(x)$ are not the simple turning point, from which eight Stokes curves stretch along the directions:
\begin{equation}
    \varphi = \frac{3}{4}\qty(\frac{(2n + 1)\pi}{2} + \vartheta - \frac{(a+ b)\pi}{3} - \frac{\arg p'(x_\ast)}{3}), \qquad n \in \mathbb{Z}.
\end{equation}
Fixing $\vartheta$ and the pair $a,b$ with $a<b$, one finds the directions the curves can stretch.
The condition \eqref{eq:curve_label} indicates that for odd $n$ we have to assign the label $(a,b)$ to the curve , and for even $n$ we assign the label $(b,a)$.
We can read the labels of the curves with the starting point $x_{\ast}$ in ascending order of $\varphi$ in the range of $-\frac{1}{4}\arg{p'(x_{\ast})}$ to $2\pi-\frac{1}{4}\arg{p'(x_{\ast})}$: $(1, 3)$, $(1, 2)$, $(3, 2)$, $(3, 1)$, $\dots$.

The label of the curve can also be read off by considering the dominant or subdominant solutions in asymptotic regions.
In addition to the subdominant solution $\phi_k$ in the sector $\mathcal{S}_k$ $(k\in\mathbb{Z})$,
we introduce the solution $\overline{\phi}_k$ of the adjoint ODE \cite{Dorey:2000ma,Dorey:2006an,Ito:2020htm} which is defined by
\begin{equation}
    \overline{\phi}_k(x,\{u_i\};\epsilon)\coloneqq\phi_k(x,\{u_i\};-\epsilon),\qquad k\in\mathbb{Z}.
    \label{eq:phi_bar}
\end{equation}
Note that $\overline{\phi}_k$ is subdominant in the sector $\mathcal{S}_{k + 3/2}$.
Obviously, from the definition \eqref{eq:phi_bar}, in the region where $\phi_k$ is subdominant, $\overline{\phi}_k$ is dominant and vice versa. 
We introduce asymptotic directions
\begin{equation}
    \ell_k := e^{\frac{2\pi i k}{N + 4}}\mathbb{R}_+,\qquad
    \overline{\ell}_k := e^{\frac{\pi i (2k +3)}{N + 4}}\mathbb{R}_+,\qquad k \in \mathbb{Z}.
    \label{eq:asymototic_direction}
\end{equation}
Along $\ell_k$ ($\overline{\ell}_k$), the solution $\phi_k$ ($\overline{\phi}_k$) is subdominant.  
More precisely, the line $\ell_k$ ($\overline{\ell}_k$) is in the middle of the sector ${\cal S}_k$ (${\cal S}_{k+3/2}$).
Now we consider the sheets on which $\phi_k$ and $\overline{\phi}_k$ live, respectively.
In the asymptotic region, the solutions $\phi_k$ and $\overline{\phi}_k$ are 
\begin{equation}
    \phi_k\simeq\exp[-\frac{\delta_k}{\epsilon}\int^x_{\infty \exp[\frac{2\pi ik}{N+2}]} y_{a_k}\dd{x'}],\qquad \overline{\phi}_k\simeq\exp[\frac{\overline{\delta}_k}{\epsilon}\int^x_{\infty \exp[\frac{\pi i(2k+3)}{N+2}]} y_{\overline{a}_k}\dd{x'}].
    \label{eq:phi_and_phibar}
\end{equation}
Here $a_k,\overline{a}_k\in\{1,2,3\}$ are the sheet's labels of $\Sigma$, and $\delta_k,\overline{\delta}_k\in\{1,e^{\pm\frac{2\pi i}{3}}\}$ ensure the asymptotic behaviors of the solutions studied in the previous subsection.
Suppose, around direction $e^{\frac{\pi i k}{2(N+4)}}\mathbb{R}_{+}$ $(k\in\mathbb{Z})$, one has the asymptotic direction $\ell_{k_1}$ and $\bar{\ell}_{k_2}$, where solutions $\phi_{k_1}$ and $\bar{\phi}_{k_2}$ become subdominant, respectively.
A Stokes curves related with solutions $\phi_{k_1}$ and $\bar{\phi}_{k_2}$ thus exist in the direction $e^{\frac{\pi i k}{2(N+4)}}\mathbb{R}_{+}$ $(k\in\mathbb{Z})$.
Since on the Stokes curves $\Im y_{a_{k_1}}=\Im y_{\overline{a}_{k_2}}$ and $\Re y_{a_{k_1}}>\Re y_{\overline{a}_{k_2}}$ are satisfied, the labels of the curves are $(\overline{a}_{k_2},a_{k_1})$.
For the case of $k=1$ and the branch cut chosen as in fig. \ref{fig:abelianization_tree_Y_11}, one finds that $k_1=0,k_2=-1,a_{k_1}=3$ and $\overline{a}_{k_2}=1$, which leads to the label of the curves $(1,3)$.

\subsubsection{The cycles associated with Y-functions}
Using the Wronskian representation of the Y-functions and the Stokes graphs, we can associate the cycles on the Riemann surface to the Y-functions. The cycle can be constructed by using the abelianization tree \cite{Gaiotto:2012rg} made of junctions and lines. 
For completeness, we review the construction of the cycles following \cite{Neitzke:2017yos}.
An abelianization tree is a junction that has three endpoints at the infinity of $\ell_{k_i}$ $(i=1,2,3)$ where the solution $\phi_{k_i}$ in the Wronskian $W[\phi_{k_1},\phi_{k_2},\phi_{k_3}]$ becomes subdominant. The line starting from the junction and ending at the infinity point of $\ell_k(\overline{\ell}_k)$ has the label $a_k(\overline{a}_k)$, the sheet label of the subdominant solution $\phi_k(\overline{\phi}_k)$.
To make sure the precision of the WKB approximation, the line with the label $a_k$ cannot cross the Stokes curve with the label $(a_k,a_l)$.

When we substitute (\ref{eq:phi_and_phibar}) into the Wronskian, one obtains the leading term in $1/\epsilon$. It depends on the end points of the integration paths of the three solutions. However, if we set these points as the same, the contributions from the integrals cancel. Then it becomes independent of the final point. We can use this property to deform the contour associated with the Y-function.

\begin{figure}[htbp]
    \centering
    \begin{minipage}[b]{0.5\linewidth}
        \centering
        \includegraphics[width=8cm]{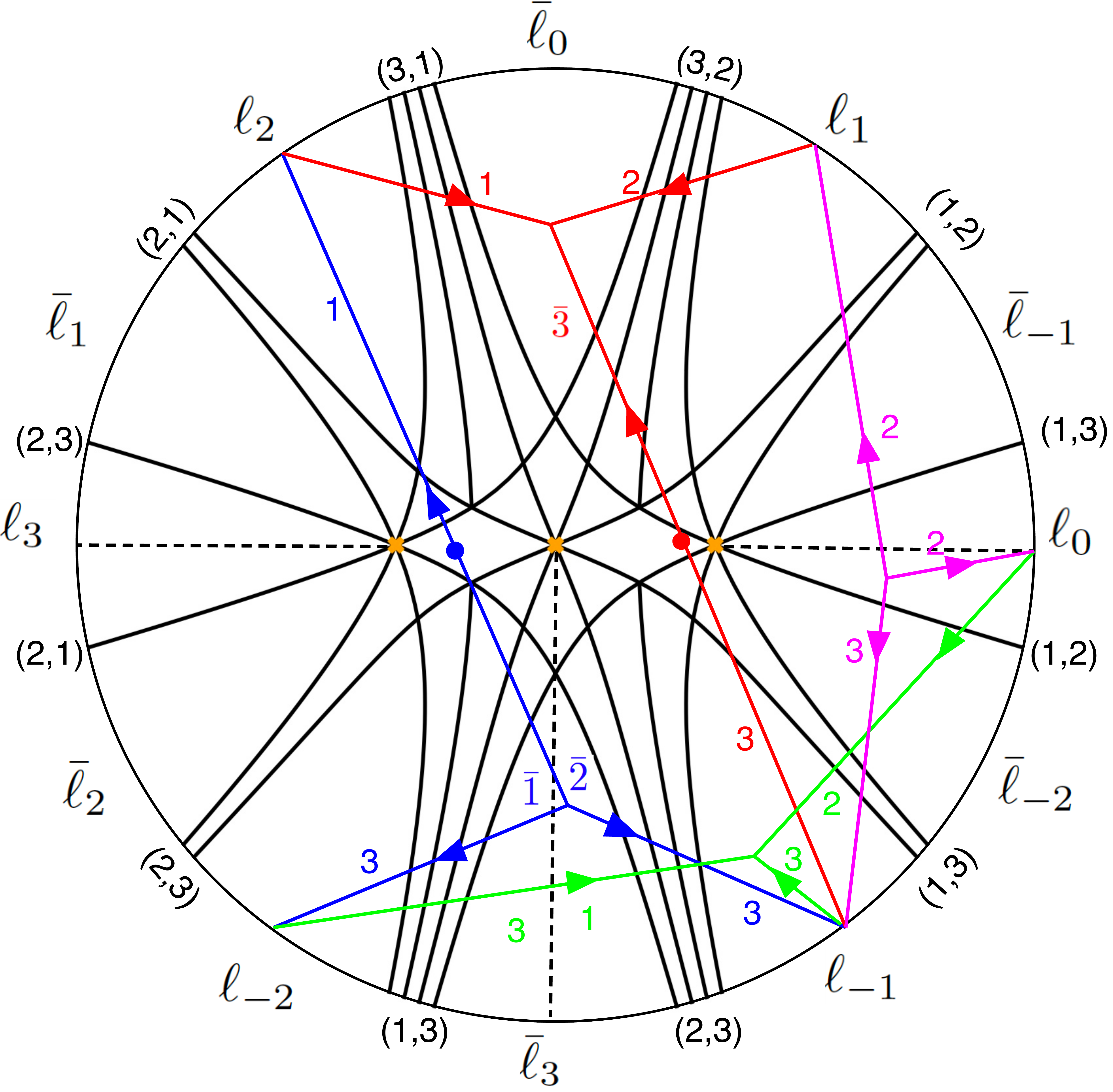}
        \subcaption{}
    \end{minipage}
    \begin{minipage}{0.05\linewidth}
        \hspace{0.5cm}
    \end{minipage}
    \begin{minipage}[b]{0.3\linewidth}
        \centering
        \begin{tikzpicture}
        \def\dv{0.07};
        \def\rt{1.3};
        \def\rv{0.3};
        
        \coordinate[] (ni) at (-2.5, 0);
        \coordinate[] (pi) at (2.5, 0);
        \coordinate[] (0i) at (0, -1.2);
        \coordinate[label=above:] (x2) at ({-\rt}, 0);
        \coordinate[label=above:] (x1) at (0, 0);
        \coordinate[label=above:] (x0) at ({\rt}, 0);
        
        \coordinate[] (x0p) at ($({\rt}, {\dv})+({\rv},0)$);
        \coordinate[] (x0n) at ($({\rt}, {-\dv})+({\rv},0)$);
        \coordinate[] (x1p) at ({\dv}, {-\rv});
        \coordinate[] (x1n) at ({-\dv}, {-\rv});
        
        \coordinate[] (x0ip) at ({\rt}, {\rv});
        \coordinate[] (x0in) at ({\rt}, {-\rv});
        \coordinate[] (x1in) at ({-\rv}, 0);
        \coordinate[] (x1vp) at (0, {\rv});
        
        \fill (x2) circle [radius=2pt, red];\draw ({-\rt}, 0.8) node[below]{$x_2$};
        \fill (x1) circle [radius=2pt, red];\draw (0, 0.8) node[below]{$x_1$};
        \fill (x0) circle [radius=2pt, red];\draw ({\rt}, 0.8) node[below]{$x_0$};
        
        \draw[line width=0.4mm, gray, dashed] (ni) -- (x2);
        \draw[line width=0.4mm, gray, dashed] (x1) -- (0i);
        \draw[line width=0.4mm, gray, dashed] (x0) -- (pi);

        \draw[line width=0.3mm, black] (x1n) to [out=175, in=270] (x1in);
        \draw[line width=0.3mm, black] (x1in) to [out=90, in=180] (x1vp);
        \draw[line width=0.3mm, black,->-] (x1vp) to [out=0, in=180] (x0in);
        \draw[line width=0.3mm, black] (x0in) to [out=0, in=265] (x0n);
        
        \draw[line width=0.3mm, dashed, black] (x0p) to [out=95, in=0] (x0ip);
        \draw[line width=0.3mm, dashed, black,->-] (x0ip) to [out=180, in=5] (x1p);
        
        \end{tikzpicture}
        \vspace{2.2cm}
        \subcaption{}
    \end{minipage}
    \caption{The stokes curve of $(A_2,A_2)$-type ODE with the potential $p(x)=x^3-x$ and the abelianization tree for $Y_{1,1}$ (a) and the cycle $\gamma_{1,1}$ (b).}
    \label{fig:abelianization_tree_Y_11}
\end{figure}
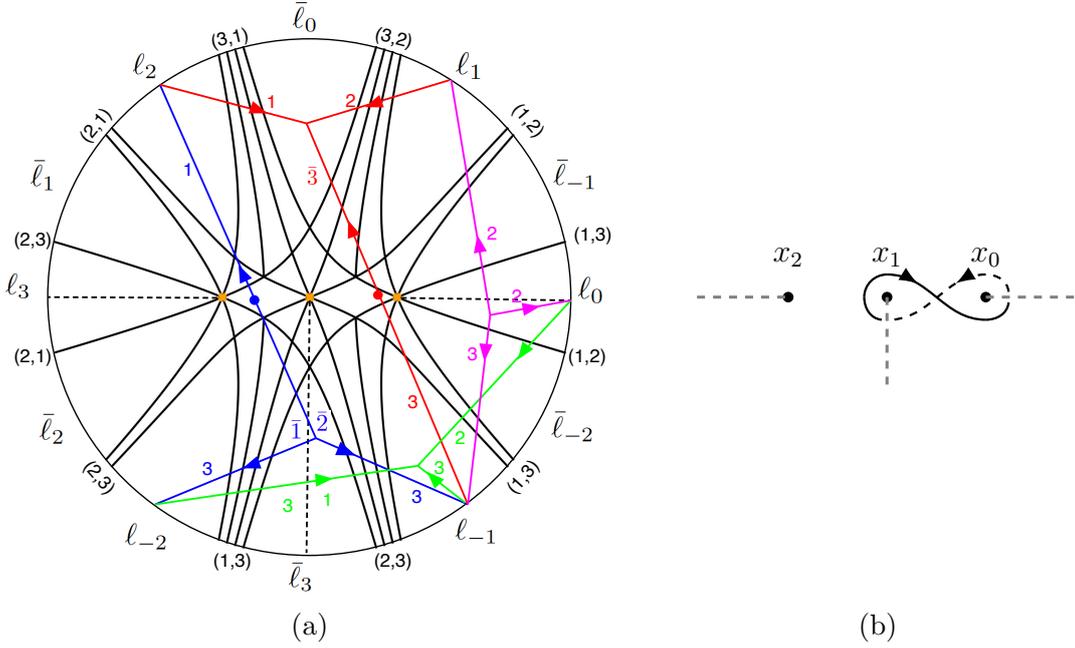

A Y-function is defined by four Wronskians, which thus provide four abelianization trees. Combining these four trees, we obtain a closed cycle associated with the Y-function.
Let us consider the cycle made of the abelianization tree associated with the Y-function.
We illustrate the procedure by taking an example of $Y_{1,1}$ of $(A_2,A_2)$-type ODE, which is given by
\begin{equation}
    Y_{1, 1} = \frac{W[\phi_{-1},\phi_1,\phi_2]W[\phi_{-2},\phi_{-1},\phi_0]}{W[\phi_{-2},\phi_{-1},\phi_2]W[\phi_{-1},\phi_0,\phi_1]}.
    \label{eq:A2A2_Y11}
\end{equation}
The Wronskian $W[\phi_{-2},\phi_{-1},\phi_0]$ is represented by the junction which ends on the infinity points of $\ell_{-2},\ell_{-1}$, and $\ell_0$ expressed by the green tree in figure \ref{fig:abelianization_tree_Y_11}.
The lines of the trees represent the integration paths of the solutions in \eqref{eq:phi_and_phibar}.
On each line, we assigned the arrows associated with the direction of the integration path.
Let us consider the abelianization tree for the Wronskian $W[\phi_{-1},\phi_1,\phi_2]$.
At the asymptotic region, the lines representing the integration paths of $\phi_1,\phi_2$, and $\phi_{-1}$ are labeled by $2,1$, and $3$, respectively.
However, the line labeled by $3$ starting from the infinity of $\ell_{-1}$ will cross the Stokes curve with the label $(3,\ast)$. We indicate the point just before the crossing as the red dotted point in the figure. After the red dotted point, we relabel the line as $\bar{3}$, which lives on the same sheet three with opposite arrows. The solution $\phi_{-1}$ is subdominant before crossing the red dot, while after the crossing, on line $\bar{3}$, the solution $\overline{\phi}_0$ becomes subdominant. 
The line with label $\overline{3}$ cannot cross the Stokes curve labeled by $(\ast,3)$.
Similarly, for the Wronskian $W[\phi_{-2},\phi_{-1},\phi_2]$, we added a point on the line with sheet labels $1$ and $\overline1$.

Since Wronskians are independent of the location of the junction, we can move the junctions of the Wronskians to the appropriate points and find that the integration path can be identified with the cycle $\gamma_{1,1}$.
In the same way, one can identify the cycles for other Y-functions.
For $Y_{1,2}$, it is convenient to draw the diagram for $Y^{[+1]}_{1,2}$ rather than $Y_{1,2}$:
\begin{equation}
    Y_{1,2}^{[+1]}(\{u_i\};\epsilon)=Y_{1,2}(\{u_i\};e^{-\frac{\pi i}{3}}\epsilon)=\frac{W[\phi_{-1},\phi_2,\phi_3]W[\phi_{-2},\phi_{-1},\phi_0]}{W[\phi_{-1},\phi_0,\phi_2]W[\phi_{-2},\phi_{-1},\phi_3]}.
\end{equation}
Then the corresponding Stokes curve is rather than that of adjoint ODE. 
The Stokes graph becomes the same as in figure \ref{fig:abelianization_tree_Y_11} but with the asymptotic directions replaced by $\ell_k=e^{\frac{\pi i(2k-1)}{N+4}}\mathbb{R}_{+}$ and $\overline{\ell}_k=e^{\frac{2\pi i(k+1)}{N+4}}\mathbb{R}_{+}$.

Repeating the same process for $Y_{2,1}^{[-1]}$ and $Y_{2,2}$, we thus found one-cycles for the Y-functions. At the leading order in $\epsilon^{-1}$ of the WKB approximation,
one obtains the following behaviors of the Y-functions 
\begin{equation}
    \begin{aligned}
        &\log Y_{1,1}\sim \epsilon^{-1}\Pi_{\gamma_{1,1}}^{(0)}, &
        &\log Y_{1,2}^{[+1]}\sim \epsilon^{-1}\Pi_{\gamma_{3,2}}^{(0)},\\
        &\log Y_{2,1}^{[-1]}\sim \epsilon^{-1}\Pi_{\gamma_{1,1}+\gamma_{2,1}}^{(0)}, &
        &\log Y_{2,2}\sim \epsilon^{-1}\Pi_{\gamma_{3,2}+\gamma_{1,2}}^{(0)},
    \end{aligned}\qquad \epsilon\to 0.
\end{equation}
Note that since $\gamma_{1,1}+\gamma_{2,1}=-\gamma_{3,1}$ and $\gamma_{3,2}+\gamma_{1,2}=-\gamma_{2,2}$, Y-functions have the $\mathbb{Z}_2$-symmetry $Y_{1,k}=Y_{2,k}$ ($k=1,2$) in the limit $\epsilon\to0$.
As we will see in the next section, the asymptotic behavior and the TBA equations together with the discontinuity structure of the WKB periods lead to the relations between the logarithm of the Y-functions and the WKB periods: 
\begin{equation}
    \begin{aligned}
        &\log Y_{1,1}= \epsilon^{-1}\Pi_{\gamma_{1,1}}, &
        &\log Y_{1,2}= \qty[\epsilon^{-1}\Pi_{\gamma_{3,2}}]^{[-1]},\\
        &\log Y_{2,1}= \qty[\epsilon^{-1}\Pi_{\gamma_{1,1}+\gamma_{2,1}}]^{[+1]}, &
        &\log Y_{2,2}= \epsilon^{-1}\Pi_{\gamma_{3,2}+\gamma_{1,2}},
    \end{aligned}
\end{equation}
which will be tested by numerical calculations in the following sections.
From the analysis of $(A_2,A_2)$ and $(A_2,A_3)$, we arrive at the formula which identifies the Y-functions with WKB periods of $(A_2,A_N)$-type ODE:
\begin{equation}
    \log Y_{a,k} = \qty[\frac{1}{\epsilon}\Pi_{\hat{\gamma}_{a,k}}]^{[a-k]},\quad \hat{\gamma}_{a,k}\coloneqq \gamma_{2-k,k}+\cdots+\gamma_{a+1-k,k},\quad a=1,2,\quad k=1,\dots,N.
    \label{eq:ident_Y_with_period}
\end{equation}
We will check these relations for the $(A_2,A_2)$ and $(A_2,A_3)$ cases. We note, however, that (\ref{eq:ident_Y_with_period}) is valid in the minimal chamber of the moduli space. Outside the minimal chamber, we need to replace them with new Y-functions obtained from the wall-crossing of the TBA equations. This is the subject of Sect. \ref{sec:wc}.

\section{TBA equations in minimal chamber}\label{sec:TBA_in_minimal}
In this section, we derive an integral equation, called the TBA equation, starting from the Y-system and the asymptotic behaviors discussed in the previous section, especially in the case of the minimal chamber.

\subsection{TBA equations in minimal chamber}
The Y-system for a pair $(\mathfrak{g},\mathfrak{h})$ of simply-laced Lie algebras $\mathfrak{g}$ and $\mathfrak{h}$ has been studied in \cite{Ravanini:1992fi}. 
Our Y-system \eqref{eq:Y-system_A2AN} corresponds to $\mathfrak{g}=A_N$ and $\mathfrak{h}=A_r$, and the Y-functions are the inverses of them. Then, 
the $(A_2, A_N)$-type Y-system \eqref{eq:Y-system_A2AN} can be rewritten as follows:
\begin{equation}
    \frac{Y_{a, k}^{[+1]}Y_{a, k}^{[-1]}}{\prod_{b = 1}^{r}Y_{b, k}^{G_{ab}}} = \frac{\prod_{b = 1}^{r}(1 + Y_{b, k}^{-1})^{G_{ab}}}{(1 + Y_{a, k - 1}^{-1})(1 + Y_{a, k + 1}^{-1})},
    \label{eq:Y-system_G}
\end{equation}
where the matrix $G_{ab}$ is the incidence matrix of $A_2$.
Here $a=1,2,k=1,\dots,N$.
The boundary conditions for Y-functions are given by
\begin{equation}
    Y_{0, k} = Y_{3, k} = 0,\qquad Y_{a, 0} = Y_{a, N + 1} = \infty,\qquad a=1,2.
    \label{eq:boundary_condition_for_Y-function}
\end{equation}
At the large and positive $\theta$, which is defined by $\theta \coloneqq -\log{\epsilon}$, the logarithm of $Y_{a, k}$ is assumed to behave as
\begin{equation}
    \log{Y_{a, k}}(\theta) \sim m_{a, k}e^{\theta},
\end{equation}
which is the leading order of \eqref{eq:ident_Y_with_period}:
\begin{equation}\label{eq:m-period}
    m_{a, k} = e^{\frac{\pi i}{3}(k-a)}\Pi_{\hat{\gamma}_{a, k}}^{(0)}.
\end{equation}
The cycle $\hat{\gamma}_{a,k}$ and $\hat{\gamma}_{3-a,k}$ are different only by the living sheets; their classical periods are related by
\begin{equation}
    \Pi_{\hat{\gamma}_{a,k}}^{(0)}=e^{-\frac{\pi i}{3}(3-2a)}\Pi_{\hat{\gamma}_{3-a,k}}^{(0)},\qquad a=1,2,\quad  k=1,\dots,N.
\end{equation}
From \eqref{eq:ident_Y_with_period}, one obtains
\begin{equation}
    m_{1,k}=m_{2,k},\qquad k=1,\dots,N.
    \label{eq:Z2sym_of_mass}
\end{equation}
\eqref{eq:Z2sym_of_mass} can be written 
\begin{equation}
    2\cos(\frac{\pi}{3})m_{a, k} = \sum_{b = 1}^{2}G_{ab}m_{b, k},\qquad a=1,2,\qquad k=1,\dots,N,
\end{equation}
which can also be derived from the leading order of the Y-system.
We can follow the standard approach to convert the Y-system to the TBA equations\cite{zamo-ADE} for these mass parameters. 
Taking the logarithm of the Y-system \eqref{eq:Y-system_G}, we get
\begin{equation}
    \begin{aligned}
        f_{a, k}^{[+1]} + f_{a, k}^{[-1]} - \sum_{b = 1}^{2}G_{ab}f_{b, k} = \sum_{b = 1}^{2}G_{ab}L_{b, k} - L_{a, k - 1} - L_{a, k + 1},
        \label{eq:Y-system_f}
    \end{aligned}
\end{equation}
where
\begin{equation}
    \begin{aligned}
        f_{a, k}(\theta) &\coloneqq \log{Y_{a, k}}(\theta) - m_{a, k}e^{\theta},\\
        L_{a, k}(\theta) &\coloneqq \log(1 + Y_{a, k}^{-1}(\theta)).
    \end{aligned}
\end{equation}
Taking the Fourier transformation
\begin{equation}
    \widetilde{f}(p) \coloneqq \mathcal{F}[f](p) =  \int_{-\infty}^{\infty}f(\theta)e^{-ip\theta}\dd{\theta},
\end{equation}
then \eqref{eq:Y-system_f} becomes
\begin{equation}
    \sum_{b = 1}^{2}\qty(2\delta_{ab}\cosh(\frac{\pi p}{3})\delta_{ab} - G_{ab})\widetilde{f}_{b, k}(p) = \sum_{b = 1}^{2}G_{ab}\widetilde{L}_{b, k}(p) - \widetilde{L}_{a, k - 1}(p) - \widetilde{L}_{a, k + 1}(p).
\end{equation}
Solving about $\widetilde{f}_{a, k}$ and taking the inverse Fourier transformation of both sides, then we get a set of the TBA equations:
\begin{equation}
    \log{Y}_{a, k}(\theta) = m_{a, k}e^{\theta} + K \star (L_{a, k} - L_{a, k - 1} - L_{a, k + 1}), \qquad a = 1, 2,
    \label{eq:tba1}
\end{equation}
where we have used $Y_{1,s}(\theta)=Y_{2,s}(\theta)$, which is a conclusion of the $\mathbb{Z}_2$ symmetry of the masses, i.e. $m_{1,s}=m_{2,s}$.
Here the kernel function is given by
\begin{equation}
    K(\theta) \coloneqq \frac{1}{2\pi}\frac{4\sqrt{3}\cosh{\theta}}{1 + 2\cosh{2\theta}}
\end{equation}
and the convolution is defined by
\begin{equation}
    K\star L=\int_{-\infty}^\infty \dd\theta^\prime K(\theta-\theta^\prime)L(\theta^\prime).
\end{equation}
In the following, we only consider the TBA equations for $a=1$ since that for $a=2$ is the copy of the one for $a=1$.
For the complex masses, it is more convenient to shift the arguments of the Y-functions, such that the leading order is real and positive:
\begin{equation}
    \log{Y_{1, k}(\theta - i\phi_{k})} = \abs{m_{1, k}}e^{\theta} + K \star \overline{L}_{1, k} - K_{k, k - 1} \star \overline{L}_{1, k - 1} - K_{k, k + 1} \star \overline{L}_{1, k + 1},
    \label{eq:tba-shifted}
\end{equation}
where $\overline{L}_{a,k}(\theta)=L_{a,k}(\theta-i\phi_k)$. 
$K_{k_1, k_2}$ is the kernel with the argument shifted in the imaginary direction, defined by
\begin{equation}
    K_{k_1, k_2}(\theta) = K(\theta - i(\phi_{k_1} - \phi_{k_2})),
\end{equation}
where $\phi_{k}$ denotes the phase of the mass:
\begin{equation}
    \phi_{k} \coloneqq \arg m_{1, k},\qquad k=1,\dots,N.
\end{equation}
It is convenient to rewrite the kernel as
\begin{equation}
K(\theta)=\frac{1}{2\pi}\frac{1}{\cosh(\theta+\frac{\pi i}{6})}+\frac{1}{2\pi}\frac{1}{\cosh(\theta-\frac{\pi i}{6})}.
\end{equation}
The kernel function $K$ has the poles at
\begin{equation}
    \theta = \pm\frac{\pi i}{3}+n\pi i,\qquad n\in\mathbb{Z}.
\end{equation}
Therefore in the case of the minimal chamber, i.e. $\abs{\phi_{k} - \phi_{k \pm 1}} < \pi/3$, the integration path of the TBA equation can be shifted without causing the wall-crossing phenomenon.
In the numerical calculation, the equations (\ref{eq:tba-shifted}) are helpful to see the convergence of the solution under the iteration.
The case where $\abs{\phi_{k} - \phi_{k \pm 1}} > \pi/3$, however, the solutions do not give the correct solution. This will be treated in the next section.

The integrable model corresponding to the ODE is characterized by the (kink) TBA equations (\ref{eq:tba-shifted}), which is described by a conformal field theory \cite{Bazhanov:1994ft}.
The effective central charge of the conformal field theory is given by 
\begin{equation}
\label{ceff}
    c_{\mathrm{eff}} \coloneqq 2\times\frac{6}{\pi^2}\sum_{k = 1}^{N}\int_{-\infty}^{\infty}\abs{m_{1, k}}\overline{L}_{1, k}(\theta)e^{\theta}\dd{\theta},
\end{equation}
where the factor $2$ is due to the summation of $a=1,2$. 
This can be evaluated by the $\theta\to -\infty$ limit of TBA equations, where the deriving term vanishes and thus leads to constant Y-functions satisfying
\begin{align}
    \log Y_{a,k}(-\infty-i\phi_k)&=\overline{L}_{a,k}(-\infty)-\overline{L}_{a,k-1}(-\infty)-\overline{L}_{a,k+1}(-\infty).
\end{align}
The solution to these equations has been found in \cite{Bazhanov:1989yk,Kuniba:1993cn} as
\begin{align}
    Y_{a,k}(-\infty-i\phi_k)= \frac{\sin\frac{\pi (k+1)}{N+4} \sin\frac{\pi (k+2)}{N+4}}{\sin\frac{\pi  k}{N+4} \sin\frac{\pi(k+3)}{N+4}}-1.
\end{align}
Using the Rogers dilogarithm identity\cite{kirillov1987identities}, the effective central charge becomes
\begin{equation}
    c_{\mathrm{eff}} = 2\times \frac{6}{\pi^2}\sum_{ k=1}^{N}\mathcal{L}\qty(\frac{1}{1 + Y_{1, k}(-\infty-i\phi_k)}) = \frac{2N(N + 1)}{N + 4},
    \label{eq:ceff}
\end{equation}
where $\mathcal{L}(t)$ is the Rogers dilogarithm function defined by
\begin{equation}
    \mathcal{L}(t) \coloneqq -\frac{1}{2}\int_0^{t}\qty(\frac{\log(1 - t')}{t'} + \frac{\log{t'}}{1 - t'})\dd{t'}.
\end{equation}
The central charge (\ref{eq:ceff}) is that of the generalized parafermion
$SU(N+1)_{3}/U(1)^N$. Its massive deformation is described by the homogeneous sine-Gordon model based on the same coset, whose TBA system is the same as (\ref{eq:tba1}) \cite{Castro-Alvaredo:1999ybz}.

\subsection{Numerical test: minimal chamber}\label{sec:Num_minimal}
We now compare the WKB periods with the Y-functions in the minimal chamber numerically. In particular, we will calculate the coefficients of the expansions of \eqref{eq:ident_Y_with_period} in $\epsilon = e^{-\theta}$ and will check 
\begin{equation}
    m_{1, k}^{(n - 1)} = e^{\frac{\pi i}{3}(1-k)(n-1)}\Pi_{\hat{\gamma}_{1, k}}^{(n)}, \qquad n = 0, 1, \dots,
\end{equation}
where $m_{1,k}^{(n-1)}$ is defined by the $e^{-\theta}$-expansions of $\log{Y_{1, k}}(\theta)$:
\begin{equation}
    \log{Y_{1, k}}(\theta) = m_{1, k}e^{\theta} + \sum_{n = 1}^{\infty}m_{1, k}^{(n)}e^{-n\theta}.
\end{equation}
Here $m_{1,k}^{(-1)}=m_{1,k}$, $m_{1,k}^{(0)}=0$, and 
$m_{1, k}^{(n)}$ with positive integer $n$ is given by
\begin{equation}
    m_{1, k}^{(n)} = k_n\int_{-\infty}^{\infty}(\overline{L}_{1, k}(\theta)e^{n(\theta - i\phi_{k})} - \overline{L}_{1, k - 1}(\theta)e^{n(\theta - i\phi_{ k - 1})} - \overline{L}_{1, k + 1}(\theta)e^{n(\theta - i\phi_{ k + 1})})\dd{\theta},
    \label{eq:m-coefficient}
\end{equation}
and
\begin{equation}
    k_n = \frac{1}{\pi}\qty(\sin(\frac{\pi}{3}n) + \sin(\frac{2\pi}{3}n)).
\end{equation}
$\Pi_{\hat{\gamma}_{1,k}}^{(n)}$ has been calculated by using the Picard-Fuchs operator discussed in section \ref{sec:WKB_ana}.  On the other hand, $m_{1,k}^{(n-1)}$ is computed by solving the TBA equations numerically.

\subsubsection{\texorpdfstring{$(A_2, A_2)$}{A2A2}}
Let us consider $(A_2,A_2)$-type ODE in the minimal chamber studied in a previous paper \cite{Ito:2021boh}.
The Y-system \eqref{eq:Y-system_A2AN} becomes
\begin{equation}
\begin{aligned}
    Y_{1,1}^{[+1]}Y_{1,1}^{[-1]}=\frac{(1+Y_{2,1})}{(1+Y_{1,2}^{-1})},\quad&Y_{2,1}^{[+1]}Y_{2,1}^{[-1]}=\frac{(1+Y_{1,1})}{(1+Y_{2,2}^{-1})},\\
    Y_{1,2}^{[+1]}Y_{1,2}^{[-1]}=\frac{(1+Y_{2,2})}{(1+Y_{1,1}^{-1})},\quad&Y_{2,2}^{[+1]}Y_{2,2}^{[-1]}=\frac{(1+Y_{1,2})}{(1+Y_{2,1}^{-1})}
    \end{aligned}
\end{equation}
and the TBA equations \eqref{eq:tba-shifted} read
\begin{equation}
    \begin{aligned}
        \log{Y_{1, 1}(\theta - i\phi_{1})} &= \abs{m_{1, 1}}e^{\theta} + K \star  \overline{L}_{1, 1} - K_{1, 2} \star \overline{L}_{1, 2},\\
        \log{Y_{1, 2}(\theta - i\phi_{2})} &= \abs{m_{1, 2}}e^{\theta} - K_{2, 1} \star \overline{L}_{1, 1} + K \star \overline{L}_{1, 2},
    \end{aligned}
    \label{eq:TBA_A2A2_minimal}
\end{equation}
with effective central charge $c_{\rm eff}=2$.
As an example, we consider the case where all branch points are aligned on the real axis:
\begin{equation}
    p(x) = -(x - 3)(x + 1)(x + 2) = -x^3 + 7x + 6,
\end{equation}
which is different from the one in the previous paper \cite{Ito:2021boh}.
The mass parameters are defined through the classical periods and are computed by \eqref{eq:classical_period_A2A2} as
\begin{equation}
    \begin{aligned}
        m_{1, 1} &= \Pi_{\hat{\gamma}_{1, 1}}^{(0)} \simeq13.14579499i,\\
        m_{1, 2} &= e^{\frac{\pi i}{3}}\Pi_{\hat{\gamma}_{1, 2}}^{(0)}\simeq1.514970717i.
    \end{aligned}
\end{equation}
In Table \ref{tab:A2A2_minimal}, we compare the coefficients $m_{1, 1}^{(n - 1)}$ to the quantum corrections $\Pi_{\gamma_{1, 1}}^{(n)}$ numerically. From the TBA equations, it is easy to find $m_{1,2}^{(n-1)}=-m_{1,1}^{(n-1)}$, which is also checked from the WKB periods.
\begin{table}[htbp]
    \centering
    {\small
    \begin{tabular}{c||c|c}
        $n$ & $\Pi_{\hat{\gamma}_{1,1}}^{(n)}$ & $m_{1, 1}^{(n - 1)}$ \\\hline
        $2$     & $0.2172157436i$   & $0.2172157436i$ \\
        $6$     & $-1.519567945i$   & $-1.519567945i$  \\
        $8$     & $-20.48661777i$   & $-20.48661776i$\\
        $12$    & $20065.20970i$    & $20065.20605i$ \\
        $14$    & $1160395.676i$    & $1160393.422i$  
    \end{tabular}}
    \caption{The quantum corrections for $p(x) = -x^3 + 7x + 6$.}
    \label{tab:A2A2_minimal}
\end{table}
One of the interesting properties of $(A_2, A_2)$-type TBA equation is that the quantum corrections for $\log Y_{1,1}$ and $\log Y_{1,2}$ become the same up to the sign, which can be checked numerically from the calculations of the WKB periods. It is worth noting that this symmetry of quantum correction is very non-trivial from the viewpoint of the Picard-Fuchs operator. 

Another characteristic of $(A_2, A_2)$-type TBA is that the correction terms vanish when $m_{1,1}=m_{1,2}$, which can be produced by considering the potential such that the zeros are symmetric with respect to the midpoint. An example of the potential is $p(x)=x(x+v)(x-v)$.
The $(A_2, A_2)$-type TBA equations \eqref{eq:TBA_A2A2_minimal} become
\begin{equation}
    \begin{aligned}
        \log{Y_{1, 1}(\theta - i\phi_{1})} &= \abs{m_{1, 1}}e^{\theta} + K \star  (\overline{L}_{1, 1} - \overline{L}_{1, 2}),\\
        \log{Y_{1, 2}(\theta - i\phi_{2})} &= \abs{m_{1, 2}}e^{\theta} - K \star (\overline{L}_{1, 1} - \overline{L}_{1, 2}).
    \end{aligned}
\end{equation}
Then one finds $\overline{L}_{a, 1} = \overline{L}_{a, 2}$, which means that the higher order terms of $e^{-\theta}$-expansion $m_{1, k}^{(n)}$, $n = 1, \dots, \infty$ vanish.
This corresponds to the case $D_0$ in \eqref{eq:PF_D0} and the Picard-Fuchs operators vanish. This example can be regarded as the third-order version of the harmonic oscillator.

\subsubsection{\texorpdfstring{$(A_2, A_3)$}{A2A3}}
Let us demonstrate another example for the third order ODE with the potential of fourth order in the minimal chamber.
The Y-system \eqref{eq:Y-function_A2AN} is
\begin{equation}
    \begin{aligned}
    Y_{a,1}^{[+1]}Y_{a,1}^{[-1]}&=\frac{\prod_{b=1}^{r}(1+Y_{b,1})^{G_{ab}}}{(1+Y_{a,2}^{-1})},\quad Y_{a,2}^{[+1]}Y_{a,2}^{[-1]}=\frac{\prod_{b=1}^{r}(1+Y_{b,2})^{G_{ab}}}{(1+Y_{a,1}^{-1})(1+Y_{a,3}^{-1})},\\ Y_{a,3}^{[+1]}Y_{a,3}^{[-1]}&=\frac{\prod_{b=1}^{r}(1+Y_{b,3})^{G_{ab}}}{(1+Y_{a,2}^{-1})},
    \end{aligned}
\end{equation}
where $a=1,2$. The TBA equation \eqref{eq:tba-shifted} can be written as
\begin{equation}
    \begin{aligned}
     \log Y_{1,1}(\theta-i\phi_{1})=&|m_{1,1}|e^{\theta}+K\star\overline{L}_{1,1}-K_{1,2}\star\overline{L}_{1,2},\\
       \log Y_{1,2}(\theta-i\phi_{2})=&|m_{1,2}|e^{\theta}-K_{2,1}\star\overline{L}_{1,1}+K\star\overline{L}_{1,2}-K_{2,3}\star\overline{L}_{1,3},\\
    \log Y_{1,3}(\theta-i\phi_{3})=&|m_{1,3}|e^{\theta}+K\star\overline{L}_{1,3}-K_{3,2}\star\overline{L}_{1,2}
    \end{aligned}
\end{equation}
with effective central charge $c_{\rm eff}=\frac{24}{7}$.
As an example, we set the potential to be
\begin{equation}
    p(x)=-(x-4)(x-1)(x+1)(x+2)=-x^4+2x^3+9x^2-2x-8.
\end{equation}
The masses $m_{1,k}$, the corrections $m_{1,k}^{(n-1)}$ and their counterpart periods $\Pi_{\hat{\gamma}_{1,k}}^{(n)}$ are shown in Table \ref{tab:A2A3_minimal}.
Again, one sees the agreements of the WKB periods and the Y-functions numerically.
\begin{table}[htbp]
    \centering
    {\small
    \begin{tabular}{c||c|c|c}
        $n$ & $\Pi_{\hat{\gamma}_{1,1}}^{(n)}$ & $e^{\frac{\pi i}{3}(1-n)}\Pi_{\hat{\gamma}_{1,2}}^{(n)}$ & $e^{\frac{2\pi i}{3}(1-n)}\Pi_{\hat{\gamma}_{1,3}}^{(n)}$ \\\hline
        $0$     & $14.29120679i$        & $5.748396528i$    & $2.197175863i$ \\
        $2$     & $0.06210586398i$      & $0.1352435882i$   & $-0.09711585835i$\\
        $6$     & $-0.002344098573i$    & $-0.2410756322i$  & $0.2410502603i$\\
        $8$     & $- 0.002115215318i$   & $-1.544858396i$   & $1.544854731i$\\
        $12$    & $0.009495408700i$     & $339.9172291i$    & $-339.9172287i$\\
        $14$    & $0.03754344682i$      & $9328.147139i$    & $-9328.147139i$ \\\hline\hline
        $n$ & $m_{1, 1}^{(n - 1)}$ & $m_{1, 2}^{(n - 1)}$ & $m_{1, 3}^{(n - 1)}$ \\\hline
        $0$     & $14.29120679i$    & $5.748396528i$    & $2.197175863i$\\
        $2$     & $0.06210586398i$  & $0.1352435882i$   & $-0.09711585835i$\\
        $6$     & $-0.002344098573i$& $-0.2410756322i$  & $0.2410502602i$\\
        $8$     & $-0.002115215317i$& $-1.544858396i$   & $1.544854731i$\\
        $12$    & $0.009495407453i$ & $339.9171779i$    & $-339.9171774i$\\
        $14$    & $0.03754338997i$  & $9328.131493i$    & $-9328.131493i$
    \end{tabular}}
    \caption{The quantum corrections for $p(x)=-x^4+2x^3+9x^2-2x-8$.}
    \label{tab:A2A3_minimal}
\end{table}

\subsection{Discontinuity of WKB period from TBA equations}
So far, we have shown the identification between the Y-function and the WKB period in the $\epsilon$-expansion \eqref{eq:ident_Y_with_period}. Thus the  WKB period and the $\epsilon$ expansion of the Y-function are the same asymptotic series of $\epsilon$. 
Moreover, since the Y-function is an analytic function on the $\epsilon$-plane, our TBA equations include more information i.e. singularities. To make a more precise statement for \eqref{eq:ident_Y_with_period}, we replace the asymptotic series with the Borel-resummed one:
\begin{equation}
    \begin{aligned}
    \log Y_{a,k}(\theta)
    = s\qty(\qty[\frac{1}{\epsilon}\Pi_{\hat{\gamma}_{a,k}}]^{[a-k]}),
    \end{aligned}
\end{equation}
where the right hand side has discontinuity on the $\epsilon$ plane as discussed in section \ref{sec:Borel_resum}.
To test this statement, we compute the discontinuity formula of the Y-functions, and then compare them with the one of the Borel resummed WKB periods. For simplicity, we will focus on the minimal chamber of $(A_2, A_2)$ case. Other cases can be studied in a similar way.
 
Let us consider the case ${\rm Arg}(m_{1,1})=\phi_1, {\rm Arg}(m_{1,2})=\phi_2$ with $-\frac{\pi}{3}<\phi_2-\phi_1<\frac{\pi}{3}$. The TBA equations are given by \eqref{eq:TBA_A2A2_minimal}.
We thus can express the WKB periods by
\begin{equation}
    \begin{aligned}
        \log Y_{1,1}(\theta) =m_{1,1}e^{\theta}&+\int_{-\infty}^\infty \dd\theta^\prime  K(\theta-\theta^{\prime}+i\phi_{1})\overline{L}_{1,1}(\theta^{\prime})\\
        &-\int_{-\infty}^\infty \dd\theta^\prime K(\theta-\theta^{\prime}+i\phi_{2})\overline{L}_{1,2}(\theta^{\prime}) +\cdots, \\
        \log Y_{1,2}(\theta-\frac{\pi i}{3})=m_{1,2}e^{\theta-\frac{\pi i}{3}}&+\int_{-\infty}^\infty \dd\theta^\prime K(\theta-\theta^{\prime}+i\phi_{2}-\frac{\pi i}{3})\overline{L}_{1,2}(\theta^{\prime})\\
        &-\int_{-\infty}^\infty \dd\theta^\prime K(\theta-\theta^{\prime}+i\phi_{1}-\frac{\pi i}{3})\overline{L}_{1,1}(\theta^{\prime}) +\cdots,
    \end{aligned}
\end{equation}
where the $\cdots$ means the residue due to the shift of $\theta$, which can be ignored in the discussion of discontinuity. 
From the pole structure of the kernels in the TBA equations, we find $\log Y_{1,1}(\theta)$ is Borel non-summable in the directions
\begin{equation}
    \theta=\pm(\frac{\pi}{3}+n\pi)-\phi_{1},\quad\pm(\frac{2\pi}{3}+n\pi)-\phi_{1},\quad \pm(\frac{\pi}{3}+n\pi)-\phi_{2},\quad\pm(\frac{2\pi}{3}+n\pi)-\phi_{2},
\end{equation}
where $n\in \mathbb{Z}_{\geq 0}$.
$\log Y_{1,2}(\theta-\frac{\pi i}{3})$ is non-summable in the direction
\begin{equation}
\label{nonsum-dirc}
\begin{aligned}
  \theta=&-\phi_{1}-n\pi,\quad(-\frac{\pi}{3}-n\pi)-\phi_{1},\quad(\frac{2\pi}{3}+n\pi)-\phi_{1},\quad \pi-\phi_{1}+n\pi,\\
  &-\phi_{2}-n\pi,\quad(-\frac{\pi}{3}-n\pi)-\phi_{2},\quad(\frac{2\pi}{3}+n\pi)-\phi_{2},\quad \pi-\phi_{2}+n\pi.
  \end{aligned}
\end{equation}
When ${\rm Arg}(m_{1,1})={\rm Arg}(m_{1,2})=\frac{\pi}{2}$, i.e. the case studied in table \ref{tab:A2A2_minimal}, one finds the directions of non-summability of Y-functions from (\ref{nonsum-dirc}), which reproduce the locations of the discontinuity of $s(\qty[\frac{1}{\epsilon}\Pi_{\hat{\gamma}_{1,1}}])(\theta)$ and $s(\qty[\frac{1}{\epsilon}\Pi_{\hat{\gamma}_{1,2}}])(\theta)$ obtained from the Borel-Pad\'e technique, See Fig.\ref{fig:discon}. Since the exact WKB periods are uniquely determined by their asymptotic behaviors and the discontinuity \cite{voros-quartic}, we thus expect our TBA equations provide the exact form of the WKB periods.

\section{Wall-crossing of TBA equations}\label{sec:wc}
In ${\cal N}=2$ supersymmetric gauge theory, the BPS spectrum can be expressed by using the central charge in the ${\cal N}=2$ SUSY algebra:
\begin{equation}
    Z_{(\vec{n}_e,\vec{n}_m)}=\vec{a}\cdot\vec{n}_{e}+\vec{a}_{D}\cdot\vec{n}_{m} ,\quad M_{(\vec{n}_e,\vec{n}_m)}=|Z_{(\vec{n}_e,\vec{n}_m)}|,
\end{equation}
where $M$ is the mass of the BPS particle with charge $(\vec{n}_e,\vec{n}_m)$. $\vec{a}$ is the vev of the scalars in the Cartan subalgebra, and $\vec{a}_D$ is its dual.
The central charge corresponds to the SW period $\Pi_\gamma$ of the low-energy effective theory, associated with the cycle $\gamma=n_e^i \alpha^i+n_m^i \beta^i$, where $\alpha^i$ and $\beta^i$ are one-cycle for $a^i$ and $a_D^i$, respectively.
Let us consider the decay process $\gamma_0\to \sum_{i=1}^p\gamma_i$. From the conservation of the charges and masses, one finds that the phase of each $\Pi_{\gamma_i}$ in the decay process has to be the same value. In other words, when the vectors of the SW periods are in parallel, which leads to the wall of marginal stability in the moduli space, the BPS particle is unstable. Let us spell out this in detail in the case of $(A_2, A_2)$ AD theory, where we have seen two independent SW periods $\Pi_{\hat{\gamma}_{1,1}}^{(0)}$ and $\Pi^{(0)}_{\hat{\gamma}_{1,2}}$.
The marginal stability walls are located at 
\begin{equation}
\label{wall-cond-1}
    {\rm Im}\qty(\frac{\Pi_{\hat{\gamma}_{1,1}}^{(0)}}{\Pi_{\hat{\gamma}_{1,2}}^{(0)}})= \abs{\frac{\Pi_{\hat{\gamma}_{1,1}}^{(0)}}{\Pi_{\hat{\gamma}_{1,2}}^{(0)}}}\sin(\phi_1-\phi_2+\frac{\pi}{3})=0,
\end{equation}
where we used $m_{1,1}=\Pi_{\hat{\gamma}_{1,1}}^{(0)}$ and $m_{1,2}=e^{\frac{\pi i}{3}}\Pi_{\hat{\gamma}_{1,2}}^{(0)}$, and $\phi_{k}$ is the phase of $m_{1,k}$.
The condition \eqref{wall-cond-1} thus leads to half of the marginal stability conditions.
Note that $\Pi^{(0)}_{\hat{\gamma}_{2,2}}=e^{\frac{2\pi i}{3}}\Pi^{(0)}_{\hat{\gamma}_{1,2}}$, which implies $m_{1,2}=e^{-\frac{i\pi}{3}} \Pi^{(0)}_{\hat{\gamma}_{2,2}}$. Then the marginal stability condition 
\begin{align}
    {\rm Im} \qty( \frac{\Pi^{(0)}_{\hat{\gamma}_{2,2}}}{\Pi^{(0)}_{\hat{\gamma}_{1,1}}})
    &=\abs{\frac{\Pi^{(0)}_{\hat{\gamma}_{2,2}}}{\Pi_{\hat{\gamma}_{1,1}}^{(0)}}} \sin(\phi_2+\frac{\pi}{3}-\phi_1)=0
\end{align}
leads to the remaining half of the marginal stability conditions. Therefore, the marginal stability walls are located at
\begin{equation}
    \phi_2-\phi_1=\pm \frac{\pi}{3},\quad \pm \frac{2\pi}{3},
\end{equation}
which is nothing but the locations of the pole in the TBA equations.

The marginal stability walls divide the moduli space into several chambers. Crossing these walls, we have to modify the TBA equation, which is called the wall-crossing of the TBA equations.
The chamber with the minimal (maximal) number of BPS spectrum is called the minimal (maximal) chamber. Note that the general structure of the wall-crossing of the BPS spectrum in the rank one gauge theory is first presented in \cite{Gaiotto:2009hg}. However, the general case for higher rank gauge theory/higher order ODE is not well known\footnote{See \cite{Chen:2011gk} for higher rank cases.}.
The TBA equations presented in the previous section are valid in the minimal chamber of moduli space. In this section, we analytically continue the TBA equations to the whole moduli space, and show the identification between new Y-functions and WKB periods.

\subsection{General procedure of wall-crossing of TBA equations}
Now we will see how the TBA equations change under the wall-crossing.
Let us consider the TBA equations \eqref{eq:tba-shifted} of $(A_2, A_N)$-type in the minimal chamber, which are written as
\begin{equation}
    \begin{aligned}
       \log Y_{1,1}(\theta-i\phi_{1})=&|m_{1,1}|e^{\theta}+K\star\overline{L}_{1,1}-K_{1,2}\star\overline{L}_{1,2},\\
       \log Y_{1,2}(\theta-i\phi_{2})=&|m_{1,2}|e^{\theta}-K_{2,1}\star\overline{L}_{1,1}+K\star\overline{L}_{1,2}-K_{2,3}\star\overline{L}_{1,3},\\
    \log Y_{1,3}(\theta-i\phi_{3})=&|m_{1,3}|e^{\theta}+K\star\overline{L}_{1,3}-K_{3,2}\star\overline{L}_{1,2}-K_{3,4}\star\overline{L}_{1,4},\cdots.
    \end{aligned}\label{eq:A2A2-TBA-min}
\end{equation}
The TBA equations (\ref{eq:A2A2-TBA-min}) are valid only in the minimal chamber, namely the region
\begin{equation}
    |\phi_k-\phi_{k\pm 1}|<\frac{\pi}{3}.
\end{equation}
When the moduli parameters change such that  $\phi_k-\phi_{k\pm 1}$ crosses $\pm \frac{\pi}{3}, \pm \frac{2\pi}{3},\cdots$, one needs to modify the TBA equations by picking the contributions of the poles of the kernels, as discussed in the previous section. This modification of TBA equations is known as the wall-crossing of the TBA equations \cite{Alday:2010vh,toledo,toledo-thesis}\footnote{This idea was originally used to calculate the excited states of the integrable model in finite volume \cite{Dorey:1996re}. However, the origins of the poles are different. In \cite{Dorey:1996re}, the pole appears when $1+Y=0$. See also \cite{gabai2021exact} for its application to the Schr\"odinger equation with centrifugal  potential.}.

Without loss of generality, we consider the situation where $\phi_2-\phi_1$ crosses $\pi/3$ while all other $|\phi_k-\phi_{k\pm 1}|<\pi/3$. Other situations of $(A_2, A_N)$ can be treated in a similar way. We thus need to pick the residue of the pole in the kernels of the first and the second TBA equations:
\begin{equation}
\label{eq:2-TBA}
    \begin{aligned}
    \log Y_{1,1}(\theta-i\phi_{1})=&|m_{1,1}|e^{\theta}+K\star\overline{L}_{1,1}-K_{1,2}\star\overline{L}_{1,2}-L_{1,2}(\theta-\frac{\pi i}{3}-i\phi_{1}),\\
    \log Y_{1,2}(\theta-i\phi_{2})=&|m_{1,2}|e^{\theta}-K_{2,1}\star\overline{L}_{1,1}+K\star\overline{L}_{1,2}-K_{2,3}\star\overline{L}_{1,3}-L_{1,1}(\theta+\frac{\pi i}{3}-i\phi_{2}),\\
    \log Y_{1,3}(\theta-i\phi_{3})=&|m_{1,3}|e^{\theta}+K\star\overline{L}_{1,3}-K_{3,2}\star\overline{L}_{1,2}-K_{3,4}\star\overline{L}_{1,4},\,
    \cdots\,,
   \end{aligned}
\end{equation}
while other TBA equations keep the same form. To obtain a closed system, we also need to shift the spectral parameter of $Y_{1,1}$ and $Y_{1,2}$ to obtain the equations for $\log Y_{1,1}(\theta+\frac{\pi i}{3}-i\phi_2)$ and $\log Y_{1,2}(\theta-\frac{\pi i}{3}-i\phi_1)$. We thus obtain a closed system with $N+2$ TBA equations.

It would be more interesting to introduce new Y-functions $Y_{1,k}^{\rm n}$ and $Y_{12}^{\rm n}$
\begin{equation}
    \begin{aligned}
        Y_{1,1}^{\rm n}(\theta)&=Y_{1,1}(\theta)\big(1+\frac{1}{Y_{1,2}(\theta-\frac{\pi i}{3})}\big), \quad Y_{1,2}^{\rm n}(\theta)=Y_{1,2}(\theta)\big(1+\frac{1}{Y_{1,1}(\theta+\frac{\pi i}{3})}\big),\\
        Y_{12}^{\rm n}(\theta)&=\frac{1+\frac{1}{Y_{1,2}(\theta-\frac{\pi i}{3})}+\frac{1}{Y_{1,1}(\theta)}}{\frac{1}{Y_{1,1}(\theta)Y_{1,2}(\theta-\frac{\pi i}{3})}},\quad Y^{\rm n}_{1,k>2}=Y_{1,k>2}
    \end{aligned}
\end{equation}
to absorb the residue in the right hand side of (\ref{eq:2-TBA}), satisfying
\begin{equation}\label{eq:oldY-newY}
    \begin{aligned}
        &\big(1+\frac{1}{Y_{1,1}(\theta)}\big)=\big(1+\frac{1}{Y_{1,1}^{{\rm n}}(\theta)}\big)\big(1+\frac{1}{Y_{12}^{{\rm n}}(\theta)}\big),\\
        &\big(1+\frac{1}{Y_{1,2}(\theta)}\big)=\big(1+\frac{1}{Y_{1,2}^{{\rm n}}(\theta)}\big)\big(1+\frac{1}{Y_{12}^{{\rm n}}(\theta+\frac{\pi i}{3})}\big),\\
        &Y_{1,1}^{{\rm n}}(\theta)Y_{1,2}^{{\rm n}}(\theta-\frac{\pi i}{3})=Y_{12}^{{\rm n}}(\theta)\big(1+\frac{1}{Y_{12}^{{\rm n}}(\theta)}\big).
    \end{aligned}
\end{equation}
From \eqref{eq:oldY-newY}, one finds the mass for the Y-function  $Y_{12}^{\rm n}(\theta)$ is $m_{12}=m_{1,1}+m_{1,2}e^{-\frac{\pi i}{3}}$, whose phase is defined by $\phi_{12}$.
The first three TBA equations (\ref{eq:2-TBA}) thus become
\begin{equation}
    \begin{aligned}
        \log Y_{1,1}^{{\rm n}}(\theta-i\phi_{1})=&|m_{1,1}|e^{\theta}+K\star\overline{L}_{1,1}^{{\rm n}}-K_{1,2}\star\overline{L}_{1,2}^{{\rm n}}+K_{1,12}^-\star\overline{L}_{12}^{{\rm n}},\\
        \log Y_{1,2}^{{\rm n}}(\theta-i\phi_{2})=&|m_{1,2}|e^{\theta}+K\star\overline{L}_{1,2}^{{\rm n}}-K_{2,1}\star\overline{L}_{1,1}^{{\rm n}}-K_{2,3}\star\overline{L}_{1,3}^{{\rm n}}-K_{2,12}^-\star\overline{L}_{12}^{{\rm n}},\\
        \log Y_{1,3}^{\rm n}(\theta-i\phi_{3})=&|m_{1,3}|e^{\theta}+K\star\overline{L}_{1,3}^{{\rm n}}-K_{3,2}\star\overline{L}_{1,2}^{{\rm n}}-K_{3,12}^{+}\star\overline{L}_{12}^{{\rm n}}-K_{3,4}\star\overline{L}_{1,4}^{{\rm n}},
    \end{aligned}\label{eq:nTBA-123}
\end{equation}
where $K^{\pm}(\theta)=K^{[\mp1]}(\theta)=K(\theta\pm \frac{\pi i}{3})$.
We have shifted the integral path associated with $Y^{\rm n}_{12}$ because the values $|\phi_{12}-\phi_{1}|, |\phi_2-\frac{\pi }{3}-\phi_{12}|$ are small enough. To obtain the TBA equations associated with $Y_{12}^{\rm n}$, we evaluate the first two TBA equations in (\ref{eq:2-TBA}) at the appropriate value and then take the summation 
\begin{equation}
    \log Y_{12}^{{\rm n}}(\theta-i\phi_{12})=|m_{12}|e^{\theta}+K\star\tilde{L}_{12}^{{\rm n}}-K_{12,3}^{-}\star\overline{L}_{1,3}+K_{12,1}^+\star\tilde{L}_{1,1}^{{\rm n}}-K_{12,2}^+\star\tilde{L}_{1,2}^{{\rm n}}.
    \label{eq:nTBA-n12}
\end{equation}
(\ref{eq:nTBA-123}),(\ref{eq:nTBA-n12}) and other $N-3$ TBA equations provide a closed system with $N+1$ TBA equations. Using the relation \eqref{eq:oldY-newY}, it is easy to rewrite the effective central charge \eqref{ceff} in terms of the new Y-functions
\begin{equation}
    c_{\rm eff}=2\times \frac{6}{\pi^2}\int^\infty_{-\infty}\Big(|m_{12}|e^\theta\overline{L}_{12}^{\rm n}+\sum_{k=1}^N|m_{1,k}|e^\theta \overline{L}_{1,k}^{\rm n}\Big)d\theta.
\end{equation}
Similarly, as in the case in the minimal chamber, one can find the constant value of the Y-function at $\theta\to -\infty$ through
\begin{equation}\label{eq:const-Y}
    \log \vec{Y}(-\infty-i\phi)=M\cdot \vec{\overline{L}}(-\infty),
\end{equation}
where $\vec{Y}$ and $\vec{\overline{L}}$ are the vector of $(Y^{\rm n}_{1,k}(-\infty-i\phi_k), Y_{12}^{\rm n}(-\infty-i\phi_{12}))^t$ and $(\overline{L}^{\rm n}_{1,k}(-\infty), \overline{L}_{12}^{\rm n}(-\infty))^t$, respectively. $M$ is a constant matrix which can be obtained by the integration of the kernel.
$M$ represents the connection matrix of the Y-functions in the TBA equations at $\theta\to -\infty$ \cite{Emery:2020qqu}. The matrix $M$ can also be obtained from the Fourier transform of the kernel matrix evaluated at zero momentum. Using the Rogers dilogarithm identity and the constant solution of the new Y-function, one finds the value of the effective central charge is the same as the one in the minimal chamber.

From the asymptotic behavior of the TBA equations, we find the new Y-functions $Y_{1,1}^{\rm n}$ and $Y_{1,2}^{\rm n}$ are related to the cycles $\hat{\gamma}_{1,1}=\gamma_{1,1}$ and $\hat{\gamma}_{1,2}=\gamma_{3,2}$. We thus find that the new Y-functions are identified with the WKB periods at the region $\pi/3<\phi_2-\phi_1<2\pi/3$:
\begin{equation}
    \log Y_{1,1}^{\rm n}(\theta)=e^{\theta}\Pi_{\hat{\gamma}_{1,1}},\quad    \log Y_{1,2}^{\rm n}(\theta)=e^{\frac{\pi i}{3}}e^{\theta}\Pi_{\hat{\gamma}_{1,2}}(\theta+\frac{\pi i}{3}).
    \label{eq:new_Yfunction}
\end{equation}
Noting the relation
\begin{equation}
    m_{12}=m_{1,1}+e^{-\frac{\pi i}{3}}m_{1,2}=\Pi^{(0)}_{\hat{\gamma}_{1,1}}+\Pi^{(0)}_{\hat{\gamma}_{1,2}},
\end{equation}
we find $Y^{\rm n}_{12}$ is related to the cycle $\hat{\gamma}_{1,1}+\hat{\gamma}_{1,2}$:
\begin{equation}
    \log Y_{12}^{\rm n}(\theta)=e^{\theta}\Pi_{\hat{\gamma}_{1,1}}(\theta) + e^{\theta}\Pi_{\hat{\gamma}_{1,2}}(\theta)= e^{\theta}\Pi_{\hat{\gamma}_{1,1}+\hat{\gamma}_{1,2}}.
\end{equation}
One has to pick the contribution of the pole in the TBA equation of $\log Y^{\rm n}_{1,2}(\theta-\frac{\pi i}{3})$ if one wants to compute $e^{\theta}\Pi_{\hat{\gamma}_{1,2}}(\theta)$ from the second equation of \eqref{eq:new_Yfunction}.
From the viewpoint of the Gauge theory, the basis of BPS states changes from $(\hat{\gamma}_{1,1}, \hat{\gamma}_{1,2},\cdots, \hat{\gamma}_{1,N})$ to  $(\hat{\gamma}_{1,1}, \hat{\gamma}_{1,2},\cdots, \hat{\gamma}_{1,N}, \hat{\gamma}_{1,1}+\hat{\gamma}_{1,2})$ in the progress of wall-crossing \cite{Gaiotto:2009hg}. We thus need to identify the WKB periods with the new Y-functions rather than the original one.
Similar redefinitions are found for further wall-crossings. We will repeat this procedure from the minimal chamber to the maximal chamber for non-trivial examples of $(A_2,A_2)$ and $(A_2,A_3)$ in the following. The first wall-crossing of $(A_2,A_2)$ TBA has been studied in \cite{Ito:2021boh}. We, however, repeat the analysis here for completeness.

\subsection{\texorpdfstring{$(A_2, A_2)$}{A2A2}}
\begin{figure}[htbp]
    \centering
    \includegraphics[width=8cm]{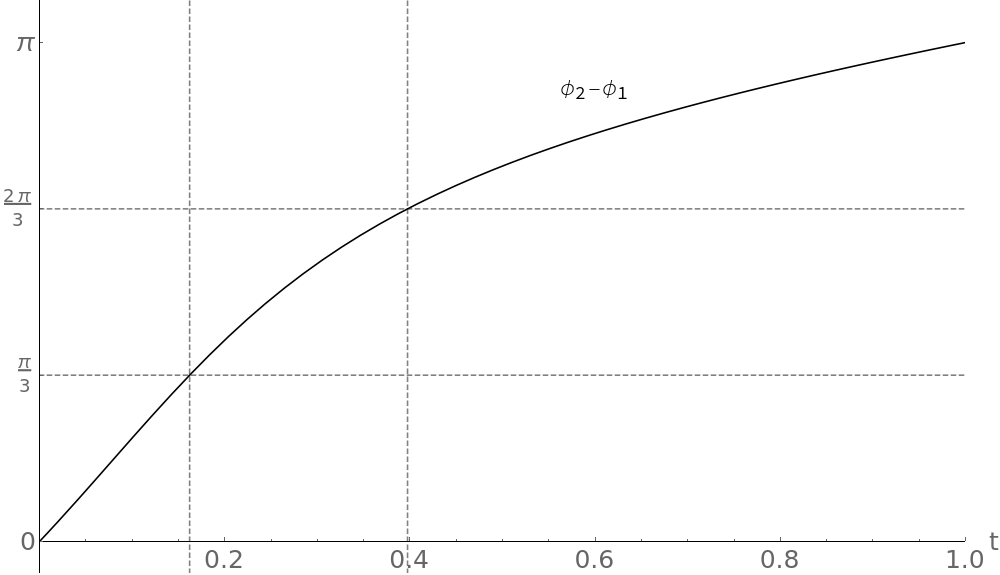}
    \caption{$t$-dependence of the phase difference $\phi_2-\phi_1$ for $0\leq t\leq1$.}
    \label{fig:A2A2_phases}
\end{figure}

In this subsection, we study the wall-crossing of the $(A_2, A_2)$ TBA equations and compare the Y-functions with the WKB periods in each chamber. Let us consider particular points in the minimal chamber and the maximal chamber with the potential $p(x)=(x-3)(x+1)(x+2)=x^3-7x-6$ and $p(x)=x^3-8$, respectively. 
These two sets of branch points are interpolated by the path
\begin{equation}\label{eq:A2A2_zeros}
    x_0(t)=3-t,\quad x_1(t)=-1+\sqrt{3}it,\quad x_2(t)=-2+t-\sqrt{3}it,\quad 0\leq t\leq1,
\end{equation}
whose potential is given by $p(x;t)=(x-x_0(t))(x-x_1(t))(x-x_2(t))$.
From this potential, we are able to compute the classical periods and masses for Y-functions through \eqref{eq:m-period} for $0\leq t\leq1$. The difference of phases $\phi_1$ and $\phi_2$ for $0\leq t\leq1$ are plotted in figure \ref{fig:A2A2_phases}.
In the path \eqref{eq:A2A2_zeros}, we thus find two walls associated with $\phi_2-\phi_1$: 
\begin{align}
    t&=0.162117..., & &\phi_2-\phi_1=\frac{\pi}{3},& &\Im\qty(\frac{\Pi_{\gamma_{3,2}}^{(0)}}{\Pi_{\gamma_{1,1}}^{(0)}})=0,\label{eq:A2A2-wc-cond1}\\
    t&=0.397459..., & &\phi_2-\phi_1=\frac{2\pi}{3},& &\Im\qty(\frac{\Pi_{\gamma_{2,2}}^{(0)}}{\Pi_{\gamma_{1,1}}^{(0)}})=0.\label{eq:A2A2-wc-cond2}
\end{align}

\subsubsection{The first wall-crossing}
The first wall-crossing occurs when $\phi_2-\phi_1$ cross $\frac{\pi}{3}$. Following the general procedure in the previous subsection, we define the new Y-functions
\begin{equation}
\begin{aligned}
    Y_{1,1}^{(1)}(\theta)&=Y_{1,1}(\theta)\big(1+\frac{1}{Y_{1,2}(\theta-\frac{\pi i}{3})}\big), \quad Y_{1,2}^{(1)}(\theta)=Y_{1,2}(\theta)\big(1+\frac{1}{Y_{1,1}(\theta+\frac{\pi i}{3})}\big),\\
    Y_{12}^{(1)}(\theta)&=\frac{1+\frac{1}{Y_{1,2}(\theta-\frac{\pi i}{3})}+\frac{1}{Y_{1,1}(\theta)}}{\frac{1}{Y_{1,1}(\theta)Y_{1,2}(\theta-\frac{\pi i}{3})}},
\end{aligned}
\end{equation}
where the superscript ${(i)}$ labels the Y-function after the $i$-th wall-crossing.
The TBA equations \eqref{eq:nTBA-123} and \eqref{eq:nTBA-n12} read
\begin{equation}
    \begin{aligned}
      \log Y_{1,1}^{(1)}(\theta-i\phi_{1})=&|m_{1,1}|e^{\theta}+K\star\overline{L}_{1,1}^{(1)}-K_{1,2}\star\overline{L}_{1,2}^{(1)}+K_{1,12}^{-}\star\overline{L}_{12}^{(1)},\\
      \log Y_{1,2}^{(1)}(\theta-i\phi_{2})=&|m_{1,2}|e^{\theta}+K\star\overline{L}_{1,2}^{(1)}-K_{2,1}\star\overline{L}_{1,1}^{(1)}-K_{2,12}^{-}\star\overline{L}_{12}^{(1)},\\
      \log Y_{12}^{(1)}(\theta-i\phi_{12})=&|m_{12}|e^{\theta}+K\star\overline{L}_{12}^{(1)}+K_{12,1}^{+}\star\overline{L}_{1,1}^{(1)}-K_{12,2}^{+}\star\overline{L}_{1,2}^{(1)}.
    \end{aligned}\label{eq:A2A2-1st-wc}
\end{equation}
The matrix $M$ introduced in \eqref{eq:const-Y} now becomes
\begin{equation}
    M^{(1)}=\left(
    \begin{array}{ccc}
     1 & 0 & 1 \\
     0 & 1 & 1 \\
     1 & 1 & 1 \\
    \end{array}
    \right),
\end{equation}
where the vector $\vec{Y}^{(1)}$ is arrayed by $(Y^{(1)}_{1,1}(-\infty-i\phi_1), Y^{(1)}_{1,2}(-\infty-i\phi_2), Y^{(1)}_{12}(-\infty-i\phi_{12}))^t$. We thus can find the constant solutions and the effective value $c_{\rm eff}=2$.

\begin{figure}[htbp]
    \centering
    \includegraphics[width=10cm]{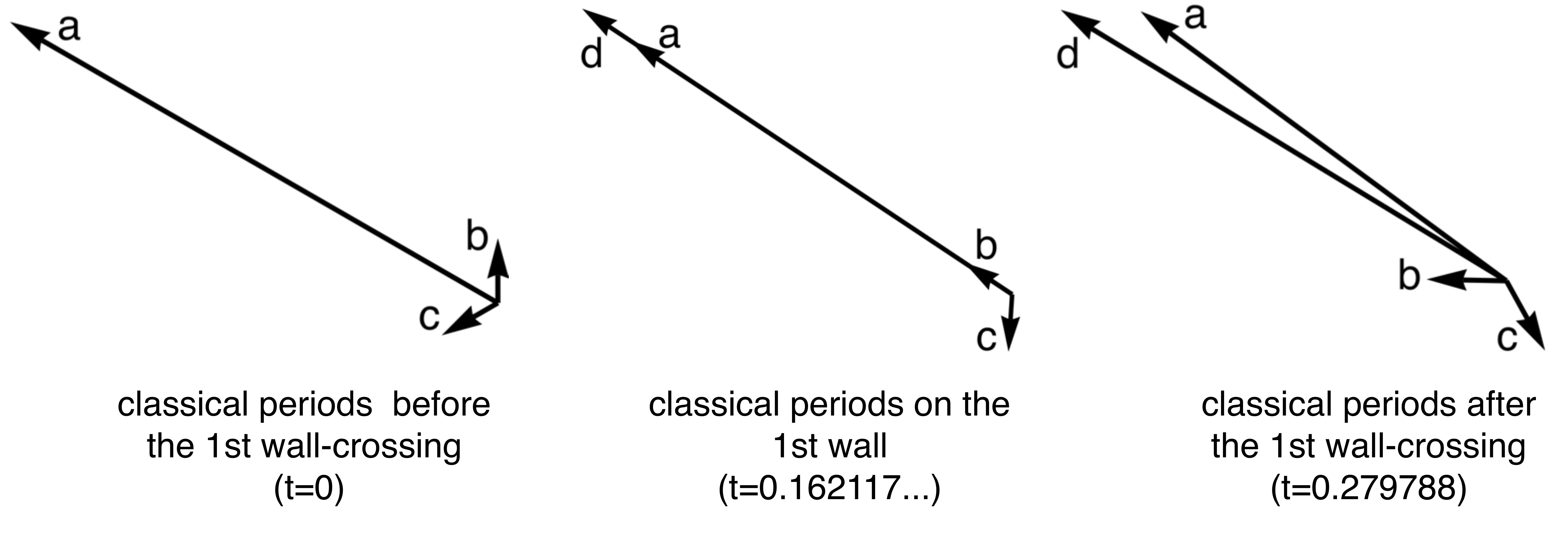}~~~~~~~
    \caption{The classical periods in the process of the first wall-crossing. The vectors of the classical periods before, middle and after the wall-crossing are shown from left to right. The arrows labeled by a,b,c,d and e represent the periods $\Pi_{\gamma_{1,1}}^{(0)},\Pi_{\gamma_{3,2}}^{(0)},\Pi_{\gamma_{1,2}}^{(0)}$ and $\Pi_{\gamma_{1,1}+\gamma_{3,2}}^{(0)}$ respectively.
   }
    \label{fig:A2A2-1stws-period-vec}
\end{figure}

The new Y-functions $Y_{1,1}^{(1)},Y_{1,2}^{(1)}$ and $Y_{12}^{(1)}$ are associated with the cycles $\hat{\gamma}_{1,1}$, $\hat{\gamma}_{1,2}$ and $\hat{\gamma}_{1,1}+\hat{\gamma}_{1,2}$ respectively by
\begin{equation}
    \log Y_{1,1}^{(1)}(\theta)=e^{\theta}\Pi_{\hat{\gamma}_{1,1}},\quad\log Y_{1,2}^{(1)}(\theta)=e^{\frac{\pi i}{3}}e^{\theta}\Pi_{\hat{\gamma}_{1,2}}(\theta+\frac{\pi i}{3}),\quad \log Y_{12}^{(1)}(\theta)=e^{\theta}\Pi_{\hat{\gamma}_{1,1}+\hat{\gamma}_{1,2}}(\theta),\label{eq:1stwc-wkbY}
\end{equation}
which is valid in the region $\pi/3<|\phi_2-\phi_1|<2\pi/3$. To test these identifications, we compare the expansion of WKB periods with the $e^{-\theta}$ expansion of the TBA equations
\begin{equation}
    \begin{aligned}
        &\log{Y_{1, k}^{(1)}}(\theta) = m_{1, k}e^{\theta} + \sum_{n = 1}^{\infty}m_{1, k}^{(1),(n)}e^{-n\theta},\\
        &\log{Y_{12}^{(1)}}(\theta) = m_{12}e^{\theta} + \sum_{n = 1}^{\infty}m_{12}^{(1),(n)}e^{-n\theta}.
    \end{aligned}
\end{equation}
To avoid confusion with many superscripts, we will omit the superscript $(1)$ or $(i)$ for the further wall-crossing on the right hand side.
In table \ref{tab:A2A2_p_t_2}, we perform the numerical comparison of the $\epsilon$ expansion of (\ref{eq:1stwc-wkbY}) at $t=\frac{1}{5}$, which shows the agreement in high precision.
\begin{table}[htbp]
    \centering
    {\small
    \begin{tabular}{c||c|c}
        $n$ & $\Pi_{\hat{\gamma}_{1,1}}^{(n)}$ & $m_{1, 1}^{(n - 1)}$ \\\hline
        $0$ & $-9.747530080 + 6.701716666i$ & $-9.747530080 + 6.701716666i$ \\
        $2$ & $0.1568931454 - 0.1561575487i$ & $0.1568931454 - 0.1561575487i$ \\
        $6$ & $1.037931258 - 0.1841377709i$ & $1.037931259 - 0.1841377696i$ \\
        $8$ & $-0.7530182822 + 12.31639555i$ & $-0.7530183059 + 12.31639556i$\\
        $12$ & $-4657.014593 + 7803.136493i$ & $-4657.014191+ 7803.135760i$\\\hline\hline
        $n$ & $e^{\frac{\pi i}{3}(1-n)}\Pi_{\hat{\gamma}_{1,2}}^{(n)}$ & $m_{1, 2}^{(n - 1)}$ \\\hline
        $0$ & $-1.280370055 - 1.004961987i$ & $-1.280370055 - 1.004961987i$ \\
        $2$ & $-0.1568931454 + 0.1561575487i$ & $-0.1568931454 + 0.1561575487i$ \\
        $6$ & $-1.037931258 + 0.1841377709i$ & $-1.037931259 + 0.1841377696i$ \\
        $8$ & $0.7530182822 - 12.31639555i$ & $0.7530183059 - 12.31639556i$\\
        $12$ & $4657.014593 - 7803.136493i$ & $4657.014191 - 7803.135760i$\\\hline\hline
        $n$ & $\Pi_{\hat{\gamma}_{1,1}+\hat{\gamma}_{1,2}}^{(n)}$ & $m_{12}^{(n - 1)}$ \\\hline
        $0$ & $-11.25803772+7.308068666i$ & $-11.25803772+7.308068666i$ \\
        $2$ & $-0.05678983152 - 0.2139522239i$ & $-0.05678983147 - 0.2139522240i $ \\
        $6$ & $0.6784336165 + 0.8068059514i$ & $0.6784336157 + 0.8068059527i$ \\
        $8$ & $10.28980229 + 6.810330737i$ & $10.28980228 + 6.810330760i$\\
        $12$ & $-9086.221728 - 131.5246971i$ & $-9086.220893 - 131.5247156i$
    \end{tabular}
    }
    \caption{The WKB periods $\Pi_{\hat{\gamma}_{1,1}}^{(n)},e^{\frac{\pi i}{3}(1-n)}\Pi_{\hat{\gamma}_{1,2}}^{(n)}$ and $\Pi_{\hat{\gamma}_{12}}^{(n)}$, and the masses $m_{1,1}^{(n-1)},m_{1,2}^{(n-1)}$ and $m_{12}^{(n-1)}$ for $t = \frac{1}{5} = 0.2$.}
    \label{tab:A2A2_p_t_2}
\end{table}
The way of the first wall-crossing can be visualized by plotting the classical periods on the complex plane, see Fig.\ref{fig:A2A2-1stws-period-vec}.  
Before the wall-crossing (left), we plot the vectors of the classical periods $\Pi_{\gamma_{1,1}}=\Pi_{\hat{\gamma}_{1,1}}$, $\Pi_{{\gamma}_{1,2}}$ and $\Pi_{\gamma_{3,2}}=\Pi_{\hat{\gamma}_{1,2}}$ corresponding to the Y-functions $Y_{1,1}, Y_{2,2}$ and $Y_{1,2}$, respectively. 
At the wall of the first wall-crossing, $\Pi_{\gamma_{1,1}}^{(0)}$ and $\Pi_{\gamma_{3,2}}^{(0)}$ are in parallel; see the middle of the figure and \eqref{eq:A2A2-wc-cond1} for the condition of the marginal stability of the first wall-crossing. 
We thus need to introduce a new period related to the cycle $\gamma_{1,1}+\gamma_{3,2}$ to obtain a closed system.
After the first wall-crossing (right), we thus need to consider  vectors of four periods, including $\Pi_{\gamma_{1,1}}+\Pi_{\gamma_{3,2}}$.

\subsubsection{The second wall-crossing}
As $t$ increases to the value of $t=0.397459...$, one arrives at the second wall-crossing with $\phi_2-\phi_1=\frac{2\pi}{3}$. We pick the poles in the first two TBA equations in \eqref{eq:A2A2-1st-wc} and introduce the new Y-functions. We find a closed TBA system:
\begin{equation}\label{A2A2-max}
    \begin{aligned}
        \log Y_{1,1}^{(2)}(\theta-i\phi_{1})=&|m_{1,1}|e^{\theta}+K\star\overline{L}_{1,1}^{(2)}-K_{1,2}\star\overline{L}_{1,2}^{(2)}+K_{1,12}^{-}\star\overline{L}_{12}^{(2)}\\
        &+(K_{1,\widetilde{12}}+K_{1,\widetilde{12}}^{-})\star\overline{L}_{\widetilde{12}}^{(2)},\\
        \log Y_{1,2}^{(2)}(\theta-i\phi_{2})=&|m_{1,2}|e^{\theta}-K_{2,1}\star\overline{L}_{1,1}^{(2)}+K\star\overline{L}_{1,2}^{(2)}-K_{2,12}^{-}\star\overline{L}_{12}^{(2)}\\
        &-(K_{2,\widetilde{12}}^{-}+K_{2,\widetilde{12}})\star\overline{L}_{\widetilde{12}}^{(2)},\\
        \log Y_{12}^{(2)}(\theta-i\phi_{12})=&|m_{12}|e^{\theta}+K_{12,1}^{+}\star\overline{L}_{1,1}^{(2)}-K_{12,2}^{+}\star\overline{L}_{1,2}^{(2)}+K\star\overline{L}_{12}^{(2)}\\
        &+(K_{12,\widetilde{12}}+K_{12,\widetilde{12}}^{+})\star\overline{L}_{\widetilde{12}}^{(2)},\\
        \log Y_{\widetilde{12}}^{(2)}(\theta-i\phi_{\widetilde{12}})=&|m_{\widetilde{12}}|e^{\theta}+(K_{\widetilde{12},1}+K_{\widetilde{12},1}^{+})\star\overline{L}_{1,1}^{(2)}-(K_{\widetilde{12},2}^{+}+K_{\widetilde{12},2})\star\overline{L}_{1,2}^{(2)}\\&+(K_{\widetilde{12},12}+K_{\widetilde{12},12}^{-})\star\overline{L}_{12}^{(2)}+3K\star\overline{L}_{\widetilde{12}}^{(2)},
    \end{aligned}
\end{equation}
where we have introduced the new Y-functions $Y_{1,1}^{(2)}, Y_{1,2}^{(2)}$, and $Y_{\widetilde{12}}^{(2)}$ defined by
\begin{equation}
    \begin{aligned}
        Y_{1,1}^{(2)}(\theta)&=Y_{1,1}^{(1)}(\theta)\big(1+\frac{1}{Y_{1,2}^{(1)}(\theta-\frac{2\pi i}{3})}\big),\quad Y_{1,2}^{(2)}(\theta)=Y_{1,2}^{(1)}(\theta)\big(1+\frac{1}{Y_{1,1}^{(1)}(\theta+\frac{2\pi i}{3})}\big),\\
        Y_{\widetilde{12}}^{(2)}(\theta)&=\frac{1+\frac{1}{Y_{1,2}^{(1)}(\theta-\frac{2\pi i}{3})}+\frac{1}{Y_{1,1}^{(1)}(\theta)}}{\frac{1}{Y_{1,1}^{(1)}(\theta)Y_{1,2}^{(1)}(\theta-\frac{2\pi i}{3})}},\quad Y_{12}^{(2)}(\theta)=Y_{12}^{(1)}(\theta).
    \end{aligned}
\end{equation}
The mass of the new Y-function $Y_{\widetilde{12}}^{(2)}(\theta)$ is
\begin{equation}
   m_{\widetilde{12}}=m_{1,1}+e^{-\frac{2\pi i}{3}}m_{1,2}=\Pi^{(0)}_{\gamma_{1,1}-\gamma_{1,2}},
\end{equation}
whose phase is defined by $\phi_{\widetilde{12}}$. 
The connection matrix $M$ and the vector $\vec{Y}^{(2)}$ become
\begin{equation}
    M^{(2)}=\left(
    \begin{array}{cccc}
     1 & 1 & 1 & 2 \\
     1 & 1 & 1 & 2 \\
     1 & 1 & 1 & 2 \\
     2 & 2 & 2 & 3 \\
    \end{array}
    \right),\qquad
    \vec{Y}^{(2)}=\mqty(
    Y^{(2)}_{1,1}(-\infty-i\phi_1)\\
    Y^{(2)}_{1,2}(-\infty-i\phi_2)\\
    Y^{(2)}_{12}(-\infty-i\phi_{12})\\
    Y^{(2)}_{\widetilde{12}}(-\infty-i\phi_{\widetilde{12}})
    ).
\end{equation}
We thus can find the constant solutions and the effective value $c_{\rm eff}=2$.

\begin{figure}[htbp]
    \centering
    \includegraphics[width=10cm]{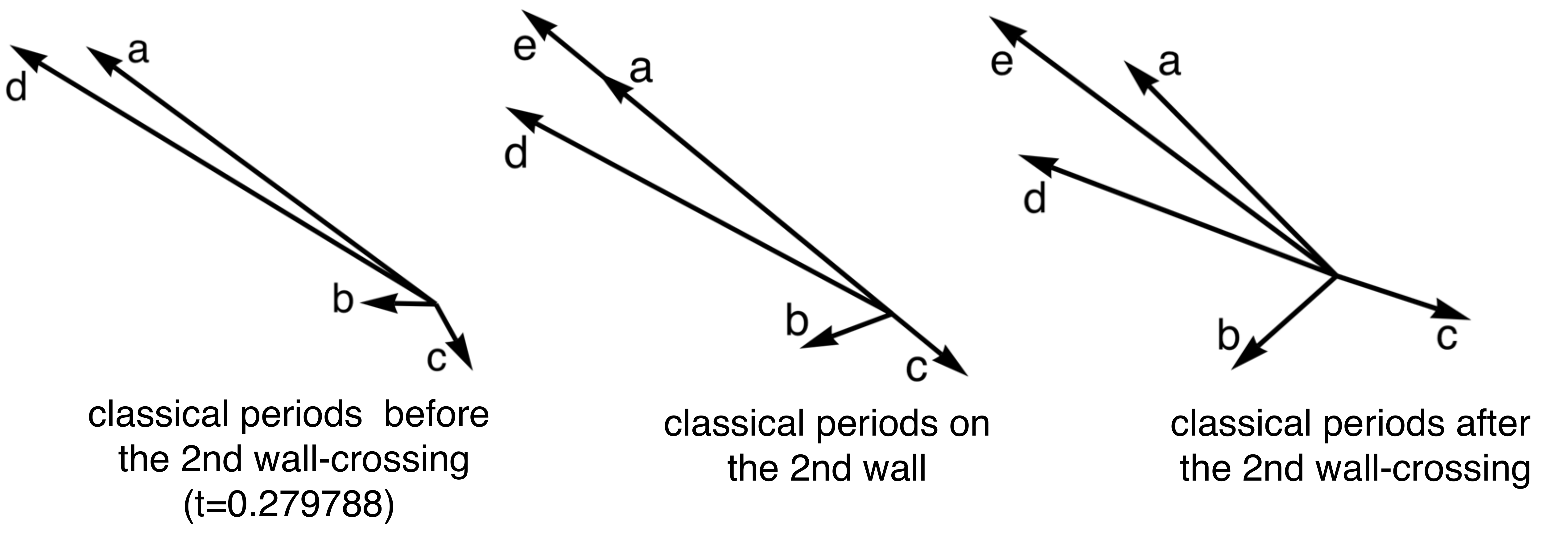}~~~~~~~
    \caption{The classical periods in the process of the second wall-crossing. The vectors of the classical periods before, middle and after the second wall-crossing are shown from left to right. The arrows labeled by $a,b,c,d$ and $e$ represent the periods $\Pi_{\gamma_{1,1}}^{(0)},\Pi_{\gamma_{3,2}}^{(0)},\Pi_{\gamma_{1,2}}^{(0)},\Pi_{\gamma_{1,1}+\gamma_{3,2}}^{(0)}$ and $\Pi_{\gamma_{1,1}-\gamma_{1,2}}^{(0)}$ respectively.
   }
    \label{fig:A2A2-2ndws-period-vec}
\end{figure}

The new Y-function $Y_{\widetilde{12}}^{(2)}(\theta)$ is thus related to the cycle $\gamma_{1,1}-\gamma_{1,2}$.
In this chamber, the WKB periods are identified with the new Y-functions as
\begin{equation}
    \begin{aligned}
        \log Y_{1,1}^{(2)}(\theta)&=e^{\theta}\Pi_{\hat{\gamma}_{1,1}},\quad \log Y_{1,2}^{(2)}(\theta)=e^{\frac{\pi i}{3}}e^{\theta}\Pi_{\hat{\gamma}_{1,2}}(\theta+\frac{\pi i}{3}),\\
        \log Y_{12}^{(2)}(\theta)&=e^{\theta}\Pi_{\hat{\gamma}_{1,1}+\hat{\gamma}_{1,2}}(\theta),\quad\log Y_{\widetilde{12}}^{(2)}(\theta)=e^{\theta}\Pi_{{\gamma}_{1,1}-{\gamma}_{1,2}}(\theta).
    \end{aligned}\label{eq:2ndwc-wkbY}
\end{equation}
To test these identifications, we compare the $\epsilon$ expansion of WKB periods with the expansion of the TBA equations:
\begin{equation}
    \begin{aligned}
        &\log{Y_{a, k}^{(2)}}(\theta) = m_{a, k}e^{\theta} + \sum_{n = 1}^{\infty}m_{a, k}^{(n)}e^{-n\theta},\\
        &\log{Y_{12}^{(2)}}(\theta) = m_{12}e^{\theta} + \sum_{n = 1}^{\infty}m_{12}^{(n)}e^{-n\theta},\\
        &\log{Y_{\widetilde{12}}^{(2)}}(\theta) = m_{\widetilde{12}}e^{\theta} + \sum_{n = 1}^{\infty}m_{\widetilde{12}}^{(n)}e^{-n\theta}. 
    \end{aligned}\label{eq:A2A2-2ndwc-TBAexp}
\end{equation}
In table \ref{tab:A2A2_p_t_0.428571}, we perform the numeric comparison of (\ref{eq:2ndwc-wkbY}) in the $\epsilon$ expansion, which shows good agreement numerically\footnote{In \cite{Dumas:2020zoz}, the authors have studied numerically the spectral coordinates obtained from the TBA-like equations and those obtained by solving the ODE for $(A_2, A_2)$.}.
\begin{table}[htbp]
    \centering
    {\small
    \begin{tabular}{c||c|c}
        $n$ & $\Pi_{\hat{\gamma}_{1,1}}^{(n)}$ & $m_{1, 1}^{(n - 1)}$ \\\hline
        $0$ & $-8.308190190 + 7.043627188i$ & $-8.308190190 + 7.043627188i$ \\
        $2$ & $-0.009010340904 - 0.1441424592i$ & $-0.009010340904 - 0.1441424592i$ \\
        $6$ & $-0.01783306296 + 0.03546352561i$ & $-0.01783306296 + 0.03546352561i$ \\
        $8$ & $-0.07558676701 + 0.1018527964i$ & $-0.07558676698 + 0.1018527964i$\\
        $12$ & $5.797603169 - 3.782968514i$ & $5.797602298 - 3.782967938i$\\\hline\hline
        $n$ & $e^{\frac{\pi i}{3}(1-n)}\Pi_{\hat{\gamma}_{1,2}}^{(n)}$ & $m_{1, 2}^{(n - 1)}$ \\\hline
        $0$ & $-0.2795945545 - 3.111293527i$ & $-0.2795945545 - 3.111293527i$ \\
        $2$ & $0.009010340904 + 0.1441424592i$ & $0.009010340904 + 0.1441424592i$ \\
        $6$ & $0.01783306296 - 0.03546352561i$ & $0.01783306296 - 0.03546352561i$ \\
        $8$ & $0.07558676701 - 0.1018527964i$ & $0.07558676698 - 0.1018527964i$\\
        $12$ & $-5.797603169 + 3.782968514i$ & $-5.797602298 + 3.782967938i$\\\hline\hline
        $n$ & $\Pi_{\hat{\gamma}_{1,1}+\hat{\gamma}_{1,2}}^{(n)}$ & $m_{12}^{(n - 1)}$ \\\hline
        $0$ & $-11.14244670+5.730116412i$ & $-11.14244670+5.730116412i$ \\
        $2$ & $-0.1293362019 - 0.06426804547i$ & $-0.1293362019 - 0.06426804547i$ \\
        $6$ & $-0.03962884557 + 0.002287877251i$ & $-0.03962884557 + 0.002287877251i$ \\
        $8$ & $0.05041372566 + 0.1163864586i$ & $0.05041372564 + 0.1163864586i$\\
        $12$ & $6.174948420 + 3.129387368i$ & $6.174947485 + 3.129386902i$\\\hline\hline
        $n$ & $\Pi_{\gamma_{1,1}-\gamma_{1,2}}^{(n)}$ & $m_{\widetilde{12}}^{(n - 1)}$ \\\hline
        $0$ & $-10.86285215+8.841409938i$ & $-10.86285215+8.841409938i$ \\
        $2$ & $-0.1383465428 - 0.2084105046i$ & $-0.1383465428 - 0.2084105046i$ \\
        $6$ & $-0.05746190853 + 0.03775140286i$ & $-0.05746190853 + 0.03775140286i$ \\
        $8$ & $-0.02517304135 + 0.2182392551i$ & $-0.02517304134 + 0.2182392550i$\\
        $12$ & $11.97255159 - 0.6535811464i$ & $11.97254978 - 0.6535810356i$
    \end{tabular}
    }
    \caption{The WKB periods $\Pi_{\hat{\gamma}_{1,1}}^{(n)},e^{\frac{\pi i}{3}(1-n)}\Pi_{\hat{\gamma}_{1,2}}^{(n)}$, $\Pi_{\hat{\gamma}_{12}}^{(n)}$ and $\Pi_{\gamma_{1,1}-\gamma_{1,2}}^{(n)}$, and the masses $m_{1,1}^{(n-1)},m_{1,2}^{(n-1)}$, $m_{12}^{(n-1)}$ and $m_{\widetilde{12}}^{(n-1)}$ for $t = \frac{3}{7}\simeq0.428571....$}
    \label{tab:A2A2_p_t_0.428571}
\end{table}
This process of the second wall-crossing can be visualized by plotting the classical period on the complex plane, see Fig.\ref{fig:A2A2-2ndws-period-vec}. 
Before the second wall-crossing (left), we plot the vectors of the classical periods $\Pi_{\gamma_{1,1}}$, $\Pi_{\gamma_{1,2}}$, $\Pi_{\gamma_{3,2}}$, $\Pi_{\gamma_{1,1}+\gamma_{3,2}}^{(0)}$ and $\Pi_{\gamma_{1,1}-\gamma_{1,2}}^{(0)}$ corresponding to the Y-functions $Y_{1,1}^{(i)}, Y_{2,2}^{(i)}$, $Y_{1,2}^{(i)}$, $Y_{12}^{(i)}$ and $Y_{\widetilde{12}}^{(i)}$, respectively. 
At the second wall, $\Pi_{\gamma_{1,1}}^{(0)}$ and $\Pi_{\gamma_{1,2}}^{(0)}$ are in parallel; see the middle of the figure and \eqref{eq:A2A2-wc-cond2} for the condition of the marginal stability of the second wall-crossing. 
We thus need to introduce a new period $\Pi_{\gamma_{1,1}-\gamma_{1,2}}^{(0)}$ related to the cycle $\gamma_{1,1}-\gamma_{1,2}$ to obtain a closed system.
In the new four TBA equations, one finds many new types of kernels, which could have led to new poles and further wall-crossing. 
We have checked numerically that no new wall-crossing occurs.
We thus conclude that the TBA equations are valid in the region $2\pi/3<\phi_2-\phi_1<\pi$ ($0.397459...<t\leq1$).

\subsubsection{Monomial potential}
When $t=1$ in \eqref{eq:A2A2_zeros}, which is in the maximal chamber, we arrive at the monomial potential
\begin{equation}
    p(x)=x^3-8,
\end{equation}
and the TBA equations are given by \eqref{A2A2-max}. Using the method explained in section \ref{sec:period-mono}, one can compute the classical periods and the masses. We find
\begin{equation}
    |m_{1,1}|=|m_{1,2}|=|m_{12}|,\quad |m_{\widetilde{12}}|=\sqrt{3}|m_{1,1}|
\end{equation}
and 
\begin{equation}
  \phi_{2}- \phi_{1}=\pi,\quad\phi_{12}- \phi_{1}=\frac{\pi}{3},\quad\phi_{\widetilde{12}}- \phi_{1}=\frac{\pi}{6}.
\end{equation}
From the TBA equations, it is easy to find the relation
\begin{equation}
    \log Y_{1,1}^{(2)}(\theta- i\phi_{1})=\log Y_{1,2}^{(2)}(\theta- i\phi_{1}-\pi i)=\log Y_{12}^{(2)}(\theta- i\phi_{1}-\frac{\pi i}{3}).
\end{equation}
We obtain a reduced TBA system
\begin{equation}
    \begin{aligned}
        \log Y_{1,1}^{(2)}(\theta-i\phi_{1})=&|m_{1,1}|e^{\theta}+3K(\theta-\theta^{\prime})\star\overline{L}_{1,1}^{(2)}\\
        &+K(\theta-\theta^{\prime}+\frac{\pi i}{6})\star\overline{L}_{\widetilde{12}}^{(2)}+K(\theta-\theta^{\prime}-\frac{\pi i}{6})\star\overline{L}_{\widetilde{12}}^{(2)},\\
        \log Y_{\widetilde{12}}^{(2)}(\theta-i\phi_{1}-\frac{\pi i}{6})=&\sqrt{3}|m_{1,1}|e^{\theta}+3K(\theta-\theta^{\prime})\star\overline{L}_{\widetilde{12}}^{(2)}\\
        &+3K(\theta-\theta^{\prime}+\frac{\pi i}{6})\star\overline{L}_{1,1}^{(2)}+3K(\theta-\theta^{\prime}-\frac{\pi i}{6})\star\overline{L}_{1,1}^{(2)}.
        \end{aligned}\label{eq:A2A2-TBA-mono}
\end{equation}
Note that the TBA equations are $D_4$-type TBA whose definition is given in Appendix \ref{sec:D4-TBA} \cite{zamo-ADE,Braden:1989bu,kl-mel-1}. From the wall-crossing, we can relate $(A_2,A_2)$-TBA and $(D_4,A_1)$-TBA, which shows the equivalence of the quantum SW curve of $(A_2,A_2)$-AD theory and $(D_4,A_1)$-AD theory at the special point in the moduli space \cite{Cecotti:2010fi,Xie:2012hs}.   
The ODE/IM correspondence for the third order ODE with monomial potential has been studied in \cite{Dorey:1999pv}, where the third order potential is related to the $D_4$ model by comparing the spectrum of NLIE and the TBA numerically. In the present work, we derived the TBA equations directly from those in the minimal chamber.

Since the monomial point is a special point after the second wall-crossing, the identifications between WKB periods and Y-functions should be the same as in \eqref{eq:2ndwc-wkbY}. In table \ref{tab:A2A2_p_t_1}, we perform the numeric comparison of \eqref{eq:2ndwc-wkbY} in the $\epsilon$ expansion for the monomial case, which shows good agreement again. 
\begin{table}[htbp]
    \centering
    {\small
    \begin{tabular}{c||c|c}
        $n$ & $\Pi_{\hat{\gamma}_{1,1}}^{(n)}$ & $m_{1, 1}^{(n - 1)}$ \\\hline
        $0$ & $-5.299916251 + 9.179724222i$ & $-5.299916251 + 9.179724222i$ \\
        $2$ & $-0.04277896287 - 0.07409533719i$ & $-0.04277896287 - 0.07409533719i$ \\
        $6$ & $0.0001166129817 - 0.0002019796092i$ & $0.0001166129817 - 0.0002019796092i$ \\
        $8$ & $-0.00003383735446 - 0.00005860801712i$ & $-0.00003383735445 - 0.00005860801710i$\\
        $12$ & $0.00001446736332 - 0.00002505820832i$ & $0.00001446736106 - 0.00002505820443i$\\\hline\hline
        $n$ & $\Pi_{\gamma_{1,1}-\gamma_{1,2}}^{(n)}$ & $m_{\widetilde{12}}^{(n - 1)}$ \\\hline
        $0$ & $-15.89974875+9.179724222i$ & $-15.89974875+9.179724222i$ \\
        $2$ & $-0.1283368886 - 0.07409533719i$ & $-0.1283368886 - 0.07409533719i$ \\
        $6$ & $0.0003498389452 - 0.0002019796092i$ & $0.0003498389452 - 0.0002019796092i$ \\
        $8$ & $-0.0001015120634 - 0.00005860801712i$ & $-0.0001015120633 - 0.00005860801710i$\\
        $12$ & $0.00004340208996 - 0.00002505820832i$ & $0.00004340208323 - 0.00002505820443i$
    \end{tabular}
    }
    \caption{The quantum corrections for $t = 1$.}
    \label{tab:A2A2_p_t_1}
\end{table}
The effective central charge of the TBA equations is evaluated as \eqref{eq:A2A2-TBA-mono}
\begin{equation}\label{eq:ceff-A2A2-mono}
    \begin{aligned}
       c_{{\rm eff}}&=2\times\frac{6}{\pi^{2}}\int\Big(3|m_{1,1}|e^{\theta}\overline{L}_{1,1}^{(2)}+|m_{\widetilde{12}}|e^{\theta}\overline{L}_{\widetilde{12}}^{(2)}\Big)d\theta=2.
    \end{aligned}
\end{equation}
Comparing the first two terms in \eqref{eq:A2A2-2ndwc-TBAexp} and using \eqref{eq:ceff-A2A2-mono}, one finds
\begin{equation}
    m_{1,1}^{(1)}m_{1,1}=\frac{\pi}{4\sqrt{3}}c_{{\rm eff}}=\frac{\pi}{2\sqrt{3}}.
\end{equation}
Note that this relation is regarded as the special case of the PNP-type relation \cite{Dunne:2014bca,Codesido:2017dns}.

\subsection{\texorpdfstring{$(A_2, A_3)$}{A2A3}}
\begin{figure}[htbp]
    \centering
    \begin{minipage}{0.7\linewidth}
        \centering
        \includegraphics[width=12cm]{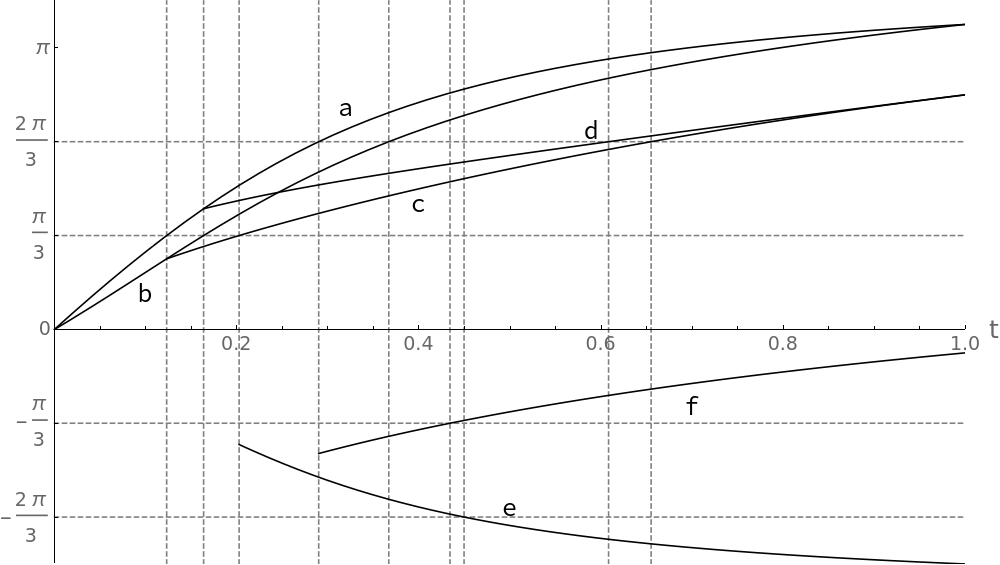}
    \end{minipage}
    \hspace{1cm}
    \begin{minipage}{0.2\linewidth}
        \begin{flushleft}
            \small{
            a: $\phi_2-\phi_1$\\
            \vspace{0.1cm}
            b: $\phi_2-\phi_3$\\
            \vspace{0.1cm}
            c: $\phi_{12}-\phi_3+\frac{\pi}{3}$\\
            \vspace{0.1cm}
            d: $\phi_{23}-\phi_1+\frac{\pi}{3}$\\
            \vspace{0.1cm}
            e: $\phi_{312}-\phi_2$\\
            \vspace{0.1cm}
            f: $\phi_{\widetilde{12}}-\phi_3-\frac{\pi}{3}$\\
            }
        \end{flushleft}
    \end{minipage}
    \caption{The differences of the phases of the masses from $t=0$ to $t=1$, which are relevant to the wall crossing. The nine dotted vertical lines show the locations of the walls.}
    \label{fig:A2A3_phases}
\end{figure}

We next study the process of wall-crossing of the $(A_2, A_3)$ TBA equations from the minimal chamber to the maximal chamber. This example contains more chambers than $(A_2,A_2)$ and shares similar features of general $(A_2,A_N)$ theory. We start with the point in the minimal chamber whose WKB curve has the branch points $\{2,1,-1,-4\}$, and the potential is given by $p(x)=x^4+2x^3-9x^2-2x+8$. We end with the monomial potential $p(x)=x^4-81$ in the maximal chamber, which has the turning points $\{3,3i,-3,-3i\}$.  We consider the path
\begin{equation}
    x_0(t)=2+t,\quad x_1(t)=1-t+3it,\quad x_2(t)=-x_1(t),\quad x_3(t)=-4+t,\quad 0\leq t\leq 1,
\end{equation}
whose potential is given by
$p(x;t)=(x-x_0(t))(x-x_1(t))(x-x_2(t))(x-x_3(t))$.
For any $t$, we compute the classical periods and the masses for the Y-functions. Moreover, we compute the differences of the phases in the kernels of the TBA equations. For the kernel $K_{k_1, k_2}^{\pm n}$, if $\phi_{k_1}-\phi_{k_2}\mp n\frac{\pi}{3}$ crosses $\pm\pi/3,\pm2\pi/3$, the wall-crossing occurs.
The differences of the relevant phases in the kernels of the TBA equations for $0\leq t\leq 1$ are plotted in figure \ref{fig:A2A3_phases}.

As $t$ increases, we find the nine walls in the moduli space, given by
\begin{align}
    t&=0.123142..., & &\phi_2-\phi_1=\frac{\pi}{3},& &\Im\qty(\frac{\Pi_{\gamma_{3,2}}^{(0)}}{\Pi_{\gamma_{1,1}}^{(0)}})=0,\label{eq:A2A3-wc-cond1}\\
    t&=0.163685..., & &\phi_2-\phi_3=\frac{\pi}{3},& &\Im\qty(\frac{\Pi_{\gamma_{3,2}}^{(0)}}{\Pi_{\gamma_{3,3}}^{(0)}})=0,\label{eq:A2A3-wc-cond2}\\
    t&=0.202635..., & &\phi_{12}-\phi_3=0,& &\Im\qty(\frac{\Pi_{\gamma_{1,1}+\gamma_{3,2}}^{(0)}}{\Pi_{\gamma_{3,3}}^{(0)}})=0,\label{eq:A2A3-wc-cond3}\\
    t&=0.290017..., & &\phi_2-\phi_1=\frac{2\pi}{3},& &\Im\qty(\frac{\Pi_{\gamma_{1,2}}^{(0)}}{\Pi_{\gamma_{1,1}}^{(0)}})=0,\label{eq:A2A3-wc-cond4}\\
    t&=0.366924..., & &\phi_2-\phi_3=\frac{2\pi}{3},& &\Im\qty(\frac{\Pi_{\gamma_{1,2}}^{(0)}}{\Pi_{\gamma_{3,3}}^{(0)}})=0,\label{eq:A2A3-wc-cond5}\\
    t&=0.434148..., & &\phi_{\widetilde{12}}-\phi_3=0,& &\Im\qty(\frac{\Pi_{\gamma_{1,1}-\gamma_{1,2}}^{(0)}}{\Pi_{\gamma_{3,3}}^{(0)}})=0,\label{eq:A2A3-wc-cond6}\\
    t&=0.449568..., & &\phi_{312}-\phi_2=-\frac{2\pi}{3},& &\Im\qty(\frac{\Pi_{\gamma_{1,1}+\gamma_{3,2}+\gamma_{3,3}}^{(0)}}{\Pi_{\gamma_{1,2}}^{(0)}})=0,\label{eq:A2A3-wc-cond7}\\
    t&=0.608205..., & &\phi_{23}-\phi_1=\frac{\pi}{3},& &\Im\qty(\frac{\Pi_{\gamma_{1,2}+\gamma_{1,3}}^{(0)}}{\Pi_{\gamma_{1,1}}^{(0)}})=0,\label{eq:A2A3-wc-cond8}\\
    t&=0.65489..., & &\phi_{12}-\phi_3=\frac{\pi}{3},& &\Im\qty(\frac{\Pi_{\gamma_{1,2}+\gamma_{2,1}}^{(0)}}{\Pi_{\gamma_{3,3}}^{(0)}})=0.\label{eq:A2A3-wc-cond9}
\end{align}
In our choice of path $0\leq t\leq 1$, no other wall-crossing occurs at the same time, so that we only need to add one new Y-function for each wall-crossing. The TBA equations in each chamber are rather complicated. In Appendix \ref{sec:newY-A2A3}, we will show only the definition of new Y-functions for completeness. The TBA equations are valid at any point in the chamber, not only on the path. Here, we  show the TBA equations at the typical chamber, which includes the symmetric potential. 
Since the symmetric potential plays an important role in the non-perturbative analysis of quantum mechanics \cite{Bender:1969si}, it would be worth writing down the TBA equations for the third order case.
In the minimal chamber, we have Y-functions $Y_{1,k}$, $k=1,2,3$, whose masses are given by $m_{1,k}$ with phase $\phi_{k}$, $k=1,2,3$. 
In the chamber after the $i$-th wall-crossing, we denote the Y-function by $Y^{(i)}$.

\paragraph{The symmetric case 1:} After the third wall-crossing, one obtains a closed system with six TBA equations with the Y-functions $Y_{1,k}^{(3)}$ and the new type Y-functions $Y^{(3)}_{12}(\theta )$, $Y^{(3)}_{23}(\theta )$ and $Y^{(3)}_{312}(\theta )$. See Appendix \ref{sec:newY-A2A3} for the definitions of the new Y-functions. The masses associated with these new Y-functions are
\begin{equation}
    m_{12}=m_{1,1}+m_{1,2}e^{-\frac{\pi i}{3}},\quad m_{23}=m_{1,3}+m_{1,2}e^{-\frac{\pi i}{3}},\quad m_{312}=m_{1,3}+m_{12},
\end{equation}
whose phases are $\phi_{12}$, $\phi_{23}$ and $\phi_{312}$, respectively.
In the symmetric case, i.e. $m_1=m_3$ and $m_{12}=m_{23}$, the six TBA equations reduce to
\begin{equation}
    \begin{aligned}
        &\log Y_{1,1}^{(3)}(\theta-i\phi_{1})=\abs{m_{1,1}}e^{\theta}+K\star\overline{L}_{1,1}^{(3)}-K_{1,2}\star\overline{L}_{1,2}^{(3)}+\qty(K_{1,12}^{-}-K_{1,12}^{+})\star\overline{L}_{12}^{(3)}+K_{1,312}^{-}\star\overline{L}_{312}^{(3)},\\
        &\log Y_{1,2}^{(3)}(\theta-i\phi_{2})=\abs{m_{1,2}}e^{\theta}-2K_{2,1}\star\overline{L}_{1,1}^{(3)}+K\star\overline{L}_{1,2}^{(3)}-2K_{2,12}^{-}\star\overline{L}_{12}^{(3)}-\qty(K_{2,312}+K_{2,312}^{-})\star\overline{L}_{312}^{(3)},\\
        &\log Y_{12}^{(3)}(\theta-i\phi_{12})=\abs{m_{12}}e^{\theta}+\qty(K_{12,1}^{+}-K_{12,3}^{-})\star\overline{L}_{1,1}^{(3)}-K_{12,2}^{+}\star\overline{L}_{1,2}^{(3)}+K\star\overline{L}_{12}^{(3)}+K_{12,312}^{+}\star\overline{L}_{312}^{(3)},\\
        &\log Y_{312}^{(3)}(\theta-i\phi_{312})=\abs{m_{312}}e^{\theta}+2K_{312,1}^{+}\star\overline{L}_{1,1}^{(3)}-\qty(K_{312,2}+K_{312,2}^{+})\star\overline{L}_{1,2}^{(3)}+2K_{312,12}^{-}\star\overline{L}_{12}^{(3)}\\
        &\qquad\qquad\qquad\qquad\qquad\qquad\quad+K\star\overline{L}_{312}^{(3)}(\theta^{\prime}-i\phi_{3}).
    \end{aligned}
\end{equation}
In Table \ref{tab:A2A3_sym1}, we compare the $\epsilon$ expansion of the Y-functions with the WKB periods, which show agreement in high precision.
\begin{table}[htbp]
    \centering
    {\small
    \begin{tabular}{c||c|c}
        $n$ & $\Pi_{\hat{\gamma}_{1,1}}^{(n)},e^{\frac{2\pi i}{3}(1-n)}\Pi_{\hat{\gamma}_{1,3}}^{(n)}$ & $m_{1, 1}^{(n - 1)},m_{1, 3}^{(n - 1)}$ \\\hline
        $0$ & $-11.39881594 + 14.23487415i$ & $-11.39881594 + 14.23487415i$ \\
        $2$ & $0.01642381444 - 0.03753315426i$ & $0.01642381444 - 0.03753315426i$ \\
        $6$ & $0.00002288822247 - 0.00002208917405i$ & $0.00002288822247 - 0.00002208917405i$ \\
        $8$ & $2.625484666\times10^{-6} + 4.618462473\times10^{-6}i$ & $2.625484666\times 10^{-6} + 4.618462473\times 10^{-6} i$\\
        $12$ & $5.362388326\times10^{-7} + 8.038711226\times10^{-7}i$ & $5.362387829\times 10^{-7} + 8.038710559\times 10^{-7}i $\\\hline\hline
        $n$ & $e^{\frac{\pi i}{3}(1-n)}\Pi_{\hat{\gamma}_{1,2}}^{(n)}$ & $m_{1, 2}^{(n - 1)}$ \\\hline
        $0$ & $-9.359933633 - 9.149856867i$ & $-9.359933633 - 9.149856867i$ \\
        $2$ & $0.004980601291 + 0.05664415536i$ & $0.004980601291 + 0.05664415536i$ \\
        $6$ & $-0.00002031318068 + 0.00002774220366i$ & $-0.00002031318068 + 0.00002774220366i$ \\
        $8$ & $-2.074149509\times10^{-6} - 4.697564517\times10^{-6}i$ & $-2.074149509\times 10^{-6} - 4.697564517\times 10^{-6} i$\\
        $12$ & $-5.119551461\times10^{-7} - 7.947194982\times10^{-7}i$ & $-5.119550997\times 10^{-7} - 7.947194331\times 10^{-7}i$\\\hline\hline
        $n$ & $\Pi_{\gamma_{1,1}+\gamma_{3,2}}^{(n)},\Pi_{\gamma_{3,2}+\gamma_{3,3}}^{(n)}$ & $m_{12}^{(n - 1)},m_{23}^{(n - 1)}$ \\\hline
        $0$ & $-24.00279124 + 17.76588602i$ & $-24.00279124 + 17.76588602i$ \\
        $2$ & $-0.03014116243 - 0.004897749341i$ & $-0.03014116243 - 0.004897749341i$ \\
        $6$ & $0.00003675708525 + 9.373658279\times10^{-6}i$ & $0.00003675708525 + 9.373658279\times 10^{-6}i$ \\
        $8$ & $5.656620120\times10^{-6} + 4.734140485\times10^{-7}i$ & $5.656620119\times 10^{-6} + 4.734140483\times 10^{-7} i$\\
        $12$ & $ -4.079860148\times10^{-7} + 8.498775356\times10^{-7}i$ & $-4.079859849\times 10^{-7} + 8.498774613\times 10^{-7}i$\\\hline\hline
        $n$ & $\Pi_{\gamma_{1,1}+\gamma_{3,2}+\gamma_{3,3}}^{(n)}$ & $m_{312}^{(n - 1)}$ \\\hline
        $0$ & $-35.40160718 + 32.00076016i$ & $-35.40160718 + 32.00076016i$ \\
        $2$ & $-0.01371734800 - 0.04243090360i$ & $-0.01371734800 - 0.04243090360i$ \\
        $6$ & $5.964530772\times10^{-5} - 1.271551577\times10^{-5}i$ & $5.964530772\times10^{-5}- 1.271551577\times10^{-5}i$ \\
        $8$ & $8.282104786\times10^{-6} + 5.091876522\times10^{-6}i$ & $8.282104785\times 10^{-6} + 5.091876521\times 10^{-6}i$\\
        $12$ & $1.282528178\times10^{-7} + 1.653748658\times10^{-6}i$ & $1.282527980\times 10^{-7} + 1.653748517\times 10^{-6} i$
    \end{tabular}
    }
    \caption{The quantum corrections for $p(x)=x^4-2(8+i)x^2+32i$.}
    \label{tab:A2A3_sym1}
\end{table}

\paragraph{The symmetric case 2:} After the sixth wall-crossing, we obtain nine TBA equations with Y-functions $Y^{(6)}_{1,k}$, $Y^{(6)}_{12}(\theta )$, $Y^{(6)}_{23}(\theta )$, $Y^{(6)}_{312}(\theta )$ and three new Y-functions $Y^{(6)}_{\widetilde{12}}(\theta )$, $Y^{6)}_{\widetilde{23}}(\theta )$ and $Y^{(6)}_{\widehat{312}}(\theta )$.
The masses of the new Y-functions are denoted by
\begin{equation}
    m_{\widetilde{12}}=m_{1,1}+e^{-\frac{2\pi i}{3}}m_{1,2},\quad m_{\widetilde{23}}=m_{1,3}+m_{1,2}e^{-\frac{2\pi i}{3}},\quad m_{\widehat{312}}=m_{1,3}+m_{\widetilde{12}}
\end{equation}
with their phases $\phi_{\widetilde{12}}$, $\phi_{\widetilde{23}}$ and $\phi_{\widehat{312}}$, respectively.
In the symmetric case, $m_1=m_3$, $m_{12}=m_{23}$ and $m_{\widetilde{12}}=m_{\widetilde{23}}$, the nine TBA equations reduce to 
\begin{equation}
    \begin{aligned}
        \log Y_{1,1}^{(6)}(\theta-i\phi_{1})=&|m_{1,1}|e^{\theta}+K\star\overline{L}_{1,1}^{(6)}-K_{1,2}\star\overline{L}_{1,2}^{(6)}\\
        &+\big(K_{1,12}^{-}-K_{1,12}^{+}\big)\star\overline{L}_{12}^{(6)}+\big(K_{1,\widetilde{12}}+2K_{1,\widetilde{12}}^{-}\big)\star\overline{L}_{\widetilde{12}}^{(6)}\\
        &+K_{1,312}^{-}\star\overline{L}_{312}^{(6)}+\big(K_{1,\widehat{312}}+K_{1,\widehat{312}}^{-}\big)\star\overline{L}_{\widehat{312}}^{(6)},\\
        \log Y_{1,2}^{(6)}(\theta-i\phi_{2})=&|m_{1,2}|e^{\theta}-2K_{2,1}\star\overline{L}_{1,1}^{(6)}+K\star\overline{L}_{1,2}^{(6)}\\
        &-2K_{2,12}^{-}\star\overline{L}_{12}^{(6)}-2\big(K_{2,\widetilde{12}}+K_{2,\widetilde{12}}^{-}\big)\star\overline{L}_{\widetilde{12}}^{(6)}\\
        &-\big(K_{2,312}+K_{2,312}^{-}\big)\star\overline{L}_{312}^{(6)}-\big(2K_{2,\widehat{312}}+K_{2,\widehat{312}}^{-}\big)\star\overline{L}_{\widehat{312}}^{(6)},\\
        \log Y_{12}^{(6)}(\theta-i\phi_{12})=&|m_{12}|e^{\theta}+\big(K_{12,1}^{+}-K_{12,1}^{-}\big)\star\overline{L}_{1,1}^{(6)}-K_{12,2}^{+}\star\overline{L}_{1,2}^{(6)}\\
        &+K\star\overline{L}_{12}^{(6)}+\big(K_{12,\widetilde{12}}+2K_{12,\widetilde{12}}^{+}\big)\star\overline{L}_{\widetilde{12}}^{(6)}\\
        &+K_{12,312}^{+}\star\overline{L}_{312}^{(6)}+2K_{12,\widehat{312}}^{+}\star\overline{L}_{\widehat{312}}^{(6)},\\
        \log Y_{312}^{(6)}(\theta-i\phi_{312})=&|m_{312}|e^{\theta}+2K_{312,1}^{+}\star\overline{L}_{1,1}^{(6)}-\big(K_{312,2}+K_{312,2}^{+}\big)\star\overline{L}_{1,2}^{(6)}\\
        &+2K_{312,12}^{-}\star\overline{L}_{12}^{(6)}+4K_{312,\widetilde{12}}\star\overline{L}_{\widetilde{12}}^{(6)}\\
        &+K\star\overline{L}_{312}^{(6)}+\big(2K_{312,\widehat{312}}+K_{312,\widehat{312}}^{+}\big)\star\overline{L}_{\widehat{312}}^{(6)},\\
        \log Y_{\widetilde{12}}^{(6)}(\theta-i\phi_{\widetilde{12}})=&|m_{\widetilde{12}}|e^{\theta}+\big(K_{\widetilde{12},1}+2K_{\widetilde{12},1}^{+}\big)\star\overline{L}_{1,1}^{(6)}-\big(K_{\widetilde{12},2}+K_{\widetilde{12},2}^{+}\big)\star\overline{L}_{1,2}^{(6)}\\
        &+\big(K_{\widetilde{12},12}+2K_{\widetilde{12},12}^{-}\big)\star\overline{L}_{12}^{(6)}+5K(\theta-\theta^{\prime})\star L_{\widetilde{12}}^{(6)}\\
        &+2K_{\widetilde{12},312}\star\overline{L}_{312}^{(6)}+\big(K_{\widetilde{12},\widehat{312}}^{+}+3K_{\widetilde{12},\widehat{312}}\big)\star\overline{L}_{\widehat{312}}^{(6)},\\
        \log Y_{\widehat{312}}^{(6)}(\theta-i\phi_{1})=&|m_{\widehat{312}}|e^{\theta}+2\big(K_{\widehat{312},1}+K_{\widehat{312},1}^{+}\big)\star\overline{L}_{1,1}^{(6)}\\
        &-\big(2K_{\widehat{312},2}+K_{\widehat{312},2}^{+}\big)\star\overline{L}_{1,2}^{(6)}+4K_{\widehat{312},12}^{-}\star\overline{L}_{12}^{(6)}\\
        &+\big(6K_{\widehat{312},\widetilde{12}}+2K_{\widehat{312},\widetilde{12}}^{-}\big)\star\overline{L}_{\widetilde{12}}^{(6)}+5K(\theta-\theta^{\prime})\star\overline{L}_{\widehat{312}}^{(6)}\\
        &+\big(2K_{\widehat{312},312}+K_{\widehat{312},312}^{-}\big)\star\overline{L}_{312}^{(6)}.
    \end{aligned}
\end{equation}
The distribution of the branch points of the symmetric potential of case $2$ is point-symmetric about the origin as in the case  of the symmetric potential $1$, but closer to that of monomial potential.

\paragraph{Monomial potential:} After the ninth wall-crossing, we finally arrive at the maximal chamber, where twelve TBA equations for the Y-functions $Y^{(9)}_{1,k}$, $Y^{(9)}_{12}$, $Y^{(9)}_{23}$, $Y^{(9)}_{312}$, $Y^{(9)}_{\widetilde{12}}$, $Y^{(9)}_{\widetilde{23}}$, $Y^{(9)}_{\widehat{312}}$ and the new Y-functions $Y^{(9)}_{3122}$, $Y^{(9)}_{231}$, $Y^{(9)}_{\overline{123}}$ appear. 
These equations are shown in Appendix \ref{sec:TBA-max}. 
For the new Y-functions $Y^{(9)}_{3122}$, $Y^{(9)}_{231}$, $Y^{(9)}_{\overline{123}}$, their masses are defined by
\begin{equation}
    m_{3122}=m_{312}+m_{1,2}e^{-\frac{2\pi i}{3}},\quad m_{231}=m_{1,1}+e^{-\frac{\pi i}{3}}m_{23},\quad m_{\overline{123}}=m_{1,3}+e^{-\frac{\pi i}{3}}m_{12},
\end{equation}
whose phases are denoted as $\phi_{3122}$, $\phi_{231}$ and $\phi_{\overline{123}}$, respectively. 
There exists a maximally symmetric point such that the potential becomes a monomial potential. 
In our example, the potential is given by
\begin{equation}
    p(x)=x^4-81,
    \label{eq:A2A3-mono}
\end{equation}
which corresponds to $t=1$. In general, for the monomial potential,  the masses satisfy the relations
\begin{equation}
    \begin{aligned}
        |m_{1,1}|=&|m_{1,3}|=|m_{12}|=|m_{23}|,\quad |m_{1,2}|=|m_{312}|,\\
        |m_{\widetilde{12}}|=&|m_{\widetilde{23}}|=|m_{231}|=|m_{\overline{123}}|,\quad |m_{\widehat{312}}|=|m_{3122}|,\\
        \frac{|m_{1,2}|}{|m_{1,1}|}=&\frac{\sin(\pi/4)}{\sin(\pi/6)},\quad \frac{|m_{\widetilde{12}}|}{|m_{1,1}|}=\frac{\sin(\pi/6)}{\sin(\pi/12)},\quad \frac{|m_{\widehat{312}}|}{|m_{1,1}|}=\frac{\sin(\pi/4)}{\sin(\pi/12)}
    \end{aligned}
\end{equation}
with the phase
\begin{equation}
    \begin{aligned}
        \phi_{1}=&\phi_{3}=\frac{7\pi}{12},\quad\phi_{12}=\phi_{23}=-\frac{11\pi}{12},\\\phi_{\widetilde{12}}=&\phi_{\widetilde{23}}=\frac{5\pi}{6},\quad\phi_{231}=\phi_{\overline{123}}=\frac{2\pi}{3},\\\phi_{2}=&-\frac{\pi}{3},\quad \phi_{312}=\frac{5\pi}{6},\quad\phi_{\widehat{312}}=\frac{3\pi}{4},\quad\phi_{3122}=\frac{11\pi}{12},
    \end{aligned}
\end{equation}
up to an overall phase. 
These relations can be shown by using \eqref{eq:classical_period_mono}.
One then observes that some of the TBA equations take the identical form, from which we find the identifications among the Y-functions as
\begin{equation}\label{eq:12-to-4-iden}
    \begin{gathered}
        \log Y_{1,1}(\theta-i\phi_{1})=\log Y_{1,3}(\theta-i\phi_{3})=\log Y_{12}(\theta-i\phi_{12})=\log Y_{23}(\theta-i\phi_{23}),\\
        \log Y_{\widetilde{12}}(\theta-i\phi_{\widetilde{12}})=\log Y_{\widetilde{23}}(\theta-i\phi_{\widetilde{23}})=\log Y_{231}(\theta-i\phi_{231})=\log Y_{\overline{123}}(\theta-i\phi_{\overline{123}}),\\
        \log Y_{1,2}(\theta-i\phi_{2})=\log Y_{312}(\theta-i\phi_{312}),\;\;
        \log Y_{\widehat{312}}(\theta-i\phi_{\widehat{312}})=\log Y_{3122}(\theta-i\phi_{3122}).
    \end{gathered}
\end{equation}
The TBA equations thus reduce to four independent TBA equations
\begin{equation}\label{eq:tba_a2a3_mono}
    \begin{aligned}
        \log Y_{1,1}^{(9)}(\theta-i\phi_{1})=&|m_1|e^{\theta}+\big(K+K_{1,12}^{-}-K_{1,12}^{+}\big)\star\overline{L}_{1,1}^{(9)}+\big(K_{1,312}^{-}-K_{1,2}\big)\star\overline{L}_{1,2}^{(9)}\\
           &+\big(K_{1,\widetilde{12}}+2K_{1,\widetilde{12}}^{-}+2K_{1,231}+K_{1,231}^{-}\big)\star\overline{L}_{\widetilde{12}}^{(9)}\\
           &+\big(K_{1,\widehat{312}}+K_{1,\widehat{312}}^{-}+2K_{1,3122}^{-}\big)\star\overline{L}_{\widehat{312}}^{(9)},\\
        \log Y_{1,2}^{(9)}(\theta-i\phi_{2})=&|m_2| e^{\theta}-\big(2K_{2,1}+2K_{2,12}^{-}\big)\star\overline{L}_{1,1}^{(9)}+\big(K-K_{2,312}-K_{2,312}^{-}\big)\star\overline{L}_{1,2}^{(9)}\\
           &-\big(2K_{2,\widetilde{12}}+2K_{2,\widetilde{12}}^{-}+4K_{2,231}\big)\star\overline{L}_{\widetilde{12}}^{(9)}\\
           &-\big(2K_{2,\widehat{312}}+K_{2,\widehat{312}}^{-}+K_{2,3122}+2K_{2,3122}^{-}\big)\star\overline{L}_{\widehat{312}}^{(9)},\\
        \log Y_{\widetilde{12}}^{(9)}(\theta-i\phi_{\widetilde{12}})=&|m_{\widetilde{12}}|e^{\theta}+\big(K_{\widetilde{12},1}+2K_{\widetilde{12},1}^{+}+2K_{\widetilde{12},12}+K_{\widetilde{12},12}^{-}-K_{\widetilde{12},12}^{+}\big)\star\overline{L}_{1,1}^{(9)}\\
          &-\big(K_{\widetilde{12},2}+K_{\widetilde{12},2}^{+}-2K_{\widetilde{12},312}\big)\star\overline{L}_{1,2}^{(9)}+\big(5K+3K_{\widetilde{12},231}+3K_{\widetilde{12},231}^{+}\big)\star L_{\widetilde{12}}^{(9)}\\
         &+\big(K_{\widetilde{12},\widehat{312}}^{+}+3K_{\widetilde{12},\widehat{312}}+K_{\widetilde{12},3122}^{-}+3K_{\widetilde{12},3122}\big)\star\overline{L}_{\widehat{312}}^{(9)},\\
        \log Y_{\widehat{312}}^{(9)}(\theta-i\phi_{1})=&|m_{\widehat{312}}|e^{\theta}+\big(2K_{\widehat{312},1}+2K_{\widehat{312},1}^{+}+4K_{\widehat{312},12}^{-}\big)\star\overline{L}_{1,1}^{(9)}\\
            &+\big(2K_{\widehat{312},312}+K_{\widehat{312},312}^{-}-2K_{\widehat{312},2}-K_{\widehat{312},2}^{+}\big)\star\overline{L}_{1,2}^{(9)}\\
            &+\big(6K_{\widehat{312},\widetilde{12}}+2K_{\widehat{312},\widetilde{12}}^{-}+6K_{\widehat{312},231}+2K_{\widehat{312},231}^{+}\big)\star\overline{L}_{\widetilde{12}}^{(9)}\\
            &+\big(5K+3K_{\widehat{312},3122}+3K_{\widehat{312},3122}^{-}\big)\star\overline{L}_{\widehat{312}}^{(9)}.
    \end{aligned}
\end{equation}
In Table 5.5, we compare the $\epsilon$ expansion of the Y-functions with the WKB periods, which show agreement in high precision.
This closed system reproduces the $E_6$ TBA in Appendix \ref{sec:E6} under the following identifications:
\begin{equation}
    \begin{aligned}
      &\log Y_{1,1}(\theta-i\phi_{1})\leftrightarrow\epsilon_{1}(\theta),\quad\log Y_{1,2}(\theta-i\phi_{2})\leftrightarrow\epsilon_{3}(\theta),\\
      &\log Y_{\widetilde{12}}(\theta-i\phi_{\widetilde{12}})\leftrightarrow\epsilon_{4}(\theta),\quad\log Y_{\widehat{312}}(\theta-i\phi_{\widehat{312}})\leftrightarrow\epsilon_{6}(\theta),
    \end{aligned}
\end{equation}
where $\epsilon_a$ are the pseudo energies of the $E_6$ TBA equations given in Appendix \ref{sec:E6}.
Integrating the kernel matrix, one obtains the connection matrix at $\theta\to -\infty$, where the Y-function becomes constant, satisfying the algebraic relation related with the connection matrix at $\theta\to -\infty$. See Appendix \ref{sec:newY-A2A3} for details of the definition of the matrix. It is worth to note that the connection matrix at the monomial potential can be obtained from the matrix after the ninth wall-crossing under the identification \eqref{eq:12-to-4-iden}, which also coincides with that of the $E_6$ TBA. See Appendix \ref{sec:newY-A2A3}.

To realize this observation in the gauge theory side, we note that the $(A_2, A_3)$ AD theory and $(E_6,A_1)$ AD theory have the common AD point $y^3+x^4=0$ and can be regarded as the equivalent theory \cite{Cecotti:2010fi,Xie:2012hs}. The ODE with the monomial potential (\ref{eq:A2A3-mono}) can be interpreted as the quantum SW curve of $E_6$-type AD theory.

In \cite{Dorey:1999pv}, the ODE/IM correspondence for the third order ODE with monomial potential has been studied. In particular, it is observed that the NLIE obtained from the solutions of the ODE has the same spectrum of the TBA system of the integrable model.
This includes the non-trivial correspondence such that cubic potential and $D_4$ TBA, quartic and $E_6$ TBA.
\begin{table}[htbp]
    \centering
    {\small
    \begin{tabular}{c||c|c}
        $n$ & $\Pi_{\hat{\gamma}_{1,1}}^{(n)},e^{\frac{2\pi i}{3}(1-n)}\Pi_{\hat{\gamma}_{1,3}}^{(n)}$ & $m_{1, 1}^{(n - 1)},m_{1, 3}^{(n - 1)}$ \\\hline
        $0$ & $-7.469227532 + 27.87553664$ & $-7.469227532 + 27.87553664i$ \\
        $2$ & $-0.009523311889 - 0.03554148383i$ & $ -0.009523311889 - 0.03554148383i$ \\
        $6$ & $-3.287667860\times10^{-7} - 8.809279480\times10^{-8}i$ & $-3.287667860\times 10^{-7} - 8.809279480\times 10^{-8} i$ \\
        $8$ & $-1.483370697\times10^{-8} + 3.974679804\times10^{-9}i$ & $-1.483370697\times 10^{-8} + 3.974679803\times10^{-9}i $\\
        $12$ & $-1.084025230\times10^{-10} + 4.045637237\times10^{-10}i$ & $-1.084025095\times 10^{-10} + 4.045636732\times 10^{-10} i$\\\hline\hline
        $n$ & $e^{\frac{\pi i}{3}(1-n)}\Pi_{\hat{\gamma}_{1,2}}^{(n)}$ & $m_{1, 2}^{(n - 1)}$ \\\hline
        $0$ & $20.40630911 - 35.34476417i$ & $20.40630911 - 35.34476417i$ \\
        $2$ & $0.02601817194 + 0.04506479572i$ & $0.02601817194 + 0.04506479572i$ \\
        $6$ & $-2.406739912\times10^{-7} + 4.168595808\times10^{-7}i$ & $-2.406739912\times 10^{-7} + 4.168595808\times 10^{-7} i$ \\
        $8$ & $-1.085902717\times10^{-8} - 1.880838678\times10^{-8}i$ & $-1.085902716\times10^{-8} - 1.880838677\times 10^{-8} i$\\
        $12$ & $2.961612006\times10^{-10} - 5.129662467\times10^{-10}i$ & $2.961611637\times10^{-10} - 5.129661827\times 10^{-10} i$\\\hline\hline
        $n$ & $\Pi_{\gamma_{1,1}-\gamma_{1,2}}^{(n)},\Pi_{\gamma_{3,3}-\gamma_{1,2}}^{(n)}$ & $m_{\widetilde{12}}^{(n - 1)},m_{\widetilde{23}}^{(n - 1)}$ \\\hline
        $0$ & $-15.89974875 + 9.179724222i$ & $-15.89974875 + 9.179724222i$ \\
        $2$ & $-0.06155965576 - 0.03554148383i$ & $-0.06155965576 - 0.03554148383i$ \\
        $6$ & $1.525811964\times10^{-7} - 8.809279480\times10^{-8}i$ & $1.525811964\times 10^{-7} - 8.809279480\times 10^{-8} i$ \\
        $8$ & $6.884347364\times10^{-9} + 3.974679804\times10^{-9}i$ & $6.884347362\times 10^{-9} + 3.974679803\times 10^{-9} i$\\
        $12$ & $-7.007249243\times10^{-10} + 4.045637237\times10^{-10}i$ & $-7.007248369\times 10^{-10} + 4.045636732\times 10^{-10} i$\\\hline\hline
        $n$ & $\Pi_{\gamma_{1,1}-\gamma_{1,2}+\gamma_{3,3}}^{(n)}$ & $m_{\widehat{312}}^{(n - 1)}$ \\\hline
        $0$ & $-23.36897628 + 37.05526087i$ & $-23.36897628 + 37.05526087i$ \\
        $2$ & $-0.07108296765 - 0.07108296765i$ & $-0.07108296765 - 0.07108296765i$ \\
        $6$ & $-1.761855896\times10^{-7} - 1.761855896\times10^{-7}i$ & $-1.761855896\times 10^{-7} - 1.761855896\times 10^{-7}i$ \\
        $8$ & $-7.949359608\times10^{-9} + 7.949359608\times10^{-9}i$ & $-7.949359605\times 10^{-9} + 7.949359605\times 10^{-9} i$\\
        $12$ & $-8.091274473\times10^{-10} + 8.091274473\times10^{-10}i$ & $-8.091273464\times 10^{-10} + 8.091273464\times 10^{-10}i$
    \end{tabular}
    }
    \caption{The quantum corrections for $p(x)=x^4-81$.}
    \label{tab:A2A3_mono}
\end{table}

\section{Conclusions and Discussion}\label{sec:con-dis}
In this paper, we have studied the correspondence between the WKB periods of the ODE and the Y-functions of the integrable model for the third order ODE.
Here the ODE is regarded as a generalization of the Schr\"odinger type ODE, where the second order derivative is replaced by the higher order derivatives.
In particular, we have studied the case of polynomial potential with cubic and quartic orders in detail.
We first studied the minimal chamber where the Y-functions from the TBA equations and the WKB periods agree with each other numerically.
We then investigated the wall-crossing of the TBA equations and rewrote them by introducing new Y-functions.
We have chosen the special path in the Coulomb branch moduli space, which connects a point in the minimal chamber to the point in the maximal chamber, where the potential becomes monomial. We have traced the change of the TBA system and finally obtained the TBA system in the maximal chamber.

The maximal chamber of the theory contains the monomial potential, where the TBA system has some extra symmetry.
We found that for $(A_2,A_2)$-type ODE, we obtain the TBA system of $D_4$-type, and for $(A_2,A_3)$, we obtain $E_6$-type. 
It is natural to expect the $E_8$-type TBA from the $(A_2,A_4)$. 
Nevertheless, it is complicated to compute the TBA equations by wall-crossing.
We need a more systematic approach to work out the structure of the wall-crossing of the TBA equations like the diagrammatic method for the $(A_1,A_N)$-type\cite{toledo-thesis}.
The cluster algebra \cite{iwaki2014exact,Cecotti:2014zga} would be helpful for this analysis. 

It is interesting to explore more general higher order differential operators. 
For example, one can introduce the monodromy around the origin, which modifies the T-/Y-system and corresponding the integrable models.
For the monomial potential case for the higher order ODE associated with the linear problem of the affine Toda field equations, the corresponding T-Q relations and the Bethe ansatz equations have been studied \cite{Ito:2020htm}. 
Then it is interesting to generalize this to the polynomial potential. See \cite{Fioravanti:2020udo,Fioravanti:2021bzq} for the case related to the second order ODE. It is also interesting to include more irregular/regular singular points in the potential, which will help us to study the four dimensional ${\cal N}=2$ super Yang-Mills theory \cite{Grassi:2019coc,Fioravanti:2019vxi,Imaizumi:2020fxf,Imaizumi:2021cxf,Grassi:2021wpw}.

When the ODE has simple turning points where the differential operator factorizes into the product of the second order and the other, it has been shown that the WKB analysis essentially reduces to the second order \cite{Honda_2015}. 
By degeneration of the WKB curve, we can obtain the ODE presented in this work. It would be nice to study the limit and the change of the TBA system.
We have found the TBA equations for the WKB periods which determine the perturbative/non-perturbative corrections in $\epsilon$. 
However, we should study the Borel resummation and their resurgence structure for a deep understanding of the theory.

We did not investigate the full structure of the marginal wall of stability. 
We expect that for each chamber surrounded by the walls, there exist integrable models.
It is important to determine them for the higher order ODE for the complete characterization of the ODE/IM correspondence. 
Through the wall-crossing, different integrable models are unified by the same ODE but with different moduli parameters. 
It is important to see how these integrable models are connected in the IM side.

\subsection*{Acknowledgements}
We would like to thank Davide Fioravanti, Daniele Gregori, Yongchao L\"u, Hao Ouyang, Marco Rossi, and Dan Xie for useful discussions. We also thank Kohei Kuroda for his collaboration in an early stage of this work.
The work of K.I. is supported in part by Grant-in-Aid for Scientific Research 21K03570, 18K03643, and 17H06463 from Japan Society for the Promotion of Science (JSPS).
The work of H.S. is supported by the grant ``Exact Results in Gauge and String Theories'' from the Knut and Alice Wallenberg foundation. H.S. would like to thank Jilin University for their (online) hospitality.

\appendix

\section{\texorpdfstring{$D_4$ and $E_6$ type TBA equations}{D4E6}}\label{sec:D4_E6_TBA}
In this Appendix, we summarize the $D_4$ and $E_6$-type TBA equations, which are used to compare with the TBA equations obtained from the $(A_2, A_2)$ and $(A_2, A_3)$-type ODE at the monomial potential.

For two-dimensional massless scattering theories with the S-matrix \cite{Braden:1989bu, kl-mel-1} associated with simply-laced Lie algebras $\mathfrak{g}=ADE$ of rank $r$, the TBA equations take the form of the integral equations for the pseudo energies $\epsilon_a(\theta)$ ($a=1,\dots, r)$ \cite{zamo-ADE},
\begin{align}
    \epsilon_a(\theta)&=m_a e^\theta -\sum_{b=1}^{r}\frac{1}{2\pi} \phi_{ab}\star L_b(\theta),
    \label{eq:tba_ade}
\end{align}
where $(m_1,m_2,\dots,m_r)$ is the Perron-Frobenius eigenvector and
\begin{align}
    L_b(\theta)=\log(1+e^{-\epsilon_b(\theta)}).
\end{align}
The kernel function is defined by
\begin{align}
    \phi_{ab}(\theta)&=\int \frac{\dd k}{2\pi} \tilde{\phi}_{ab}(k)e^{ik\theta},\\
    \tilde{\phi}_{ab}(k)&=-2\pi\qty( 2I \cosh\qty(\frac{\pi k}{h})-G)^{-1}_{ac} G_{cb}.\label{eq:ker-til}
\end{align}
Here $h$ is the Coxeter number of ${\mathfrak g}$, $G_{ab}$ is the incidence matrix and $I$ the identity matrix of rank $r$.
The kernel functions $\phi_{ab}(\theta)$ can be also expressed in terms of the $S$-matrix as
\begin{align}
    \phi_{ab}(\theta)&=-i\frac{d}{d\theta}\log S_{ab}(\theta).
\end{align}
We write down the kernel functions for $\mathfrak{g}=D_4$ and $E_6$ explicitly and compare the TBA equations with those of the monomial point in the maximal chamber, which have been obtained in Section \ref{sec:wc}.

\subsection{\texorpdfstring{$D_4$}{D4}}\label{sec:D4-TBA}
For the Lie algebra $D_4$ with the Coxeter number $h=6$, we label the particles by
$a=1,2$, $s, s'$ and their S-matrices are expressed as \cite{Braden:1989bu}
\begin{equation}
    \begin{aligned}
        S_{11}&=\{1\}\{5\}, \\
        S_{12}&=S_{21}=\{2\}\{4\},\\
        S_{22}&=\{1\}\{5\}\{3\}\{3\},\\
        S_{ss}&=S_{s's'}=\{1\}\{5\}, \\
        S_{ss'}&=\{3\}, \\
        S_{s'1}&=S_{s1}=S_{1s}=S_{1s'}=\{3\}, \\
        S_{s'2}&=S_{s2}=S_{2s}=S_{2s'}=\{2\}\{4\}.
    \end{aligned}
\end{equation}
Here we have defined
\begin{align}
    \{x\}&= (x+1)(x-1)
\end{align}
and 
\begin{align}
    (x)&=\frac{\sinh\qty(\frac{\theta}{2}+\frac{i\pi x}{2h})}{\sinh\qty(\frac{\theta}{2}-\frac{i\pi x}{2h})}.
\end{align}
For the mass parameters satisfying $m_1=m_s=m_{s'}$, the pseudo energies also have the ${\mathbb Z}_3$-symmetry: 
\begin{align}
    \epsilon_1=\epsilon_s=\epsilon_{s'}.
\end{align}
Then, the TBA system reduces to
\begin{align}
    \epsilon_1&=m_1 e^\theta-\frac{1}{2\pi}(\phi_{11}+2\phi_{1s})\star L_1-\frac{1}{2\pi}\phi_{12}\star L_2,\\
    \epsilon_2&=m_2 e^\theta-3\frac{1}{2\pi}\phi_{21}\star L_1-\frac{1}{2\pi}\phi_{22}\star L_2,
\end{align}
where
\begin{align}
    \frac{1}{2\pi}\phi_{11}(\theta)&=\frac{1}{2\pi}\phi_{1s}(\theta)=-\frac{1}{2\pi}\frac{4\sqrt{3}\cosh(\theta)}{1+2\cosh(2\theta)}=-K(\theta),\\
    \frac{1}{2\pi}\phi_{12}(\theta)&=-\frac{1}{2\pi}6\cosh(2\theta)\sech(3\theta)=-K(\theta+\frac{i\pi}{6})-K(\theta-\frac{i\pi}{6}),\\
    \frac{1}{2\pi}\phi_{22}(\theta)&=-\frac{1}{2\pi}\frac{12\sqrt{3}\cosh(\theta)}{1+2\cosh(2\theta)}=-3K(\theta).
\end{align}
The TBA system is now  the same as (\ref{eq:A2A2-TBA-mono}).

\subsection{\texorpdfstring{$E_6$}{E6}}\label{sec:E6}
We next consider the $E_6$-type TBA equations, where the
S-matrices are given by \cite{Braden:1989bu}
\begin{equation}
    \begin{aligned}
        S_{11}&=\{1\}\{7\},\\
        S_{21}&=\{5\}\{11\},\quad S_{22}=\{1\}\{7\},\\
        S_{31}&=4,\quad S_{32}=4,\quad S_{33}=1 \ 5,\\
        S_{41}&=\{4\}\{6\}\{10\},\quad S_{42}=\{2\}\{6\}\{8\},\\
        S_{43}&=3 \ 5,\quad S_{44}=\{1\}\{3\}\{5\}\{7\}^2\{9\},\\
        S_{51}&=\{2\}\{6\}\{8\},\quad S_{52}=\{4\}\{6\}\{10\},\quad S_{53}=3 \ 5,\\
        S_{54}&=\{3\}\{5\}^2\{7\}\{9\}\{11\},\quad S_{55}=\{1\}\{3\}\{5\}\{7\}^2\{9\},\\
        S_{61}&= 3\ 5,\quad S_{62}=3\ 5,\quad S_{63}=2 \ 4\ 6,\\
        S_{64}&= 2 \ 4^2\ 6,\quad S_{65}=2\ 4^2\ 6,\quad S_{66}=1 \ 3^2\ 5^3,
    \end{aligned}
\end{equation}
where we have defined
\begin{align}
    n:=\{n\}\{12-n\}.
\end{align}
For the masses satisfying $m_1=m_2$ and $m_4=m_5$, the pseudo energies have  
the ${\mathbb Z}_2$ symmetry:
\begin{align}\label{eq:E6-folding}
\epsilon_1&=\epsilon_2,\quad \epsilon_4=\epsilon_5.
\end{align}
Then the TBA equations (\ref{eq:tba_ade}) reduce to
\begin{equation}
    \begin{aligned}
        \epsilon_1&=m_1 e^\theta-\frac{1}{2\pi}\big[(\phi_{11}+\phi_{12})\star L_1+\phi_{13}\star L_3+(\phi_{14}+\phi_{15})\star L_4+\phi_{16}\star L_6\big],\\
        \epsilon_3&=m_3 e^\theta-\frac{1}{2\pi}\big[(\phi_{31}+\phi_{32})\star L_1+\phi_{33}\star L_3+(\phi_{34}+\phi_{35})\star L_4+\phi_{36}\star L_6\big],\\
        \epsilon_4&=m_4 e^\theta-\frac{1}{2\pi}\big[(\phi_{41}+\phi_{42})\star L_1+\phi_{43}\star L_3+(\phi_{44}+\phi_{45})\star L_4+\phi_{46}\star L_6\big],\\
        \epsilon_6&=m_6 e^\theta-\frac{1}{2\pi}\big[(\phi_{61}+\phi_{62})\star L_1+\phi_{63}\star L_3+(\phi_{64}+\phi_{65})\star L_4+\phi_{66}\star L_6\big].
    \end{aligned}\label{eq:tba_e6}
\end{equation}
The explicit form of the kernels is shown to be
\begin{align}
    \phi_{11}+\phi_{12}&=\frac{-4(3+\sqrt{3})\cosh(3\theta)-6\sech(2\theta)}{1+2\cosh(4\theta)},\nonumber\\
    \phi_{13}&=-\frac{2\sqrt{2} \cosh(\theta) (3+\sqrt{3} +\sqrt{3}\sech(2\theta)}{\sqrt{3}+2\cosh(2\theta)},\nonumber\\
    \phi_{14}+\phi_{15}&=\frac{16\sqrt{2} ( (3+2\sqrt{3}) \cosh(3\theta)-3\cosh(\theta) \sech(2\theta))}{4-8 \cosh(4\theta)},\nonumber\\
    \phi_{16}&=\frac{-4(3+2\sqrt{3})\cosh(3\theta)-6\sech(2\theta)}{1+2\cosh(4\theta)}, \\
    \phi_{31}+\phi_{32}&=2\phi_{13},\quad
    \phi_{33}=\phi_{11}+\phi_{12}\quad 
    \phi_{34}+\phi_{35}=
    2\phi_{16},\nonumber\\
    \phi_{36}&=-2\sqrt{3}((-3+2\sqrt{3})\cosh(\theta)+(3+2\sqrt{3})\cosh(5\theta))\sech(6\theta),\\
    \phi_{41}+\phi_{42}&=\phi_{14}+\phi_{15}\quad
    \phi_{43}=\phi_{16}, \nonumber\\
    \phi_{44}+\phi_{45}&=-\frac{4(9+5\sqrt{3}) \cosh(3\theta)+18\sech(\theta)}{1+2\cosh(4\theta)},\nonumber\\
    \phi_{46}&=-\sqrt{2}((-9+5\sqrt{3})\cosh(\theta)+(9+5\sqrt{3})\cosh(5\theta))\sech(6\theta),\\
    \phi_{61}+\phi_{62}&=\phi_{34}+\phi_{35},\quad
    \phi_{63}=\phi_{36},\quad
    \phi_{64}+\phi_{65}=2\phi_{64},\quad
    \phi_{66}=\phi_{44}+\phi_{45}.
\end{align}
The TBA equations (\ref{eq:tba_e6}) are shown to agree with (\ref{eq:tba_a2a3_mono}).

\section{\texorpdfstring{New Y-functions of ($A_2, A_3$) case}{A2A3Y}}\label{sec:newY-A2A3}
In this Appendix, we show the definition of new Y-functions in the process of the wall-crossing of $(A_2, A_3)$. We start with the minimal chamber, where three independent Y-functions $Y_{1,k}$ exist. We denote by $Y^{(i)}$ the new Y-functions after the $i$-th wall-crossing. The TBA equations after the $i$-th wall-crossing take the form
\begin{align}
    \log Y^{(i)}_A=m_A e^{\theta}+ K^{(i)}_{AB}\star L^{(i)}_B,
\end{align}
whose effective central charge is evaluated as
\begin{equation}\label{eq:ceff-general}
    c_{\mathrm{eff}} \coloneqq \frac{6}{\pi^2}\sum_{A}\int_{-\infty}^{\infty}m_A{L}^{(i)}_A(\theta)e^{\theta}\dd{\theta}= \frac{6}{\pi^2}\sum_{ A}\mathcal{L}\qty(\frac{1}{1 + Y^{(i)}_{A}(-\infty)}).
\end{equation}
In the limit $\theta\rightarrow -\infty$, the Y-functions becomes the constants and the TBA equations reduce to the equations
\begin{align}\label{eq:YML}
    \log Y^{(i)}_A(-\infty)&=M^{(i)}_{AB}L^{(i)}_B(-\infty),
\end{align}
where $M^{(i)}_{AB}:=\int_{-\infty}^{\infty}d\theta K^{(i)}_{AB}(\theta)$ is the connection matrix, which shows the connectivity of the Y-functions in the TBA equations at $\theta\to -\infty$. The connection matrix is an integer valued matrix. When one goes from the minimal chamber to the maximal one, the size of the  matrix increases. At the maximal chamber, we observe that the TBA equations shows the maximal connectivity. Solving \eqref{eq:YML}, one obtains the constant solution of the Y-functions. Substituting these constant solutions into \eqref{eq:ceff-general}, one obtains the value of the effective central charge.

\begin{figure}[hpbt]
    \centering
    \includegraphics[width=10cm]{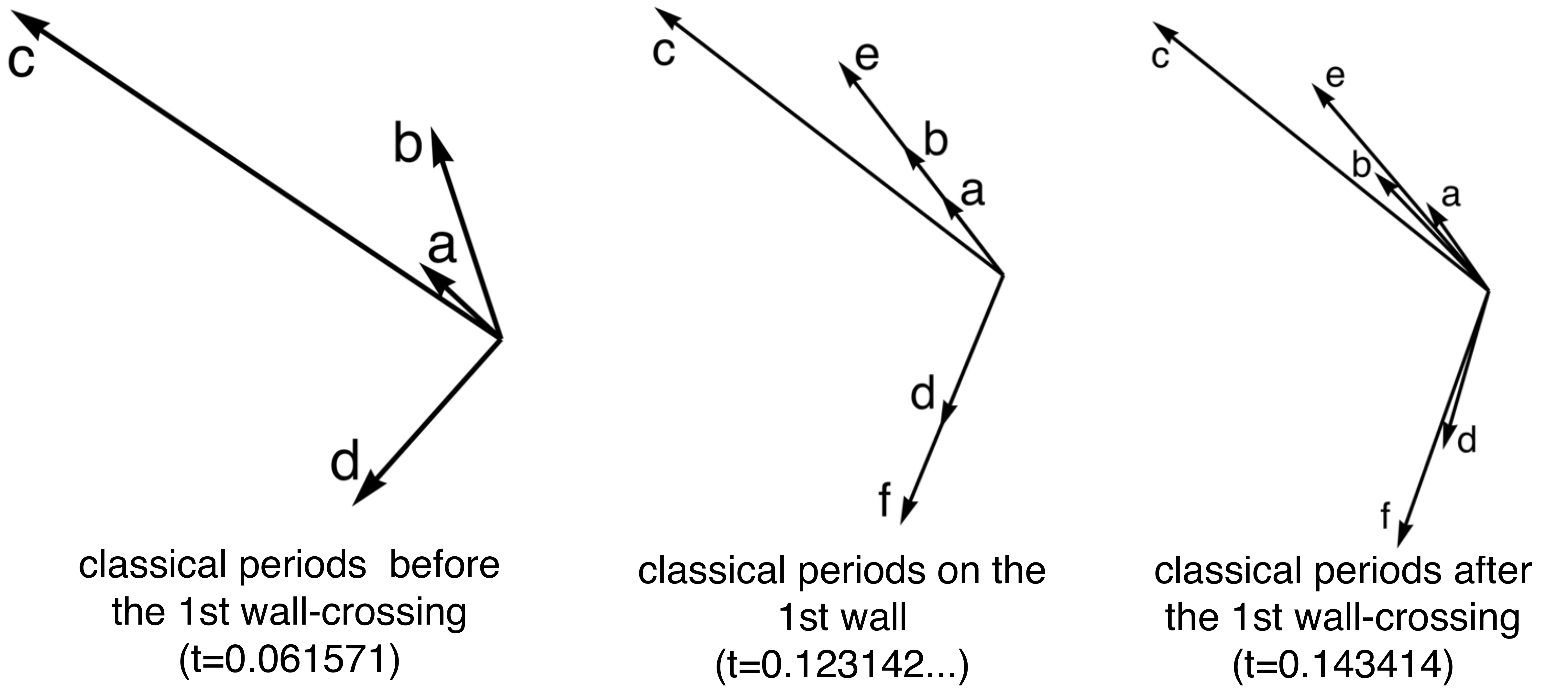}
    \caption{The classical periods in the wall-crossing process. The vectors of the classical periods before, middle and after the wall-crossing are shown from the left to right. The arrows labeled by $a,b,c,d$ and $e$ represent the periods $\Pi_{\gamma_{1,1}}^{(0)},\Pi_{\gamma_{3,2}}^{(0)},\Pi_{\gamma_{3,3}}^{(0)},\Pi_{\gamma_{1,2}}^{(0)}$ and $\Pi_{\gamma_{1,1}+\gamma_{3,2}}^{(0)}$ respectively.}
    \label{fig:A2A3-1stwc-period-vec}
\end{figure}

\paragraph{The first wall-crossing} The first wall-crossing occurs when $\phi_2-\phi_1$ crosses $\pi/3$, at $t=0.123142...$, while all other absolute value of the phases difference are smaller than  $\pi/3$. At this wall, the vectors of classical periods $\Pi_{\gamma_{1,1}}^{(0)}$ and $\Pi_{\gamma_{3,2}}^{(0)}$, corresponding to the Y-functions $Y_{1,1}$ and $Y_{1,2}$ respectively, are in parallel. After the wall-crossing, one new BPS particle, namely a new Y-function, be produced. 
See Fig.\ref{fig:A2A3-1stwc-period-vec}.
The Y-functions after the first wall-crossing are given by
\begin{equation}
    \begin{aligned}
        Y_{1,1}^{(1)}(\theta)&=Y_{1,1}(\theta)\big(1+\frac{1}{Y_{1,2}(\theta-\frac{\pi i}{3})}\big), \quad Y_{1,2}^{(1)}(\theta)=Y_{1,2}(\theta)\big(1+\frac{1}{Y_{1,1}(\theta+\frac{\pi i}{3})}\big),\\
        Y_{12}^{(1)}(\theta)&=\frac{1+\frac{1}{Y_{1,2}(\theta-\frac{\pi i}{3})}+\frac{1}{Y_{1,1}(\theta)}}{\frac{1}{Y_{1,1}(\theta)Y_{1,2}(\theta-\frac{\pi i}{3})}},\quad
        Y^{(1)}_{1,3}(\theta)=Y_{1,3}(\theta),
    \end{aligned}
\end{equation}
where the Y-functions without superscript is the original ones before the wall-crossing. For the vector 
\begin{equation}
    \vec{Y}^{(1)}=(Y^{(1)}_{1,1}(-\infty-i\phi_1), Y^{(1)}_{1,2}(-\infty-i\phi_2), Y^{(1)}_{1,3}(-\infty-i\phi_3), Y^{(1)}_{12}(-\infty-i\phi_{12}))^t,
\end{equation}
the matrix $M$ becomes
\begin{equation}
    M^{(1)}=\left(
    \begin{array}{cccc}
     1 & 0 & 0 & 1 \\
     0 & 1 & -1 & 1 \\
     0 & -1 & 1 & -1 \\
     1 & 1 & -1 & 1 \\
    \end{array}
    \right),
\end{equation}
which leads to the effective central charge $c_{\rm eff}=24/7$.

\begin{figure}[htbp]
    \centering
    \includegraphics[width=10cm]{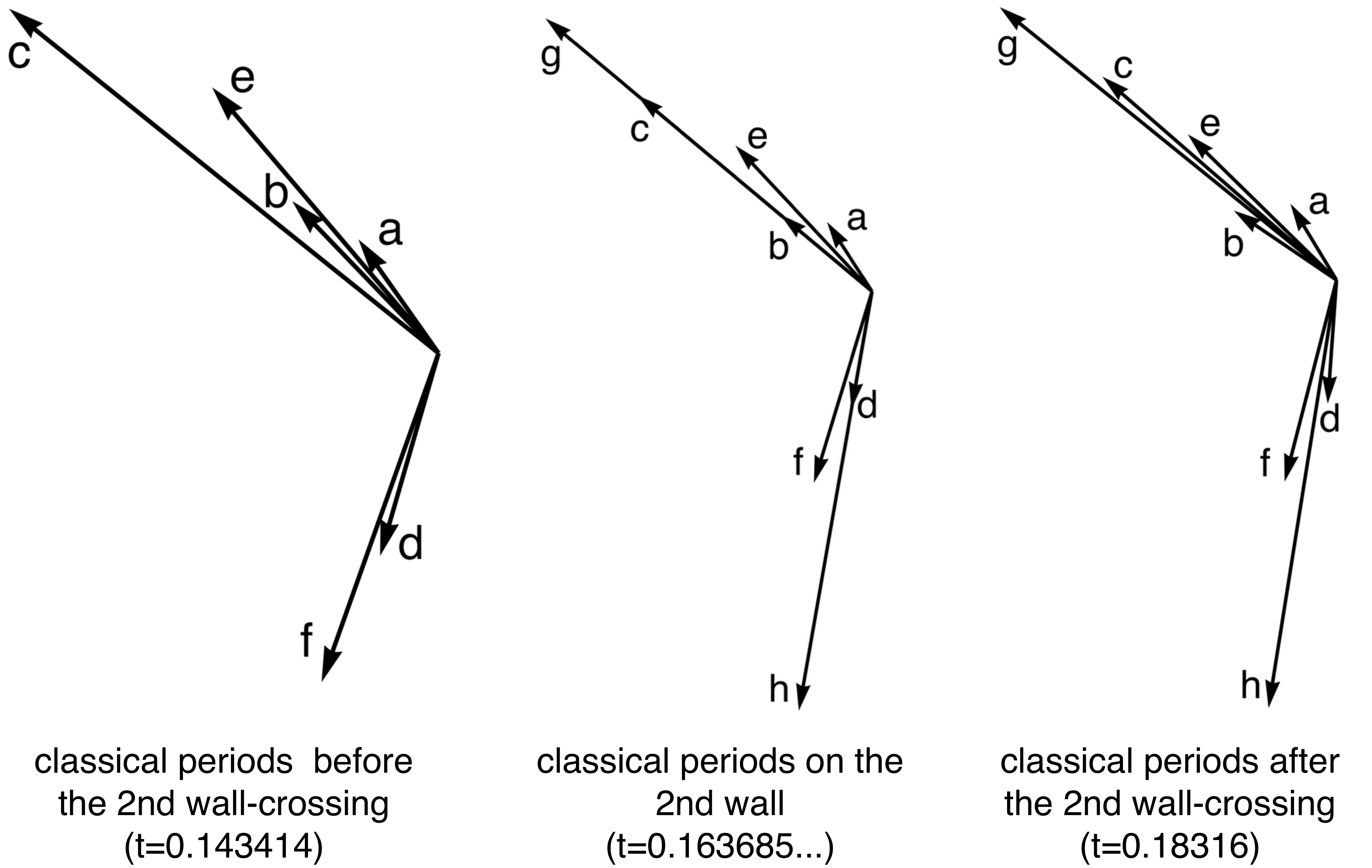}~~~~~~~
    \caption{The classical periods in the 2nd wall-crossing process. The vectors of the classical periods before, middle and after the 2nd wall-crossing are shown from left to right. The arrows labeled by $a,b,c,d,e,f,g$ and $h$ represent the periods $\Pi_{\gamma_{1,1}}^{(0)},\Pi_{\gamma_{3,2}}^{(0)},\Pi_{\gamma_{3,3}}^{(0)},\Pi_{\gamma_{1,2}}^{(0)},\Pi_{\gamma_{1,1}+\gamma_{3,2}}^{(0)},\Pi_{\gamma_{2,1}+\gamma_{1,2}}^{(0)},\Pi_{\gamma_{3,2}+\gamma_{3,3}}^{(0)}$ and $\Pi_{\gamma_{1,2}+\gamma_{1,3}}^{(0)}$ respectively.}
    \label{fig:A2A3-2ndwc-period-vec}
\end{figure}

\paragraph{The second wall-crossing} The second wall-crossing occurs when $\phi_2-\phi_3$ crosses $\pi/3$ at $t=0.163685...$. The process can be realized by plotting the classical periods, see  Fig.\ref{fig:A2A3-2ndwc-period-vec}. 
The Y-functions after the second wall-crossing are defined by
\begin{equation}
    \begin{aligned}
       Y_{1,3}^{(2)}(\theta)&=Y_{1,3}^{(1)}(\theta)\big(1+\frac{1}{Y_{1,2}^{(1)}(\theta-\frac{\pi i}{3})}\big),\quad Y_{1,2}^{(2)}(\theta)=Y_{1,2}^{(1)}(\theta)\big(1+\frac{1}{Y_{1,3}^{(1)}(\theta+\frac{\pi i}{3})}\big),\\
       Y_{23}^{(2)}(\theta)&=\frac{1+\frac{1}{Y_{1,2}^{(1)}(\theta-\frac{\pi i}{3})}+\frac{1}{Y_{1,3}^{(1)}(\theta)}}{\frac{1}{Y_{1,3}^{(1)}(\theta)Y_{1,2}^{(1)}(\theta-\frac{\pi i}{3})}},\quad Y_{\rm others}^{(2)}(\theta)=Y_{\rm others}^{(1)}(\theta).
    \end{aligned}
\end{equation}
For the vector
\begin{equation}
   \vec{Y}^{(2)}= \mqty(
   Y^{(2)}_{1,1}(-\infty-i\phi_1)\\
   Y^{(2)}_{1,2}(-\infty-i\phi_2)\\
   Y^{(2)}_{1,3}(-\infty-i\phi_3)\\
   Y^{(2)}_{12}(-\infty-i\phi_{12})\\
   Y^{(2)}_{23}(-\infty-i\phi_{23})),
\end{equation}
the connection matrix then becomes
\begin{equation}
    M^{(2)}=\left(
    \begin{array}{ccccc}
     1 & 0 & 0 & 1 & 0 \\
     0 & 1 & 0 & 1 & 1\\
     0 & 0 & 1 & -1 & 1 \\
     1 & 1 & -1 & 1 & 0 \\
     0 & 1 & 1 & 0 & 1 \\
    \end{array}
    \right).
\end{equation}
 We thus can find the constant solutions of new Y-functions and the same value of the effective central charge.

\begin{figure}[hpbt]
    \centering
    \includegraphics[width=10cm]{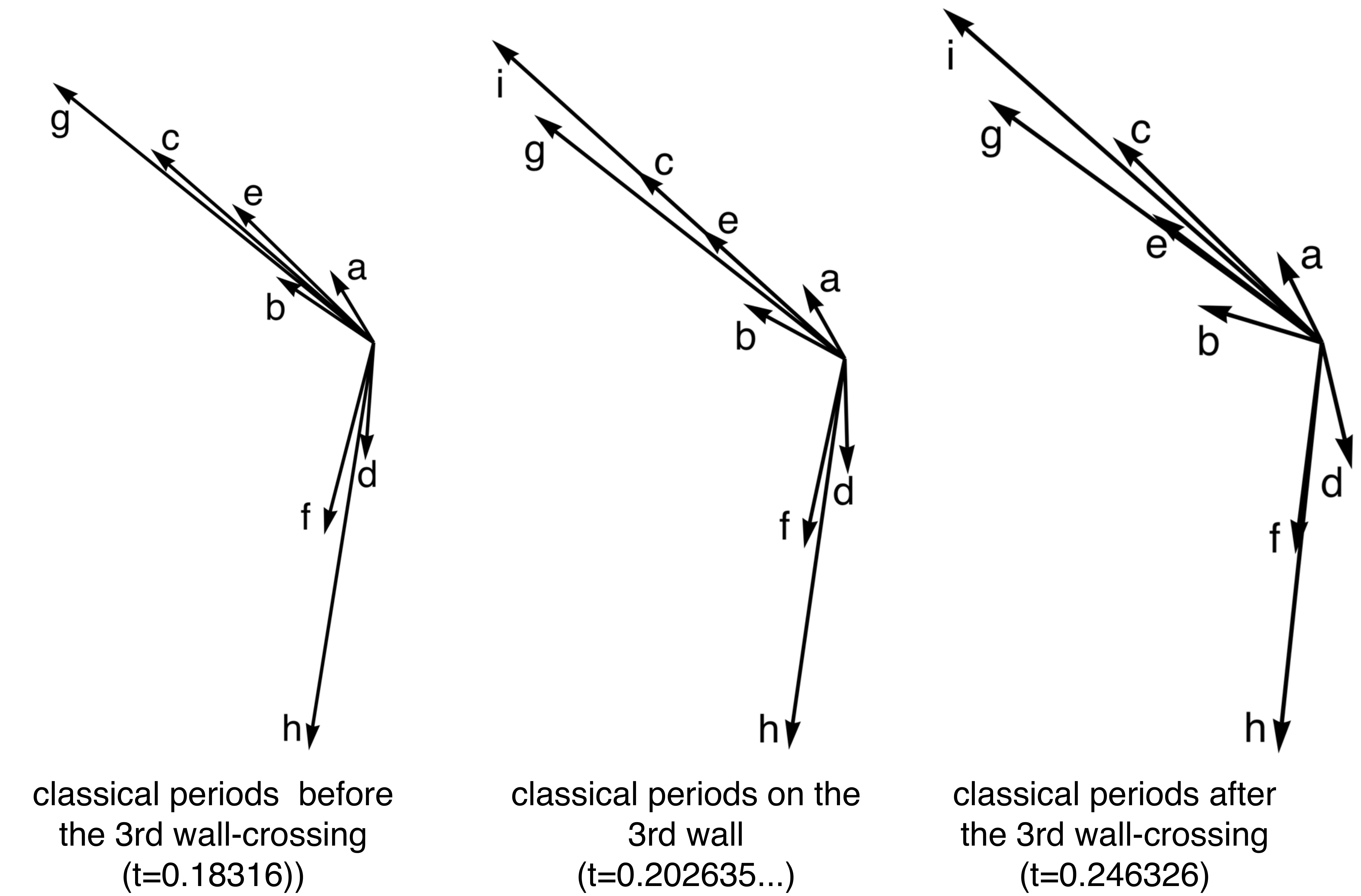}~~~~~~~
    \caption{The classical periods in the 2nd wall-crossing process. The vectors of the classical periods before, middle and after the 2nd wall-crossing are shown in turn. The arrows labeled by $a,b,c,d,e,f,g,h$ and $i$ represent the periods $\Pi_{\gamma_{1,1}}^{(0)},\Pi_{\gamma_{3,2}}^{(0)},\Pi_{\gamma_{3,3}}^{(0)},\Pi_{\gamma_{1,2}}^{(0)},\Pi_{\gamma_{1,1}+\gamma_{3,2}}^{(0)},\Pi_{\gamma_{2,1}+\gamma_{1,2}}^{(0)},\Pi_{\gamma_{3,2}+\gamma_{3,3}}^{(0)},\Pi_{\gamma_{1,2}+\gamma_{1,3}}^{(0)}$ and $\Pi_{\gamma_{1,1}+\gamma_{3,2}+\gamma_{3,3}}^{(0)}$ respectively.}
    \label{fig:A2A3-3rdwc-period-vec}
\end{figure}

\paragraph{The third wall-crossing} After the second wall-crossing, the phase $\phi_{12}-\phi_3+\pi/3$ appear in the kernel of TBA equations. The third wall-crossing occurs when $\phi_{12}-\phi_3+\pi/3$ crosses $\pi/3$ at $t=0.202635...$. The process can be realized by plotting the classical periods, see Fig.\ref{fig:A2A3-3rdwc-period-vec}.
The new functions after the third wall-crossing are
\begin{equation}
    \begin{aligned}
    Y_{1,3}^{(3)}(\theta)&=Y_{1,3}^{(2)}(\theta)\big(1+\frac{1}{Y_{12}^{(2)}(\theta)}\big), \quad Y_{12}^{(3)}(\theta)=Y_{12}^{(2)}(\theta)\big(1+\frac{1}{Y_{1,3}^{(2)}(\theta)}\big),\\
    Y_{312}^{(3)}(\theta)&=\frac{1+\frac{1}{Y_{1,3}^{(2)}(\theta)}+\frac{1}{Y_{12}^{(2)}(\theta)}}{\frac{1}{Y_{1,3}^{(2)}(\theta)Y_{12}^{(2)}(\theta)}},\quad Y_{{\rm others}}^{(3)}(\theta)=Y_{{\rm others}}^{(2)}(\theta).
    \end{aligned}
\end{equation}
For the vector
\begin{equation}
    \vec{Y}^{(3)}=\mqty(
    Y^{(3)}_{1,1}(-\infty-i\phi_1)\\
    Y^{(3)}_{1,2}(-\infty-i\phi_2)\\
    Y^{(3)}_{1,3}(-\infty-i\phi_3)\\
    Y^{(3)}_{12}(-\infty-i\phi_{12})\\
    Y^{(3)}_{23}(-\infty-i\phi_{23})\\
    Y^{(3)}_{312}(-\infty-i\phi_{312})),
\end{equation}
the connection matrix is
\begin{equation}
    M^{(3)}=\left(
    \begin{array}{cccccc}
     1 & 0 & 0 & 1 & 0 & 1 \\
     0 & 1 & 0 & 1 & 1 & 1 \\
     0 & 0 & 1 & 0 & 1 & 1 \\
     1 & 1 & 0 & 1 & 0 & 1 \\
     0 & 1 & 1 & 0 & 1 & 1 \\
     1 & 1 & 1 & 1 & 1 & 1 \\
    \end{array}
    \right).
\end{equation}
 We thus can find the same effective central charge.

In the following, we will only show the definition of the new Y-functions and the connection matrix $M$ for each wall-crossing progress:
\paragraph{The fourth wall-crossing} The fourth wall-crossing occurs when $\phi_2-\phi_1$ crosses $2\pi/3$ at $t=0.290017...$. We thus need to introduce a new Y-function related with $Y_{1,1}^{(3)}$ and $Y_{1,2}^{(3)}$:
\begin{gather}
    \begin{aligned}
        Y_{1,1}^{(4)}(\theta)&=Y_{1,1}^{(3)}(\theta)\big(1+\frac{1}{Y_{1,2}^{(3)}(\theta-\frac{2\pi i}{3})}\big),\quad Y_{1,2}^{(4)}(\theta)=Y_{1,2}^{(3)}(\theta)\big(1+\frac{1}{Y_{1,1}^{(3)}(\theta+\frac{2\pi i}{3})}\big),\\
        Y_{\widetilde{12}}^{(4)}(\theta)&=\frac{1+\frac{1}{Y_{1,2}^{(3)}(\theta-\frac{2\pi i}{3})}+\frac{1}{Y_{1,1}^{(3)}(\theta)}}{\frac{1}{Y_{1,1}^{(3)}(\theta)Y_{1,2}^{(3)}(\theta-\frac{2\pi i}{3})}},\quad Y_{{\rm others}}^{(4)}(\theta)=Y_{{\rm others}}^{(3)}(\theta),
    \end{aligned}\\
    M^{(4)}=\left(
    \begin{array}{ccccccc}
     1 & 1 & 0 & 1 & 0 & 1 & 2 \\
     1 & 1 & 0 & 1 & 1 & 1 & 2 \\
     0 & 0 & 1 & 0 & 1 & 1 & 0 \\
     1 & 1 & 0 & 1 & 0 & 1 & 2 \\
     0 & 1 & 1 & 0 & 1 & 1 & 1 \\
     1 & 1 & 1 & 1 & 1 & 1 & 2 \\
     2 & 2 & 0 & 2 & 1 & 2 & 3 \\
    \end{array}
    \right),\qquad
    \vec{Y}^{(4)}=
    \mqty(
    Y^{(4)}_{1,1}(-\infty-i\phi_1)\\ 
    Y^{(4)}_{1,2}(-\infty-i\phi_2)\\
    Y^{(4)}_{1,3}(-\infty-i\phi_3)\\
    Y^{(4)}_{12}(-\infty-i\phi_{12})\\
    Y^{(4)}_{23}(-\infty-i\phi_{23})\\
    Y^{(4)}_{312}(-\infty-i\phi_{312})\\
    Y^{(4)}_{\widetilde{12}}(-\infty-i\phi_{\widetilde{12}})).
\end{gather}

\paragraph{The fifth wall-crossing} The fifth wall-crossing occurs when $\phi_2-\phi_3$ crosses $2\pi/3$ at $t=0.366924...$, which leads to introduce a new Y-function related with $Y_{1,2}^{(4)}$ and $Y_{1,3}^{(4)}$:
\begin{gather}
    \begin{aligned}
        Y_{1,3}^{(5)}(\theta)&=Y_{1,3}^{(4)}(\theta)\big(1+\frac{1}{Y_{1,2}^{(4)}(\theta-\frac{2\pi i}{3})}\big),\quad Y_{1,2}^{(5)}(\theta)=Y_{1,2}^{(4)}(\theta)\big(1+\frac{1}{Y_{1,3}^{(4)}(\theta+\frac{2\pi i}{3})}\big),\\
        Y_{\widetilde{23}}^{(5)}(\theta)&=\frac{1+\frac{1}{Y_{1,2}^{(4)}(\theta-\frac{2\pi i}{3})}+\frac{1}{Y_{1,3}^{(4)}(\theta)}}{\frac{1}{Y_{1,3}^{(4)}(\theta)Y_{1,2}^{(4)}(\theta-\frac{2\pi i}{3})}},\quad Y_{{\rm others}}^{(5)}(\theta)=Y_{{\rm others}}^{(4)}(\theta),
    \end{aligned}\\
    M^{(5)}=\left(
        \begin{array}{cccccccc}
         1 & 1 & 0 & 1 & 0 & 1 & 2 & 1 \\
         1 & 1 & 1 & 1 & 1 & 1 & 2 & 2 \\
         0 & 1 & 1 & 0 & 1 & 1 & 0 & 2 \\
         1 & 1 & 0 & 1 & 0 & 1 & 2 & 1 \\
         0 & 1 & 1 & 0 & 1 & 1 & 1 & 2 \\
         1 & 1 & 1 & 1 & 1 & 1 & 2 & 2 \\
         2 & 2 & 0 & 2 & 1 & 2 & 3 & 2 \\
         1 & 2 & 2 & 1 & 2 & 2 & 2 & 3 \\
        \end{array}
    \right),\qquad
    \vec{Y}^{(5)}=\mqty(
    Y^{(5)}_{1,1}(-\infty-i\phi_1)\\ 
    Y^{(5)}_{1,2}(-\infty-i\phi_2)\\
    Y^{(5)}_{1,3}(-\infty-i\phi_3)\\
    Y^{(5)}_{12}(-\infty-i\phi_{12})\\
    Y^{(5)}_{23}(-\infty-i\phi_{23})\\
    Y^{(5)}_{312}(-\infty-i\phi_{312})\\
    Y^{(5)}_{\widetilde{12}}(-\infty-i\phi_{\widetilde{12}})\\
    Y^{(5)}_{\widetilde{23}}(-\infty-i\phi_{\widetilde{23}})).
\end{gather}

\paragraph{The sixth wall-crossing} The sixth wall-crossing occurs when $\phi_{\widetilde{12}}-\phi_3$ crosses $0$ at $t=0.434148...$. This process of wall-crossing leads to introduce a new Y-function related with $Y_{\widetilde{12}}^{(5)}$ and $Y_{1,3}^{(5)}$:
\begin{gather}
    \begin{aligned}
        Y_{1,3}^{(6)}(\theta)=&Y_{1,3}^{(5)}(\theta)\big(1+\frac{1}{Y_{\widetilde{12}}^{(5)}(\theta)}\big),\quad Y_{\widetilde{12}}^{(6)}(\theta)=Y_{\widetilde{12}}^{(5)}(\theta)\big(1+\frac{1}{Y_{1,3}^{(5)}(\theta)}\big),\\
        Y_{\widehat{312}}^{(6)}(\theta)=&\frac{1+\frac{1}{Y_{1,3}^{(5)}(\theta)}+\frac{1}{Y_{\widetilde{12}}^{(5)}(\theta)}}{\frac{1}{Y_{1,3}^{(5)}(\theta)}\frac{1}{Y_{\widetilde{12}}^{(5)}(\theta)}},\quad Y_{{\rm others}}^{(6)}(\theta)=Y_{{\rm others}}^{(5)}(\theta),
    \end{aligned}\\
    M^{(6)}=\left(
    \begin{array}{ccccccccc}
     1 & 1 & 0 & 1 & 0 & 1 & 2 & 1 & 2 \\
     1 & 1 & 1 & 1 & 1 & 1 & 2 & 2 & 3 \\
     0 & 1 & 1 & 0 & 1 & 1 & 1 & 2 & 2 \\
     1 & 1 & 0 & 1 & 0 & 1 & 2 & 1 & 2 \\
     0 & 1 & 1 & 0 & 1 & 1 & 1 & 2 & 2 \\
     1 & 1 & 1 & 1 & 1 & 1 & 2 & 2 & 3 \\
     2 & 2 & 1 & 2 & 1 & 2 & 3 & 2 & 4 \\
     1 & 2 & 2 & 1 & 2 & 2 & 2 & 3 & 4 \\
     2 & 3 & 2 & 2 & 2 & 3 & 4 & 4 & 5 \\
    \end{array}
    \right),\qquad
    \vec{Y}^{(6)}=\mqty(
    Y^{(6)}_{1,1}(-\infty-i\phi_1)\\ 
    Y^{(6)}_{1,2}(-\infty-i\phi_2)\\
    Y^{(6)}_{1,3}(-\infty-i\phi_3)\\
    Y^{(6)}_{12}(-\infty-i\phi_{12})\\
    Y^{(6)}_{23}(-\infty-i\phi_{23})\\
    Y^{(6)}_{312}(-\infty-i\phi_{312})\\
    Y^{(6)}_{\widetilde{12}}(-\infty-i\phi_{\widetilde{12}})\\
    Y^{(6)}_{\widetilde{23}}(-\infty-i\phi_{\widetilde{23}})\\
    Y^{(6)}_{\widehat{312}}(-\infty-i\phi_{\widehat{312}})).
\end{gather}

\paragraph{The seventh wall-crossing} The seventh wall-crossing occurs when $\phi_{{312}}-\phi_2$ crosses $-2\pi/3$ at $t=0.449568...$, which leads to new Y-function related with $Y_{312}^{(6)}$ and $Y_{1,2}^{(6)}$:
\begin{gather}
    \begin{aligned}
        Y_{1,2}^{(7)}(\theta)=&Y_{1,2}^{(6)}(\theta)\big(1+\frac{1}{Y_{312}^{(6)}(\theta+\frac{2\pi i}{3})}\big),\quad Y_{312}^{(7)}(\theta)=Y_{312}^{(6)}(\theta)\big(1+\frac{1}{Y_{1,2}^{(6)}(\theta-\frac{2\pi i}{3})}\big),\\Y_{3122}^{(7)}(\theta)=&\frac{1+\frac{1}{Y_{2}^{(6)}(\theta-\frac{2\pi i}{3})}+\frac{1}{Y_{312}^{(6)}(\theta)}}{\frac{1}{Y_{1,2}^{(6)}(\theta-\frac{2\pi i}{3})}\frac{1}{Y_{312}^{(6)}(\theta)}},\quad Y_{{\rm others}}^{(7)}(\theta)=Y_{{\rm others}}^{(6)}(\theta),
    \end{aligned}\\
    M^{(7)}=\left(
    \begin{array}{cccccccccc}
     1 & 1 & 0 & 1 & 0 & 1 & 2 & 1 & 2 & 2 \\
     1 & 1 & 1 & 1 & 1 & 2 & 2 & 2 & 3 & 3 \\
     0 & 1 & 1 & 0 & 1 & 1 & 1 & 2 & 2 & 2 \\
     1 & 1 & 0 & 1 & 0 & 1 & 2 & 1 & 2 & 2 \\
     0 & 1 & 1 & 0 & 1 & 1 & 1 & 2 & 2 & 2 \\
     1 & 2 & 1 & 1 & 1 & 1 & 2 & 2 & 3 & 3 \\
     2 & 2 & 1 & 2 & 1 & 2 & 3 & 2 & 4 & 4 \\
     1 & 2 & 2 & 1 & 2 & 2 & 2 & 3 & 4 & 4 \\
     2 & 3 & 2 & 2 & 2 & 3 & 4 & 4 & 5 & 6 \\
     2 & 3 & 2 & 2 & 2 & 3 & 4 & 4 & 6 & 5 \\
    \end{array}
    \right),\qquad
    \vec{Y}^{(7)}=\mqty(
    Y^{(7)}_{1,1}(-\infty-i\phi_1)\\ 
    Y^{(7)}_{1,2}(-\infty-i\phi_2)\\
    Y^{(7)}_{1,3}(-\infty-i\phi_3)\\
    Y^{(7)}_{12}(-\infty-i\phi_{12})\\
    Y^{(7)}_{23}(-\infty-i\phi_{23})\\
    Y^{(7)}_{312}(-\infty-i\phi_{312})\\
    Y^{(7)}_{\widetilde{12}}(-\infty-i\phi_{\widetilde{12}})\\
    Y^{(7)}_{\widetilde{23}}(-\infty-i\phi_{\widetilde{23}})\\
    Y^{(7)}_{\widehat{312}}(-\infty-i\phi_{\widehat{312}})\\
    Y^{(7)}_{{3122}}(-\infty-i\phi_{3122})).
\end{gather}

\paragraph{The eighth wall-crossing} The eighth wall-crossing occurs when $\phi_{{23}}-\phi_1$ crosses $\pi/3$ at $t=0.608205...$. We introduce new Y-function related with $Y_{23}^{(7)}$ and $Y_{1,1}^{(7)}$:
\begin{gather}
    \begin{aligned}
        Y_{1,1}^{(8)}(\theta)&=Y_{1,1}^{(7)}(\theta)\big(1+\frac{1}{Y_{23}^{(7)}(\theta-\frac{\pi i}{3})}\big),\quad Y_{23}^{(8)}(\theta)=Y_{23}^{(7)}(\theta)\big(1+\frac{1}{Y_{1,1}^{(7)}(\theta+\frac{\pi i}{3})}\big),\\
        Y_{231}^{(8)}(\theta)&=\frac{1+\frac{1}{Y_{23}^{(7)}(\theta-\frac{\pi i}{3})}+\frac{1}{Y_{1,1}^{(7)}(\theta)}}{\frac{1}{Y_{1,1}^{(7)}(\theta)Y_{23}^{(7)}(\theta-\frac{\pi i}{3})}},\quad Y_{{\rm others}}^{(8)}(\theta)=Y_{{\rm others}}^{(7)}(\theta),
    \end{aligned}\\
    M^{(8)}=\left(
    \begin{array}{ccccccccccc}
     1 & 1 & 0 & 1 & 1 & 1 & 2 & 1 & 2 & 2 & 2 \\
     1 & 1 & 1 & 1 & 1 & 2 & 2 & 2 & 3 & 3 & 2 \\
     0 & 1 & 1 & 0 & 1 & 1 & 1 & 2 & 2 & 2 & 1 \\
     1 & 1 & 0 & 1 & 0 & 1 & 2 & 1 & 2 & 2 & 1 \\
     1 & 1 & 1 & 0 & 1 & 1 & 1 & 2 & 2 & 2 & 2 \\
     1 & 2 & 1 & 1 & 1 & 1 & 2 & 2 & 3 & 3 & 2 \\
     2 & 2 & 1 & 2 & 1 & 2 & 3 & 2 & 4 & 4 & 3 \\
     1 & 2 & 2 & 1 & 2 & 2 & 2 & 3 & 4 & 4 & 3 \\
     2 & 3 & 2 & 2 & 2 & 3 & 4 & 4 & 5 & 6 & 4 \\
     2 & 3 & 2 & 2 & 2 & 3 & 4 & 4 & 6 & 5 & 4 \\
     2 & 2 & 1 & 1 & 2 & 2 & 3 & 3 & 4 & 4 & 3 \\
    \end{array}
    \right),\quad
    \vec{Y}^{(8)}=\mqty(
    Y^{(8)}_{1,1}(-\infty-i\phi_1)\\ 
    Y^{(8)}_{1,2}(-\infty-i\phi_2)\\
    Y^{(8)}_{1,3}(-\infty-i\phi_3)\\
    Y^{(8)}_{12}(-\infty-i\phi_{12})\\
    Y^{(8)}_{23}(-\infty-i\phi_{23})\\
    Y^{(8)}_{312}(-\infty-i\phi_{312})\\
    Y^{(8)}_{\widetilde{12}}(-\infty-i\phi_{\widetilde{12}})\\
    Y^{(8)}_{\widetilde{23}}(-\infty-i\phi_{\widetilde{23}})\\
    Y^{(8)}_{\widehat{312}}(-\infty-i\phi_{\widehat{312}})\\
    Y^{(8)}_{{3122}}(-\infty-i\phi_{3122})\\
    Y^{(8)}_{{231}}(-\infty-i\phi_{231})).
\end{gather}

\paragraph{The ninth wall-crossing} The ninth wall-crossing occurs when $\phi_{{12}}-\phi_3$ crosses $\pi/3$ at $t=0.65489...$. One thus needs to introduce new Y-function related with $Y_{12}^{(8)}$ and $Y_{1,3}^{(8)}$:
\begin{gather}
    \begin{aligned}
        Y_{1,3}^{(9)}(\theta)&=Y_{1,3}^{(8)}(\theta)\big(1+\frac{1}{Y_{12}^{(8)}(\theta-\frac{\pi i}{3})}\big),\quad Y_{12}^{(9)}(\theta)=Y_{12}^{(8)}(\theta)\big(1+\frac{1}{Y_{1,3}^{(8)}(\theta+\frac{\pi i}{3})}\big),\\
        Y_{\overline{123}}^{(9)}(\theta)&=\frac{1+\frac{1}{Y_{12}^{(8)}(\theta-\frac{\pi i}{3})}+\frac{1}{Y_{1,3}^{(8)}(\theta)}}{\frac{1}{Y_{1,3}^{(8)}(\theta)Y_{12}^{(8)}(\theta-\frac{\pi i}{3})}},\quad Y_{{\rm others}}^{(9)}(\theta)=Y_{{\rm others}}^{(8)}(\theta),
    \end{aligned}\\
    M^{(9)}=\left(
        \begin{array}{cccccccccccc}
         1 & 1 & 0 & 1 & 1 & 1 & 2 & 1 & 2 & 2 & 2 & 1 \\
         1 & 1 & 1 & 1 & 1 & 2 & 2 & 2 & 3 & 3 & 2 & 2 \\
         0 & 1 & 1 & 1 & 1 & 1 & 1 & 2 & 2 & 2 & 1 & 2 \\
         1 & 1 & 1 & 1 & 0 & 1 & 2 & 1 & 2 & 2 & 1 & 2 \\
         1 & 1 & 1 & 0 & 1 & 1 & 1 & 2 & 2 & 2 & 2 & 1 \\
         1 & 2 & 1 & 1 & 1 & 1 & 2 & 2 & 3 & 3 & 2 & 2 \\
         2 & 2 & 1 & 2 & 1 & 2 & 3 & 2 & 4 & 4 & 3 & 3 \\
         1 & 2 & 2 & 1 & 2 & 2 & 2 & 3 & 4 & 4 & 3 & 3 \\
         2 & 3 & 2 & 2 & 2 & 3 & 4 & 4 & 5 & 6 & 4 & 4 \\
         2 & 3 & 2 & 2 & 2 & 3 & 4 & 4 & 6 & 5 & 4 & 4 \\
         2 & 2 & 1 & 1 & 2 & 2 & 3 & 3 & 4 & 4 & 3 & 2 \\
         1 & 2 & 2 & 2 & 1 & 2 & 3 & 3 & 4 & 4 & 2 & 3 \\
        \end{array}
    \right),\quad
    \vec{Y}^{(9)}=\mqty(
    Y^{(9)}_{1,1}(-\infty-i\phi_1)\\ 
    Y^{(9)}_{1,2}(-\infty-i\phi_2)\\
    Y^{(9)}_{1,3}(-\infty-i\phi_3)\\
    Y^{(9)}_{12}(-\infty-i\phi_{12})\\
    Y^{(9)}_{23}(-\infty-i\phi_{23})\\
    Y^{(9)}_{312}(-\infty-i\phi_{312})\\
    Y^{(9)}_{\widetilde{12}}(-\infty-i\phi_{\widetilde{12}})\\
    Y^{(9)}_{\widetilde{23}}(-\infty-i\phi_{\widetilde{23}})\\
    Y^{(9)}_{\widehat{312}}(-\infty-i\phi_{\widehat{312}})\\
    Y^{(9)}_{{3122}}(-\infty-i\phi_{3122})\\
    Y^{(9)}_{{231}}(-\infty-i\phi_{231})\\
    Y^{(9)}_{\overline{123}}(-\infty-i\phi_{\overline{123}})).
\end{gather}

\paragraph{Monomial potential} At the monomial potential, one finds the identification \eqref{eq:12-to-4-iden}. Under this identification, we find the connection matrix
\begin{equation}
    M^{\mathrm{mono}}=\left(
    \begin{array}{cccc}
     3 & 2 & 6 & 4 \\
     4 & 3 & 8 & 6 \\
     6 & 4 & 11 & 8 \\
     8 & 6 & 16 & 11 \\
    \end{array}
    \right),\quad 
    \vec{Y}^{\mathrm{mono}}=\mqty(
    Y^{(9)}_{1,1}(-\infty-i\phi_1)\\ 
    Y^{(9)}_{1,2}(-\infty-i\phi_2)\\
    Y^{(9)}_{\widetilde{12}}(-\infty-i\phi_{\widetilde{12}})\\
    Y^{(9)}_{\widehat{312}}(-\infty-i\phi_{\widehat{312}})),
\end{equation}
which coincides with $-\frac{1}{2\pi}\tilde{\phi}(k=0)$ in \eqref{eq:ker-til} for $E_6$ under the identification \eqref{eq:E6-folding}.

\section{TBA equations in maximal chamber}\label{sec:TBA-max}
The TBA equations at the maximal chamber for $(A_2, A_3)$ case are
\begin{align}
    \log Y_{1,1}^{(9)}(\theta-i\phi_{1})=&\abs{m_{1,1}}e^{\theta}+K\star\overline{L}_{1,1}^{(9)}-K_{1,2}\star\overline{L}_{1,2}^{(9)}+K_{1,12}^{-}\star\overline{L}_{12}^{(9)}-K_{1,23}^{+}\star\overline{L}_{23}^{(9)}\notag\\
        &+\bigl(K_{1,\widetilde{12}}+K_{1,\widetilde{12}}^{-}\bigr)\star\overline{L}_{\widetilde{12}}^{(9)}+K_{1,\widetilde{23}}^{-}\star\overline{L}_{\widetilde{23}}^{(9)}+K_{1,312}^{-}\star\overline{L}_{312}^{(9)}\notag\\
        &+\bigl(K_{1,\widehat{312}}+K_{1,\widehat{312}}^{-}\bigr)\star\overline{L}_{\widehat{312}}^{(9)}+2K_{1,3122}^{-}\star\overline{L}_{3122}^{(9)}\notag\\
        &+\bigl(K_{1,231}+K_{1,231}^{-}\bigr)\star\overline{L}_{231}^{(9)}+K_{1,\overline{123}}\star\overline{L}_{\overline{123}}^{(9)},\\
    \log Y_{1,2}^{(9)}(\theta-i\phi_{2})=&\abs{m_{1,2}}e^{\theta}-K_{2,1}\star\overline{L}_{1,1}^{(9)}+K\star\overline{L}_{1,2}^{(9)}-K_{2,3}\star\overline{L}_{1,3}^{(9)}\notag\\
        &-K_{2,12}^{-}\star\overline{L}_{12}^{(9)}-K_{2,23}^{-}\star\overline{L}_{23}^{(9)}-\bigl(K_{2,\widetilde{12}}+K_{2,\widetilde{12}}^{-}\bigr)\star\overline{L}_{\widetilde{12}}^{(9)}\notag\\
        &-\bigl(K_{2,\widetilde{23}}+K_{2,\widetilde{23}}^{-}\bigr)\star\overline{L}_{\widetilde{23}}^{(9)}-\bigl(K_{2,312}+K_{2,312}^{-}\bigr)\star\overline{L}_{312}^{(9)}\notag\\
        &-\bigl(2K_{2,\widehat{312}}+K_{2,\widehat{312}}^{-}\bigr)\star\overline{L}_{\widehat{312}}^{(9)}-\bigl(K_{2,3122}+2K_{2,3122}^{-}\bigr)\star\overline{L}_{3122}^{(9)}\notag\\
        &-2K_{2,231}\star\overline{L}_{231}^{(9)}-2K_{2,\overline{123}}\star\overline{L}_{\overline{123}}^{(9)},\\
    \log Y_{1,3}^{(9)}(\theta-i\phi_{3})=&\abs{m_{1,3}}e^{\theta}-K_{3,2}\star\overline{L}_{1,2}^{(9)}+K\star\overline{L}_{1,3}^{(9)}-K_{3,12}^{+}\star\overline{L}_{12}^{(9)}+K_{3,23}^{-}\star\overline{L}_{23}^{(9)}\notag\\
        &+K_{3,\widetilde{12}}^{-}\star\overline{L}_{\widetilde{12}}^{(9)}+\bigl(K_{3,\widetilde{23}}+K_{3,\widetilde{23}}^{-}\bigr)\star\overline{L}_{\widetilde{23}}^{(9)}+K_{3,312}^{-}\star\overline{L}_{312}^{(9)}\notag\\
        &+\bigl(K_{3,\widehat{312}}+K_{3,\widehat{312}}^{-}\bigr)\star\overline{L}_{\widehat{312}}^{(9)}+2K_{3,3122}^{-}\star\overline{L}_{3122}^{(9)}\notag\\
        &+K_{3,231}\star\overline{L}_{231}^{(9)}+\bigl(K_{3,\overline{123}}+K_{3,\overline{123}}^{-}\bigr)\star L_{\overline{123}}^{(9)},\\
    \log Y_{12}^{(9)}(\theta-i\phi_{12})=&\abs{m_{12}}e^{\theta}+K_{12,1}^{+}\star\overline{L}_{1,1}^{(9)}-K_{12,2}^{+}\star\overline{L}_{1,2}^{(9)}-K_{12,3}^{-}\star\overline{L}_{1,3}^{(9)}+K\star\overline{L}_{12}^{(9)}\notag\\
        &+\bigl(K_{12,\widetilde{12}}+K_{12,\widetilde{12}}^{+}\bigr)\star\overline{L}_{\widetilde{12}}^{(9)}+K_{12,\widetilde{23}}^{+}\star\overline{L}_{\widetilde{23}}^{(9)}+K_{12,312}^{+}\star\overline{L}_{312}^{(9)}\notag\\
        &+2K_{12,\widehat{312}}^{+}\star\overline{L}_{\widehat{312}}^{(9)}+\bigl(K_{12,3122}+K_{12,3122}^{+}\bigr)\star\overline{L}_{3122}^{(9)}\notag\\
        &+K_{12,231}^{+}\star\overline{L}_{231}^{(9)}+\bigl(K_{12,\overline{123}}^{+}-K_{12,\overline{123}}^{-}\bigr)\star\overline{L}_{\overline{123}}^{(9)},\\
    \log Y_{23}^{(9)}(\theta-i\phi_{23})=&\abs{m_{23}}e^{\theta}-K_{23,1}^{-}\star\overline{L}_{1,1}^{(9)}-K_{23,2}^{+}\star\overline{L}_{1,2}^{(9)}+K_{23,3}^{+}\star\overline{L}_{1,3}^{(9)}+K\star\overline{L}_{23}^{(9)}\notag\\
        &+K_{23,\widetilde{12}}^{+}\star\overline{L}_{\widetilde{12}}^{(9)}+\bigl(K_{23,\widetilde{23}}^{+}+K_{23,\widetilde{23}}\bigr)\star\overline{L}_{\widetilde{23}}^{(9)}+K_{23,312}^{+}\star\overline{L}_{312}^{(9)}\notag\\
        &+2K_{23,\widehat{312}}^{+}\star\overline{L}_{\widehat{312}}^{(9)}+\bigl(K_{23,3122}+K_{23,3122}^{+}\bigr)\star\overline{L}_{3122}^{(9)}\notag\\
        &+\bigl(K_{23,231}^{+}-K_{23,231}^{-}\bigr)\star\overline{L}_{231}^{(9)}+K_{23,\overline{123}}^{+}\star\overline{L}_{\overline{123}}^{(9)},\\
    \log Y_{312}^{(9)}(\theta-i\phi_{312})=&\abs{m_{312}}e^{\theta}+K_{312,1}^{+}\star\overline{L}_{1,1}^{(9)}-\bigl(K_{312,2}+K_{312,2}^{+}\bigr)\star\overline{L}_{1,2}^{(9)}+K_{312,3}^{+}\star\overline{L}_{1,3}^{(9)}\notag\\
        &+K_{312,12}^{-}\star\overline{L}_{12}^{(9)}+K_{312,23}^{-}\star\overline{L}_{23}^{(9)}+2K_{312,\widetilde{12}}\star\overline{L}_{\widetilde{12}}^{(9)}\notag\\
        &+2K_{312,\widetilde{23}}\star\overline{L}_{\widetilde{23}}^{(9)}+K\star\overline{L}_{312}^{(9)}+\bigl(2K_{312,\widehat{312}}+K_{312,\widehat{312}}^{+}\bigr)\star\overline{L}_{\widehat{312}}^{(9)}\notag\\
        &+\bigl(K_{312,3122}^{-}+2K_{312,3122}\bigr)\star\overline{L}_{3122}^{(9)}+\bigl(K_{312,231}^{+}+K_{312,231}\bigr)\star\overline{L}_{231}^{(9)}\notag\\
        &+\bigl(K_{312,\overline{123}}^{+}+K_{312,\overline{123}}\bigr)\star L_{\overline{123}}^{(9)},\\
    \log Y_{\widetilde{12}}^{(9)}(\theta-i\phi_{\widetilde{12}})=&\abs{m_{\widetilde{12}}}e^{\theta}+\bigl(K_{\widetilde{12},1}+K_{\widetilde{12},1}^{+}\bigr)\star\overline{L}_{1,1}^{(9)}-\bigl(K_{\widetilde{12},2}+K_{\widetilde{12},2}^{+}\bigr)\star\overline{L}_{1,2}^{(9)}\notag\\
        &+K_{\widetilde{12},3}^{+}\star\overline{L}_{1,3}^{(9)}+\bigl(K_{\widetilde{12},12}+K_{\widetilde{12},12}^{-}\bigr)\star\overline{L}_{12}^{(9)}+K_{\widetilde{12},23}^{-}\star\overline{L}_{23}^{(9)}\notag\\
        &+3K(\theta-\theta^{\prime})\star L_{\widetilde{12}}^{(9)}+2K_{\widetilde{12},\widetilde{23}}\star\overline{L}_{\widetilde{23}}^{(9)}+2K_{\widetilde{12},312}\star\overline{L}_{312}^{(9)}\notag\\
        &+\bigl(K_{\widetilde{12},\widehat{312}}^{+}+3K_{\widetilde{12},\widehat{312}}\bigr)\star\overline{L}_{\widehat{312}}^{(9)}+\bigl(K_{\widetilde{12},3122}^{-}+3K_{\widetilde{12},3122}\bigr)\star\overline{L}_{3122}^{(9)}\notag\\
        &+\bigl(2K_{\widetilde{12},231}+K_{\widetilde{12},231}^{+}\bigr)\star\overline{L}_{231}^{(9)}+\bigl(2K_{\widetilde{12},\overline{123}}^{+}+K_{\widetilde{12},\overline{123}}\bigr)\star\overline{L}_{\overline{123}}^{(9)},\\
    \log Y_{\widetilde{23}}^{(9)}(\theta-i\phi_{\widetilde{23}})=&\abs{m_{\widetilde{23}}}e^{\theta}+K_{\widetilde{23},1}^{+}\star\overline{L}_{1,1}^{(9)}-\big(K_{\widetilde{23},2}^{+}+K_{\widetilde{23},2}\big)\star\overline{L}_{1,2}^{(9)}\notag\\
        &+\big(K_{\widetilde{23},3}^{+}+K_{\widetilde{23},3}\big)\star\overline{L}_{1,3}^{(9)}+K_{\widetilde{23},12}^{-}\star\overline{L}_{12}^{(9)}+\big(K_{\widetilde{23},23}+K_{\widetilde{23},23}^{-}\big)\star\overline{L}_{23}^{(9)}\notag\\
        &+2K_{\widetilde{23},\widetilde{12}}\star\overline{L}_{\widetilde{12}}^{(9)}+3K(\theta-\theta^{\prime})\star L_{\widetilde{23}}^{(9)}+2K_{\widetilde{23},312}\star\overline{L}_{312}^{(9)}\notag\\
        &+\big(K_{\widetilde{23},\widehat{312}}^{+}+3K_{\widetilde{23},\widehat{312}}\big)\star\overline{L}_{\widehat{312}}^{(9)}+\big(3K_{\widetilde{23},3122}+K_{\widetilde{23},3122}^{-}\big)\star\overline{L}_{3122}^{(9)}\notag\\
        &+\big(2K_{\widetilde{23},231}^{+}+K_{\widetilde{23},231}\big)\star\overline{L}_{231}^{(9)}+\big(K_{\widetilde{23},\overline{123}}^{+}+2K_{\widetilde{23},\overline{123}}\big)\star\overline{L}_{\overline{123}}^{(9)},\\
    \log Y_{\widehat{312}}^{(9)}(\theta-i\phi_{1})=&\abs{m_{\widehat{312}}}e^{\theta}+\big(K_{\widehat{312},1}+K_{\widehat{312},1}^{+}\big)\star\overline{L}_{1,1}^{(9)}-\big(2K_{\widehat{312},2}+K_{\widehat{312},2}^{+}\big)\star\overline{L}_{1,2}^{(9)}\notag\\
        &+\big(K_{\widehat{312},3}+K_{\widehat{312},3}^{+}\big)\star\overline{L}_{1,3}^{(9)}+2K_{\widehat{312},12}^{-}\star\overline{L}_{12}^{(9)}+2K_{\widehat{312},23}^{-}\star\overline{L}_{23}^{(9)}\notag\\
        &+\big(3K_{\widehat{312},\widetilde{12}}+K_{\widehat{312},\widetilde{12}}^{-}\big)\star\overline{L}_{\widetilde{12}}^{(9)}+\big(3K_{\widehat{312},\widetilde{23}}+K_{\widehat{312},\widetilde{23}}^{-}\big)\star\overline{L}_{\widetilde{23}}^{(9)}\notag\\
        &+\big(2K_{\widehat{312},312}+K_{\widehat{312},312}^{-}\big)\star\overline{L}_{312}^{(9)}+5K(\theta-\theta^{\prime})\star\overline{L}_{\widehat{312}}^{(9)}\notag\\
        &+\big(3K_{\widehat{312},3122}+3K_{\widehat{312},3122}^{-}\big)\star\overline{L}_{3122}^{(9)}+\big(3K_{\widehat{312},231}+K_{\widehat{312},231}^{+}\big)\star\overline{L}_{231}^{(9)}\notag\\
        &+\big(3K_{\widehat{312},\overline{123}}+K_{\widehat{312},\overline{123}}^{+}\big)\star\overline{L}_{\overline{123}}^{(9)},\\
    \log Y_{3122}^{(9)}(\theta-i\phi_{3122})=&\abs{m_{3122}}e^{\theta}+2K_{3122,1}^{+}\star\overline{L}_{1,1}^{(9)}-\big(K_{3122,2}+2K_{3122,2}^{+}\big)\star\overline{L}_{1,2}^{(9)}\notag\\
        &+2K_{3122,3}^{+}\star\overline{L}_{1,3}^{(9)}+\big(K_{3122,12}+K_{3122,12}^{-}\big)\star\overline{L}_{12}^{(9)}\notag\\
        &+\big(K_{3122,23}+K_{3122,23}^{-}\big)\star\overline{L}_{23}^{(9)}+\big(K_{3122,\widetilde{12}}^{+}+3K_{3122,\widetilde{12}}\big)\star\overline{L}_{\widetilde{12}}^{(9)}\notag\\
        &+\big(K_{3122,\widetilde{23}}^{+}+3K_{3122,\widetilde{23}}\big)\star\overline{L}_{\widetilde{23}}^{(9)}+\big(K_{3122,312}^{+}+2K_{3122,312}\big)\star\overline{L}_{312}^{(9)}\notag\\
        &+\big(3K_{3122,\widehat{312}}^{+}+3K_{3122,\widehat{312}}\big)\star\overline{L}_{\widehat{312}}^{(9)}+\big(3K_{3122,231}^{+}+K_{3122,231}\big)\star\overline{L}_{231}^{(9)}\notag\\
        &+5K(\theta-\theta^{\prime})\star\overline{L}_{3122}^{(9)}+\big(3K_{3122,\overline{123}}^{+}+K_{3122,\overline{123}}\big)\star\overline{L}_{\overline{123}}^{(9)},\\
    \log Y_{231}^{(9)}(\theta-i\phi_{231})=&|m_{231}|e^{\theta}+\big(K_{231,1}+K_{231,1}^{+}\big)\star\overline{L}_{1,1}^{(9)}-2K_{231,2}\star\overline{L}_{1,2}^{(9)}\notag\\
        &+K_{231,3}\star\overline{L}_{1,3}^{(9)}+K_{231,12}^{-}\star\overline{L}_{12}^{(9)}+\big(K_{231,23}^{-}-K_{231,23}^{+}\big)\star\overline{L}_{23}^{(9)}\notag\\
        &+\big(K_{231,312}^{-}+K_{231,312}\big)\star\overline{L}_{312}^{(9)}+\big(2K_{231,\widetilde{12}}+K_{231,\widetilde{12}}^{-}\big)\star\overline{L}_{\widetilde{12}}^{(9)}\notag\\
        &+\big(K_{231,\widetilde{23}}+2K_{231,\widetilde{23}}^{-}\big)\star\overline{L}_{\widetilde{23}}^{(9)}+\big(3K_{231,\widehat{312}}+K_{231,\widehat{312}}^{-}\big)\star\overline{L}_{\widehat{312}}^{(9)}\notag\\
        &+\big(3K_{231,3122}^{-}+K_{231,3122}\big)\star\overline{L}_{3122}^{(9)}+3K(\theta-\theta^{\prime})\star\overline{L}_{231}^{(9)}\notag\\
        &+2K_{231,\overline{123}}\star\overline{L}_{\overline{123}}^{(9)},\\
    \log Y_{\overline{123}}^{(9)}(\theta-i\phi_{\overline{123}})=&|m_{\overline{123}}|e^{\theta}+K_{\overline{123},1}\star\overline{L}_{1,1}^{(9)}-2K_{\overline{123},2}\star\overline{L}_{1,2}^{(9)}+\big(K_{\overline{123},3}^{+}+K_{\overline{123},3}\big)\star\overline{L}_{1,3}^{(9)}\notag\\
        &+\big(K_{\overline{123},12}^{-}-K_{\overline{123},12}^{+}\big)\star\overline{L}_{12}^{(9)}+K_{\overline{123},23}^{-}\star\overline{L}_{23}^{(9)}\notag\\
        &+\big(2K_{\overline{123},\widetilde{12}}^{-}+K_{\overline{123},\widetilde{12}}\big)\star\overline{L}_{\widetilde{12}}^{(9)}+\big(2K_{\overline{123},\widetilde{23}}+K_{\overline{123},\widetilde{23}}^{-}\big)\star\overline{L}_{\widetilde{23}}^{(9)}\notag\\
        &+\big(K_{\overline{123},312}^{-}+K_{\overline{123},312}\big)\star\overline{L}_{312}^{(9)}+\big(3K_{\overline{123},\widehat{312}}+K_{\overline{123},\widehat{312}}^{-}\big)\star\overline{L}_{\widehat{312}}^{(9)}\notag\\
        &+\big(3K_{\overline{123},3122}^{-}+K_{\overline{123},3122}\big)\star\overline{L}_{3122}^{(9)}+2K_{\overline{123},231}\star\overline{L}_{231}^{(9)}\notag\\
        &+3K(\theta-\theta^{\prime})\star\overline{L}_{\overline{123}}^{(9)}\,.
\end{align}




\providecommand{\href}[2]{#2}\begingroup\raggedright\endgroup

\end{document}